\documentclass[%
superscriptaddress,
preprint, onecolumn,
preprintnumbers,
amsmath,amssymb,
aps,
floatfix,
longbibliography
]{revtex4-1}

\usepackage{graphicx}
\usepackage{hyperref}
\usepackage{caption}
\usepackage{subcaption}
\usepackage{setspace}
\usepackage{amssymb}
\usepackage{rotating}

\begin{document}


\title{Flow and Deformation due to Periodic Loading in a Soft Porous Material}

\author{Matilde Fiori}
\affiliation{
    Department of Engineering Science, University of Oxford, Oxford, OX1 3PJ, UK }
\author{Satyajit Pramanik}
\affiliation{
    Department of Mathematics, Indian Institute of Technology Guwahati, Guwahati -- 781039, Assam, India }
\author{Christopher W. MacMinn}
\affiliation{
    Department of Engineering Science, University of Oxford, Oxford, OX1 3PJ, UK }
\email{christopher.macminn@eng.ox.ac.uk}

\date{\today}

\begin{abstract}
Soft porous materials, such as biological tissues and soils, are exposed to periodic deformations in a variety of natural and industrial contexts. The detailed flow and mechanics of these deformations have not yet been systematically investigated. Here, we fill this gap by identifying and exploring the complete parameter space associated with periodic deformations in the context of a 1D model problem. We use large-deformation poroelasticity to consider a wide range of loading periods and amplitudes. We identify two distinct mechanical regimes, distinguished by whether the loading period is slow or fast relative to the characteristic poroelastic timescale. We develop analytical solutions for slow loading at any amplitude and for infinitesimal amplitude at any period. We use these analytical solutions and a full numerical solution to explore the localisation of the deformation near the permeable boundary as the period decreases and the emergence of nonlinear effects as the amplitude increases. We show that large deformations lead to asymmetry between the loading and unloading phases of each cycle in terms of the distributions of porosity and fluid flux.
\end{abstract}

\maketitle

\section{Introduction}\label{Intro}

Soft porous materials are common in nature and industry; examples include biological cells and soft tissues, soils and sediments, and paper products and fabrics. In many scenarios, these materials are exposed to periodic loading. For example, soft tissues in the body can experience pulsating loads from the surrounding blood vessels, or, on a larger scale, can be cyclically loaded during their basic mechanical function. The former scenario has attracted great interest recently as a potential driver of transport in brain tissue \cite{Franceschini2006, kedarasetti-fbcns-2022, bojarskaite-natcomms-2023} and the latter is important for load-bearing and transport in cartilage~\cite{ZhangLiHai, riches2002, Mauck2003,Sengers2004, Ferguson2004, Schmidt2010, DiDomenico2017, cacheux-arxiv-2022} and bone~\cite{piekarski-nature-1977, zhang-jmps-1994, MANFREDINI1999, NGUYEN2010, Witt2014}. Periodic loads are also commonly applied in regenerative medicine to improve cell differentiation in scaffolds via mechanotransduction \cite{Mauck2000, Haj2009, grenier2005, Butler2000, GAUVIN2011, Peroglio2018, Kim1999, amrollahi2016}. Soils and sediments experience periodic loading due to seismicity \cite{GENNA1989, Li2004, POPESCU2006648, Bonazzi2021}, vehicle traffic~\cite{HU2011, Ni2015, NI2022}, and ocean waves and tides \cite{yamamoto-jfm-1978, madsen-geotechnique-1978, karim-intjsolidstruct-2002, cheng-book-2016, Trefry2019}.
%
%
%
%
From a poromechanical point of view, periodic loading is fundamentally different from steady loading because the long-time response is inherently oscillatory and therefore time-dependent. Most previous work on periodic loading has focused on internal stress and/or pressure profiles, on macroscopic observables such as surface motion or net inflow or outflow, or on solute concentration profiles.

Periodic loading due to seismicity is a classical topic in poroelasticity~\cite{BiotA,BiotB}. Seismicity involves frequencies that are high enough for inertia to play a dominant role. As a result, seismic response is typically dominated by the propagation of compressional and shear acoustic waves~\cite[\textit{e.g.},][]{BiotA, BiotB, Li2004, Gajo2012, Liu2019}. In contrast, ocean waves and tides are typically associated with a low enough frequency that inertia can be ignored~\cite[\textit{e.g.},][]{yamamoto-jfm-1978,madsen-geotechnique-1978,cheng-book-2016}. Instead, these studies typically focus on the pressure and stress profiles within seabed sediments in response to periodic fluctuations in hydrostatic pressure. The sediment is usually taken to be semi-infinite, the associated deformations are assumed to be small, and the response is often dominated by compressibility. Tidal forcing in coastal aquifers often has similar features~\cite[\textit{e.g.},][]{Trefry2019}.


Our primary motivation here is tissue mechanics, where inertia and compressibility are usually negligible but moderate to large deformations are common. In this regime, poroelasticity is physically diffusive with a characteristic poroelastic relaxation time.
%
%
For bone and cartilage, linear poroelasticity has been used to model the macroscopic mechanical response and/or the distribution of pore pressure during small periodic deformations \citep[\textit{e.g.},][]{zhang-jmps-1994, MANFREDINI1999, riches2002, KAMEO2008, chen2018}. \citet{zhang-jmps-1994} showed that the magnitude and distribution of pore pressure depend strongly on the loading period, introducing the ratio of the loading period to the poroelastic relaxation time as a key dimensionless control parameter. For cartilage and hydrogel scaffolds, both linear and nonlinear poroelasticity have been used to model the impact of periodic deformations on the transport of solutes, typically by comparing the concentration profile at the end of loading across a small set of different loading conditions~\citep[\textit{e.g.},][]{Mauck2003, Ferguson2004, Sengers2004, gardiner2007, Urciuolo2008, ZhangLiHai, Vaughan2013}. Several of these studies noted that faster loading (shorter loading period) is associated with larger fluid velocities that are localised near the surface, whereas slower loading (longer loading period) is associated with intermediate fluid velocities that penetrate more deeply~\cite{gardiner2007, Urciuolo2008, KAMEO2008, DiDomenico2017, Vaughan2013}. \citet{gardiner2007} further noted that, for small deformations, the magnitude of the fluid velocity is roughly proportional to the loading amplitude, whereas the penetration depth is relatively insensitive to amplitude. Despite this extensive previous work, many basic features of flow and deformation due to periodic loading have not yet been systematically studied, in part because most of the above studies have focused on relatively narrow regions of the associated parameter space and/or on relatively small sets of specific results. For example, kinematic and constitutive nonlinearities (characteristic of large deformations and nonlinear constitutive behavior, respectively) become increasingly important as the amplitude grows, but the emergence of these nonlinearities has not been investigated. Moreover, the impact of the deformation on the magnitude and profile of the fluid flux --- directly relevant to the transport of solutes--- have not been examined.

Here, we study the periodic loading of a soft porous material over a wide range of loading periods (from very slow to very fast) and amplitudes (from infinitesimal to moderate/large) in the context of a simple one-dimensional model problem. Following \citet{Macminn2016}, our large-deformation poroelastic model is kinematically rigorous and includes both deformation-dependent permeability and nonlinear elasticity. We characterise the motion of the fluid and the solid throughout the loading cycle, focusing on the evolution of porosity and fluid flux, which are particularly relevant for the transport of solutes. We develop a series of analytical solutions that describe loading with small amplitude but arbitrary period (\S\ref{analytical_early}--\ref{analytical_stokes}) and loading with large period but arbitrary amplitude (\S\ref{analytical_qs}). We then solve the full problem numerically and compare with our analytical results. We use these solutions to examine the transition from very slow loading to very fast loading and from infinitesimal to large amplitudes, as well as the role of the initial porosity. Finally, we discuss the relevance of these results to some specific biological and biomedical scenarios.

\section{Theoretical model}\label{model}

Our theoretical model is based on large-deformation poroelasticity (also referred to as biphasic theory in biomedical communities), here following the development in \citet{Macminn2016}.

\subsection{Model problem}

\begin{figure}[tp]
    \centering
    \includegraphics[width=0.5\textwidth]{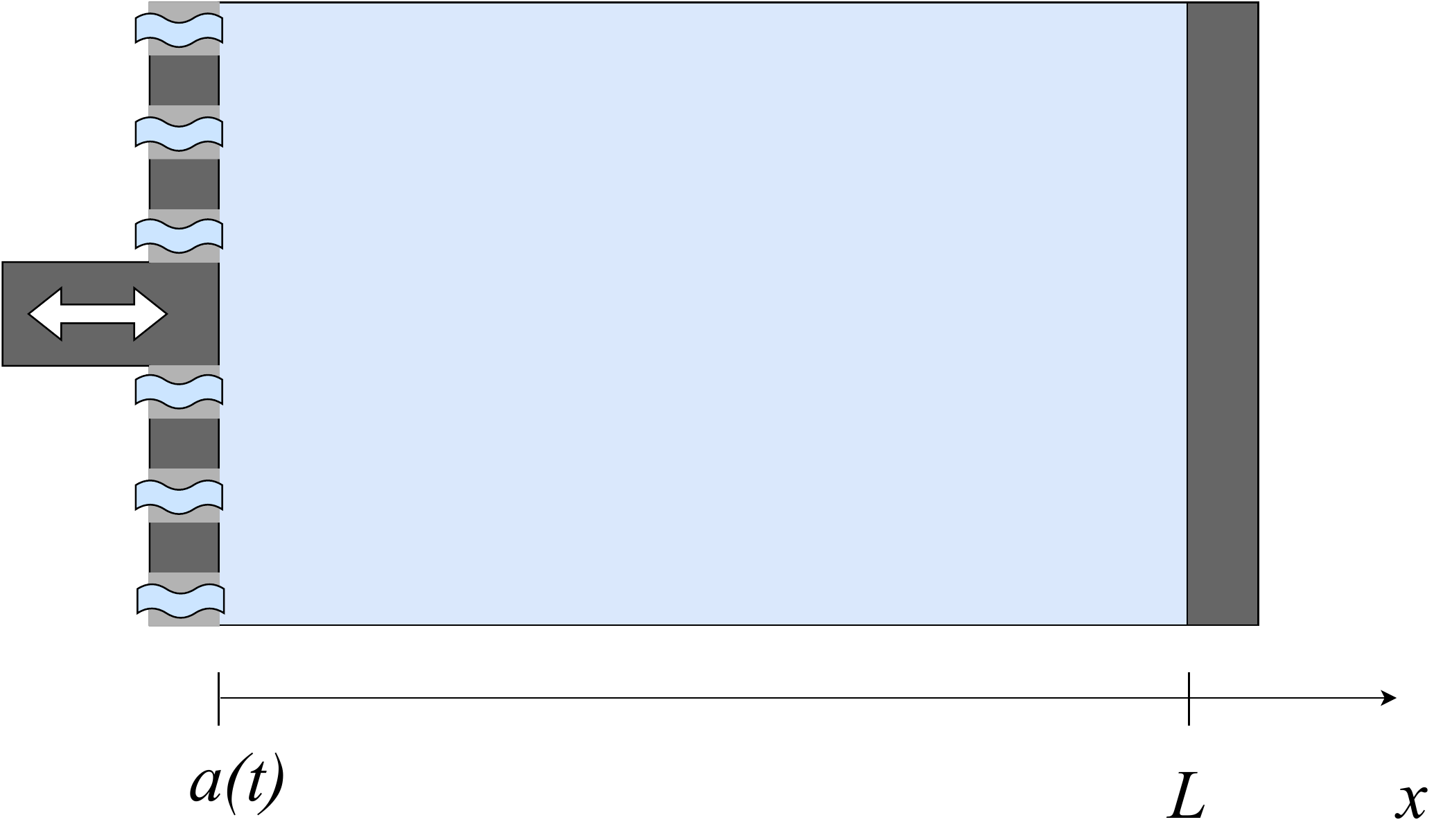}
    \caption{We consider a 1D sample of soft porous material of relaxed length $L$, subject to a periodic, displacement-driven loading at its left boundary (white arrows). The left boundary is permeable, thus allowing fluid flow in or out (blue squiggles) to accommodate the loading. The right boundary is impermeable and fixed in place. \label{system} }
\end{figure}

We consider a one-dimensional sample of soft porous material of relaxed length $L$ and relaxed porosity (fluid fraction) $\phi_{f,0}$. The left boundary of the material is permeable and located at $x=a(t)$ (moving); the right boundary is impermeable and located at $x=L$ (fixed in place) (figure~\ref{system}).
We impose a periodic, displacement-driven loading via the position of the left boundary, 
\begin{equation}\label{eq:a(t)}
    a(t)= \frac{A}{2} \left[1-\cos\left(\frac{2\pi t}{T}\right)\right],
\end{equation}
where $A$ and $T$ are the amplitude and period of the loading, respectively. Macroscopically, the deformation is strictly compressive in the sense that $a(t)\geq{}0$.

We consider deformations ranging from small to large macroscopic nominal strain ($-0.4\%$ to $-20\%$ or $0.004\leq{}A/L\leq{}0.2$), with commensurate macroscopic changes in bulk (total) volume. We assume that the fluid and the solid phases are individually incompressible, so that the total volume of solid is constant and any change in bulk volume must correspond to a change in total pore (fluid) volume via a rearrangement of the pore structure and an influx or an efflux of fluid at the left boundary. 

\subsection{Kinematics}\label{kinematics}

We work in an Eulerian reference frame, in which the solid displacement field is $\mathbf{u_s}=\mathbf{x}-\mathbf{X}(\mathbf{x},t)$, with $\mathbf{X}(\mathbf{x},t)$ the reference position of the material point that at time $t$ occupies the position $\mathbf{x}$. We choose $\mathbf{X}(\mathbf{x},0)\equiv\mathbf{x}$ so that $\mathbf{u_{s}}(\mathbf{x},0)=0$, in which case the reference configuration is the relaxed configuration. We denote the true volume fractions of fluid and solid by $\phi_f$ and $\phi_{s}$, respectively, where $\phi_{f}+\phi_{s}=1$. The flow and deformation are uniaxial, such that
\begin{equation}\label{eq:displacements}
    \mathbf{u_s} = u_s(x,t)\mathbf{\hat{e}_x}, \;
    \mathbf{v_s} = v_s(x,t)\mathbf{\hat{e}_x}, \; 
    \mathbf{v_f} = v_f(x,t)\mathbf{\hat{e}_x}, 
\end{equation} 
where $u_s$, $v_s$, and $v_f$ are the $x$-components of the solid displacement field and the solid and fluid velocity fields, respectively, and $\mathbf{\hat{e}_x}$ is the unit vector in the $x$-direction.

In 1D, the local state of deformation is fully characterised by the Jacobian determinant $J= (1-\partial{u_s}/\partial{x})^{-1}$, which measures the local current volume per unit reference volume. For incompressible constituents and uniform initial porosity $\phi_{f,0}$, the local change in volume is linked to the change in porosity according to
\begin{equation}\label{porosity-def}
    J(x,t) = \frac{1-\phi_{f,0}}{1-\phi_f} \quad\to\quad \frac{\partial{u_s}}{\partial{x}} =\frac{\phi_f-\phi_{f,0}}{1-\phi_{f,0}}.
\end{equation}
Continuity for this 1D system can be written
\begin{equation}\label{continuity}
    \frac{\partial{\phi_f}}{\partial{t}} +\frac{\partial}{\partial{x}}(\phi_f v_f) = 0 \quad\mathrm{and}\quad \frac{\partial{\phi_f}}{\partial{t}} -\frac{\partial}{\partial{x}}{[(1-\phi_f)v_s]} = 0,
\end{equation}
which together imply that the total flux $q=\phi_fv_f+(1-\phi_f)v_s$ is uniform in space, $\partial{q}/\partial{x}=0$.

\subsection{Darcy's law}

We assume that the movement of the fluid relative to the solid skeleton is described by Darcy's law,
\begin{equation}\label{darcylaw}
    \phi_f(v_f-v_s) = -\frac{k(\phi_f)}{\mu}\frac{\partial{p}}{\partial{x}},
\end{equation}
where $k(\phi_f)$ is the permeability of the solid skeleton, $\mu$ is the dynamic viscosity of the fluid, and $p$ is the fluid (pore) pressure, and where we have neglected gravity. We have taken the permeability $k(\phi_f)$ to be a function of porosity only. For simplicity, we use a normalised Kozeny-Carman relation for deformation-dependent permeability \cite{Macminn2016}: 
\begin{equation}\label{kc-permeability}
    k(\phi_f) = k_0\,\frac{(1-\phi_{f,0})^2}{\phi_{f,0}^3}\, \frac{\phi_f^3}{(1-\phi_f)^2},
\end{equation}
where $k_0\equiv{}k(\phi_{f,0})$ is the reference permeability. Kozeny-Carman permeability is a common choice for gels and soft tissues~\cite{Sacco2011, MALANDRINO2014, Rahbari2016, Gao2022}. We compare it with a simpler power-law formulation in Appendix~\ref{appendix-constitutive}, showing that the two are qualitatively and quantitatively similar and are expected to produce similar behavior.

\subsection{Fluid flow}

Combining equations~\eqref{continuity} and \eqref{darcylaw}, we arrive at the nonlinear flow equations:
\begin{equation}\label{conservation-q}
    \frac{\partial{\phi_f}}{\partial{t}} +\frac{\partial}{\partial{x}}\bigg[{\phi_f q}-(1-\phi_f) \frac{k(\phi_f)}{\mu}\frac{\partial{p}}{\partial{x}}\bigg]=0 \quad\mathrm{and}\quad \frac{\partial{q}}{\partial{x}}=0,
\end{equation}
where the total flux $q$ is again
\begin{equation}\label{q-vf-vs}
    q\equiv\phi_f v_f + (1-\phi_f)v_s.
\end{equation}
The fluid velocity and the solid velocity are then given by
\begin{equation}
    v_f =q-\frac{1-\phi_f}{\phi_f}\,\frac{k(\phi_f)}{\mu}\frac{\partial{p}}{\partial{x}} \quad\mathrm{and}\quad v_s=q+\frac{k(\phi_f)}{\mu}\frac{\partial{p}}{\partial{x}}.
\end{equation}
and the local fluid flux is
\begin{equation}\label{flux-qf}
    q_f=\phi_f v_f.
\end{equation}

\subsection{Mechanical equilibrium}

The true Cauchy total stress $\boldsymbol{\sigma}$ is supported jointly by the fluid phase and the solid phase. The total stress can be decomposed into a contribution from the fluid pressure $p$ and a contribution from Terzaghi's effective stress $\boldsymbol{\sigma'}$,
\begin{equation}\label{terzaghi}
    \boldsymbol{\sigma}=\boldsymbol{\sigma'}-p\mathbf{I},
\end{equation}
where we adopt the sign convention of tension being positive. Neglecting inertia and body forces, mechanical equilibrium can be written
\begin{equation}\label{div-sigma}
    \boldsymbol{\nabla}\cdot\boldsymbol{\sigma} = \boldsymbol{\nabla} \cdot\boldsymbol{\sigma'}-\boldsymbol{\nabla}p=0.
\end{equation}
In 1D, equation~\eqref{div-sigma} implies that
\begin{equation}\label{sigma-p}
    \frac{\partial{\sigma'}}{\partial x}=\frac{\partial{p}}{\partial{x}},
\end{equation}
where $\sigma^\prime$ is the $xx$ component of $\boldsymbol{\sigma}^\prime$.

\subsection{Elasticity law}\label{Hencky}

The effective stress is the portion of the total stress that contributes to deformation of the solid skeleton. We take the solid skeleton to be elastic, with no viscous or dissipative behaviours. For confined compression in 1D, as considered here, any elasticity law can be written in the form $\sigma^\prime=\sigma^\prime(\phi_f)$. Thus, equation~\eqref{conservation-q} can be rewritten as a nonlinear advection-diffusion equation:
\begin{equation}\label{eq:conservation-q-Df}
    \frac{\partial{\phi_f}}{\partial{t}} +\frac{\partial}{\partial{x}}\bigg[{\phi_f q}-D_f(\phi_f)\frac{\partial{\phi_f}}{\partial{x}}\bigg]=0 \quad\mathrm{and}\quad \frac{\partial{q}}{\partial{x}}=0,
\end{equation}
where the nonlinear composite constitutive function
\begin{equation}\label{eq:Df}
    D_f(\phi_f)=(1-\phi_f)\frac{k(\phi_f)}{\mu}\frac{\mathrm{d}\sigma^\prime}{\mathrm{d}\phi_f}
\end{equation}
is the poroelastic diffusivity.

Hencky elasticity is a simple nonlinear hyperelasticity model that considers logarithmic strains (``true strains'' or ``Hencky strains'') to capture kinematic nonlinearity \cite{hencky}, and which is commonly used to model soft rubbers and foams~\cite{Hencky33, Anand1979, xiao2002} and sometimes for soft biological tissues \cite{MARCHESSEAU2010185, Fraldi2018}. Hencky elasticity is convenient for our present purposes because (i)~it reduces to linear elasticity for small strains and (ii)~it uses the same two elastic parameters as linear elasticity. It is straightforward to replace Hencky elasticity in the present formulation with a different elastic behavior, such as a Neo-Hookean model, as appropriate for the problem/material of interest. Neo-Hookean elasticity is commonly used as a simple model for soft tissues~(\textit{e.g.} \cite{Ehlers2009, Sengers2004}); in the present context, we expect Hencky elasticity to provide a qualitatively similar mechanical response (see Appendix~\ref{appendix-constitutive}).

For a uniaxial deformation, the relevant component of the effective stress tensor for Hencky elasticity is \cite{Macminn2016, Auton2018}
\begin{equation}\label{sigma-hencky}
    \sigma^\prime= \mathcal{M}\frac{\ln(J)}{J}=\mathcal{M} \bigg(\frac{1-\phi_f}{1-\phi_{f,0}}\bigg) \ln\bigg(\frac{1-\phi_{f,0}}{1-\phi_f}\bigg),
\end{equation}
where $\mathcal{M}$ is the $p$-wave or oedometric modulus. With appropriate initial and boundary conditions, Equations~\eqref{eq:conservation-q-Df}, \eqref{eq:Df}, and \eqref{sigma-hencky} form a closed model for the evolution of the porosity.

\subsection{Initial and boundary conditions}\label{IBCs}

Finally, we specify appropriate initial and boundary conditions for the solid skeleton and for the fluid. As noted above, we locate the left and right boundaries of the solid at $x=a(t)$ and $x=L$, respectively.

\subsubsection{Initial conditions}

Equation~\eqref{eq:a(t)} suggests that $a(0)=0$. The solid is therefore initially relaxed and the initial porosity is uniform and equal to the relaxed porosity,
\begin{equation}\label{ICs}
   u_s(x,0)=0 \quad\mathrm{and}\quad \phi_f(x,0)=\phi_{f,0}.
\end{equation}

\subsubsection{Left boundary}

For $t>0$, we apply a displacement-controlled mechanical loading at the left boundary, which is therefore a moving boundary (see equation~\ref{eq:a(t)}). The associated boundary conditions are
\begin{subequations}\label{bc-left}
    \begin{equation}
        u_s(a,t)=a(t) \quad\mathrm{and}\quad v_s(a,t)=\frac{\mathrm{d}a}{\mathrm{d}t}.
    \end{equation}  
The left boundary is also permeable, so we take
    \begin{equation}
        p(a,t)=0.
    \end{equation}
\end{subequations}

\subsubsection{Right boundary}\label{s:bc-right}

The right boundary is impermeable and fixed in place, such that
\begin{equation}\label{bc-right}
    u_s(L,t)=v_s(L,t)=v_f(L,t)=0.
\end{equation}
This condition and the requirement that $q$ be uniform in space (see the end of \S\ref{kinematics} and equation~\ref{conservation-q}) together imply that $q\equiv0$, meaning that there is no net flow through any cross-section. Equation~\eqref{q-vf-vs} then requires that
\begin{equation}\label{vf-vs}
    v_f= -\frac{(1-\phi_f)}{\phi_f} v_s,
\end{equation}
meaning that the fluid and the solid always locally move in opposite directions.

\subsection{Linear poroelasticity}\label{linear-poro}

For comparison with the fully nonlinear model, we linearise the relations above to arrive at linear poroelasticity, which is valid for infinitesimal deformations, $(\phi_f-\phi_{f,0})/(1-\phi_{f,0})=\partial{u_s}/\partial{x}\ll{}1$. In this limit, equation~\eqref{conservation-q} reduces to the linear-poroelastic diffusion equation,
\begin{equation}\label{conservation-lin}
    \frac{\partial{\phi_f}}{\partial{t}} -\frac{\partial}{\partial{x}}\bigg(D_{f,0}\frac{\partial{\phi_f}}{\partial{x}}\bigg)\approx 0,
\end{equation}
where Hencky elasticity reduces to linear elasticity,
\begin{equation}\label{sigma-linear}
    \sigma^\prime\approx \mathcal{M}\,\frac{\partial{u_s}}{\partial{x}}=\mathcal{M}\left(\frac{\phi_f-\phi_{f,0}}{1-\phi_{f,0}}\right),
\end{equation}
and $D_{f,0}=k_0\mathcal{M}/\mu$ is the constant linear-poroelastic diffusivity. With appropriate initial and boundary conditions, Equation~\eqref{conservation-lin} is a closed linear model for the evolution of the porosity.

In the linear poroelastic model, the initial conditions and the boundary conditions for the right boundary are again equations~\eqref{ICs} and \eqref{bc-right}, respectively. The linearised boundary conditions for the left boundary are
\begin{equation}\label{bc-linear}
    u_s(0,t)\approx a(t) \quad\mathrm{and}\quad v_s(0,t) \approx \frac{\mathrm{d}a}{\mathrm{d}t},
\end{equation}
where $a(t)$ is again given by equation~\eqref{eq:a(t)}. Note that these conditions are linearized relative to equations~\eqref{bc-left} by virtue of being applied at $x=0$ rather than at $x=a(t)$.

\subsection{Scaling and summary}\label{scaling}

We make the above problem dimensionless via the scaling
\begin{equation}
\begin{split}
    \tilde{x}=\frac{x}{L},\;
    \tilde{u}_s=\frac{u_s}{L},\;
    \tilde{t}=\frac{t}{T_{\mathrm{pe}}},\;
    \tilde{\sigma}^\prime=\frac{\sigma'}{\mathcal{M}},\;
    \tilde{p}=\frac{p}{\mathcal{M}},\; \tilde{k}=\frac{k(\phi)}{k_0},\; \tilde{v}_f=\frac{v_f}{L/T_{\mathrm{pe}}}, \; \tilde{v}_s=\frac{v_s}{L/T_{\mathrm{pe}}},
\end{split}
\end{equation}
where $T_{\mathrm{pe}}=L^2/D_{f,0}=\mu{}L^2/(k_0\mathcal{M})$ is the classical poroelastic timescale, which is the characteristic diffusion time for the relaxation of pressure over a distance $L$. Now taking $q\equiv{}0$, as required by the boundary conditions (see \S\ref{s:bc-right}), the nonlinear flow equation can be rewritten in dimensionless form as
\begin{equation}\label{conservation-q-scaled}
    \frac{\partial{\phi_f}}{\partial{\tilde{t}}} -\frac{\partial}{\partial{\tilde{x}}}\bigg[\tilde{D}_f(\phi_f)\frac{\partial{\phi_f}}{\partial{\tilde{x}}}\bigg]=0
\end{equation}
with nonlinear-poroelastic diffusivity
\begin{equation}
    \tilde{D}_f=\frac{D_f}{D_{f,0}}=(1-\phi_f)\tilde{k}(\phi_f)\frac{\mathrm{d}\tilde{\sigma}^\prime}{\mathrm{d}\phi_f},
\end{equation}
elasticity law
\begin{equation}
    \tilde{\sigma}^\prime=\bigg(\frac{1-\phi_f}{1-\phi_{f,0}}\bigg) \ln\bigg(\frac{1-\phi_{f,0}}{1-\phi_f}\bigg),
\end{equation}
initial conditions
\begin{equation}
   \tilde{a}(0)=0, \;
   \phi_f(\tilde{x},0)= \phi_{f,0}, \; 
    \tilde{v}_f(\tilde{x},0)=\tilde{v}_s(\tilde{x},0)=0,
\end{equation}
left boundary conditions
\begin{equation}\label{leftB}
    \tilde{u}_s(\tilde{a},\tilde{t})=\tilde{a}(\tilde{t})= \frac{\tilde{A}}{2} \Bigg[1-\cos\left(\frac{2\pi\tilde{t}}{\tilde{T}}\right)\Bigg]
    \,,\,\,\tilde{v}_s(\tilde{a},\tilde{t})=\frac{\mathrm{d} \tilde{a}}{\mathrm{d}\tilde{t}}\,,\,\,\mathrm{and}\quad \tilde{p}(\tilde{a},\tilde{t})=0,
\end{equation}
and right boundary conditions
\begin{equation}
    \tilde{u}_s(1,\tilde{t}) =\tilde{v}_s(1,\tilde{t}) =\tilde{v}_f(1,\tilde{t})=0,
\end{equation}
where $\tilde{A}=A/L$ and $\tilde{T}=T/T_{\mathrm{pe}}$. We consider only dimensionless quantities below, dropping the tildes for convenience.

The above 1D model describes flow and mechanics in a poroelastic material subject to periodic loading. The kinematics are rigorous and nonlinear, the elasticity law is Hencky elasticity, and the permeability law is the Kozeny-Carman relation. The full and linearised problems share the same three dimensionless control parameters: the dimensionless amplitude and period of the loading, $A$ and $T$, and the relaxed porosity $\phi_{f,0}$.

\section{Analytical solutions}\label{analytical-slns}

We next develop three different analytical solutions to the linear-poroelastic problem, which are valid for small deformations ($A\ll{}1$), and to the full problem for slow deformations ($T\gg{}1$) at any amplitude (\textit{i.e.}, the quasi-static limit). As formulated in \S\ref{linear-poro}, the linear-poroelastic problem implies linearised kinematics, linear elasticity, and constant permeability, and thus a constant and uniform poroelastic diffusivity. We also solve the full problem numerically in general.

\subsection{Average porosity}\label{avg-porosity}

We begin by deriving some basic kinematic results for the average porosity. The macroscopic total volume at any instant is $1-a(t)$, whereas the total volume of solid is constant and equal to $1-\phi_{f,0}$. As a result, the total volume of fluid is $\phi_{f,0}-a(t)$ and the spatially averaged porosity is
\begin{equation}\label{phi_avg}
    \langle \phi_f \rangle (t) = \frac{\phi_{f,0}-a(t)}{1-a(t)}.
\end{equation}
The average of $\langle\phi_f\rangle(t)$ in time over any integer number of loading cycles is then
\begin{equation}
    \langle\overline{\phi_f}\rangle = \frac{1}{mT}\int_{nT}^{(n+m)T} \langle{\phi_f}\rangle \,\mathrm{d}t = 1 - \frac{1-\phi_{f,0}}{\sqrt{1-A}} 
\end{equation}
for any $n\geq0$ and integer $m\geq1$. Note that both the spatial and overall averages are negative because the loading has a nonzero mean (\textit{i.e.}, $\bar{a}=A/2$), so the material is on average compressed.

\subsection{Linear poroelasticity: Early-time solution}\label{analytical_early}

The linear problem posed in section~\ref{linear-poro} can be rewritten as a bounded linear diffusion problem for the displacement. When the loading begins, information about the motion of the left boundary propagates into the domain via poroelastic diffusion. At early times, before this information has had time to reach the right boundary, the response is the same as if the material were semi-infinite in the $x$ direction. The corresponding semi-infinite diffusion problem involves applying the right boundary conditions at $x\to\infty$. This early-time (``$\mathrm{et}$'') solution can be derived via Laplace transform and written as a convolution integral,
\begin{equation}\label{us-et}
    u_{s,\mathrm{et}}(x,t)=\frac{\pi{}A}{T}\int_0^t\,\mathrm{erfc}\left(\frac{x}{2\sqrt{\tau}}\right)\sin\left[\frac{2\pi(t-\tau)}{T}\right]\,\mathrm{d}\tau.
\end{equation}
The corresponding porosity field can be derived from equation~\eqref{us-et} via equation~\eqref{porosity-def}, and is given by
\begin{equation}\label{phi_et}
    {\phi_{f,\mathrm{et}}}(x,t)=\phi_{f,0}-(1-\phi_{f,0})\frac{\pi{}A}{T}\int_0^t\,\frac{1}{\sqrt{\pi\tau}}\exp\left(-\frac{x^2}{4\tau}\right)\sin\left[\frac{2\pi(t-\tau)}{T}\right]\,\mathrm{d}\tau.
\end{equation}
This solution is valid until the deformation spans the domain. Introducing the penetration length $\delta_\mathrm{et}(t)$ as the distance from the left boundary over which the change in porosity has exceeded a threshold, the nature of the problem and the structure of the solution suggest that $\delta_\mathrm{et}(t)\sim{}2\sqrt{t}$. We confirm this reasoning in Appendix~\ref{s:et_and_vf}. Thus, the above solution is valid for $\delta_\mathrm{et}\lesssim{}1\,\to\,t\lesssim{}1/4$.

Having originated from linear poroelasticity, the above solution is limited to small deformations ($A\ll1$). Naively, this solution is valid for any loading period $T$; however, sufficiently small values of $T$ also violate the assumption of small deformations because deformation of size $\sim{}A$ are localised to a region of size $\sim\delta_\mathrm{et}(t)$ at early times. As a result, we expect the maximum strain near the left boundary to be roughly of size $\mathrm{max}[a(t)/\delta_\mathrm{et}(t)]\sim{}A/\sqrt{T}$. More precisely, equation~\eqref{phi_et} can be used to show that the extreme values of ${\phi_{f,\mathrm{et}}}$ will always occur at the left boundary and that the evolution of ${\phi_{f,\mathrm{et}}}$ at the left boundary is given by 
\begin{equation}
    {\phi_{f,\mathrm{et}}}(0,t)=\phi_{f,0}-(1-\phi_{f,0})A\sqrt{\frac{\pi}{T}}\,\,\mathcal{I}(t/T),
\end{equation}
where
\begin{equation}
    \mathcal{I}(y)=C(2\sqrt{y})\sin(2\pi{}y)-S(2\sqrt{y})\cos(2\pi{}y)
\end{equation}
and $C$ and $S$ are the Fresnel cosine and sine integrals, respectively. The extreme values of ${\phi_{f,\mathrm{et}}}$ are then given by
\begin{equation}\label{phi_et_min}
    \mathrm{min}({\phi_{f,\mathrm{et}}}) =\phi_{f,0}-(1-\phi_{f,0})A\sqrt{\frac{\pi}{T}}\,\,\mathcal{I}_\mathrm{max},
\end{equation}
and
\begin{equation}\label{phi_et_max}
    \mathrm{max}({\phi_{f,\mathrm{et}}}) =\phi_{f,0}-(1-\phi_{f,0})A\sqrt{\frac{\pi}{T}}\,\,\mathcal{I}_\mathrm{min},
\end{equation}
where $\mathcal{I}_\mathrm{max}=\mathcal{I}(698/1909)\approx{}0.9491$ and $\mathcal{I}_\mathrm{min}=\mathcal{I}(310/353)\approx{}-0.5406$ are the maximum and minimum values of $\mathcal{I}$, respectively. Strictly, the above solution is non-physical for parameter combinations for which the porosity decreases to 0 or increases to 1, corresponding to $\mathrm{min}({\phi_{f,\mathrm{et}}})=0$ and $\mathrm{max}({\phi_{f,\mathrm{et}}})=1$, respectively. In practise, these solutions become inaccurate for much less extreme parameter combinations as kinematic and constitutive nonlinearities become increasingly important. \citet{hewitt2016} showed that very fast monotonic compression of a soft porous material can lead to extreme localisation near the piston in the form of a ``bloated'' low-porosity boundary layer, the formation and evolution of which depends sensitively on the particular constitutive functions $k(\phi_f)$ and $\sigma^\prime(\phi_f)$. We avoid these extreme loading conditions in the present study, focusing instead on scenarios that are likely to have more biological relevance. The above results confirm that the maximum local strain is indeed of size $A/\sqrt{T}$, suggesting that the above solution and the linear model in general are valid for $A\ll{}\sqrt{T}$. A similar condition can be derived by noting that, in the linear-poroelastic case, the boundary conditions at the left are applied at $x=0$ rather than at $x=a(t)$ (see eq.~\ref{bc-linear}). Thus, the validity of the linear-poroelastic model requires that the error in this linearisation, which is $\sim{}A$, must be negligible relative to the poroelastic diffusion length associated with the deformation, which is $\sim\sqrt{T}$.

\subsection{Linear poroelasticity: Full solution}\label{analytical_sov}

The original bounded linear diffusion problem can be solved analytically via separation of variables. The resulting linear-poroelastic (``$\mathrm{lpe}$'') displacement field is
\begin{equation}\label{us-lpe}
    u_{s,\mathrm{lpe}}(x,t) = a(t)(1-x) - \sum_{n=1}^{\infty} 2A \sin{(n\pi x)} \frac{[2 e^{-n^2\pi^2t}-2\cos{(\frac{2\pi t}{T})}+n^2\pi T \sin{(\frac{2\pi t}{T})}]}{n \pi (4+ n^4 \pi^2 T^2)}.
\end{equation}
The corresponding porosity field can be derived from equation~\eqref{us-lpe} via equation~\eqref{porosity-def} and is given by
\begin{equation}\label{porosity-lpe}
    \phi_{f,\mathrm{lpe}}(x,t)=\phi_{f,0}- (1-\phi_{f,0}) \left\{ a(t)+ \sum_{n=1}^{\infty} 2A \cos{(n\pi x)} \frac{[2 e^{-n^2\pi^2t}-2\cos{(\frac{2\pi t}{T})}+n^2\pi T \sin{(\frac{2\pi t}{T})}]}{ (4+ n^4 \pi^2 T^2)}\right\}.
\end{equation}
The corresponding fluid velocity field can be derived by taking the time derivative of equation~\eqref{us-lpe} to obtain the solid velocity and then using equation~\eqref{vf-vs} to arrive at
\begin{equation}\label{vf-lpe}
    v_{f,\mathrm{lpe}}(x,t)=-\frac{(1-\phi_{f,0})}{\phi_{f,0}}\left\{\dot a(t)(1-x) - \sum_{n=1}^{\infty} 4 A \sin{(n\pi x)} \frac{[ - n^2\pi e^{-n^2\pi^2t}+\frac{2}{T}\sin{(\frac{2\pi t}{T})}+n^2\pi \cos{(\frac{2\pi t}{T})}]}{n  (4+ n^4 \pi^2 T^2)}\right\}.
\end{equation}
Like the early-time solution, this solution provides a good approximation for $A\ll1$ and for $A\ll{}\sqrt{T}$. Also like the early-time solution, this solution provides general insight into the poromechanical response of the system to periodic loading. The expression for ${\phi_{f,\mathrm{lpe}}}$ can be divided into three parts:
\begin{enumerate}
    \item a uniform quasi-static part proportional to $a(t)$, which is the linearised form of the nonlinear quasi-static solution derived below;
    \item an early-time transient that decays exponentially in time at a rate that is independent of $A$ and $T$; and
    \item a periodic forced response with period $T$.
\end{enumerate}
The early-time transient captures the early-time solution derived above, which spans the domain after a time $t\approx{}1/4$ and then decays exponentially relative to the periodic forced response that dominates the solution thereafter. Going forward, we focus on this periodic regime. We show in Appendix~\ref{quasi-steady} that the same reasoning also applies for large deformations.

\subsection{Linear poroelasticity: Response to very fast loading (Stokes' second problem)}\label{analytical_stokes}

As noted above, the response at early times will be confined to a region of size $\sim\sqrt{t}$, spreading diffusively until the entire domain is engaged, at which point the response will evolve exponentially toward its periodic regime. However, the oscillations in the periodic regime will be confined to a region of size $\sim\sqrt{T}$ (or $\sim{}A$ if larger, but recall that linear poroelasticity requires that $A\ll{}\sqrt{T}$). Thus, if the period is sufficiently small (\textit{i.e.}, $\sqrt{T}\ll1$), the material near the piston will oscillate while the far field exists in a state of static compression. This response to very fast loading is well known from Stokes's classical ``second problem'', in which oscillations diffuse into a semi-infinite domain with an amplitude that decays exponentially in space, much like an evanescent wave. The corresponding analytical solution to linear poroelasticity for the periodic response to very fast loading (``$\mathrm{vf}$'') is
\begin{equation}
    u_{s,\mathrm{vf}}(x,t) = \frac{A}{2}\left[ 1 -x -\exp\left(-x\sqrt\frac{\pi}{T}\right)\cos\left(\frac{2\pi{}t}{T}-x\sqrt\frac{\pi}{T}\right)\right]
\end{equation}
and
\begin{equation}
\begin{split}
    \phi_{f,\mathrm{vf}}(x,t) = \phi_{f,0} - (1-\phi_{f,0})\frac{A}{2}\bigg\{ 1-&\sqrt{\frac{\pi}{T}}\exp\left(-x\sqrt\frac{\pi}{T}\right) \\ &\left[\cos\left(\frac{2\pi{}t}{T}-x\sqrt\frac{\pi}{T}\right)-\sin\left(\frac{2\pi{}t}{T}-x\sqrt\frac{\pi}{T}\right)\right]\bigg\},
\end{split}
\end{equation}
where the first two terms in $u_{s,\mathrm{vf}}$ and in $\phi_{f,\mathrm{vf}}$ give the static far-field compression, which is also the (linearised) overall mean compression. This solution confirms that the oscillations will be increasingly localised near the piston as $T$ decreases, featuring near-piston oscillations with an amplitude proportional to $1/\sqrt{T}$ that decay exponentially in space over a characteristic distance $\sqrt{T}$. This solution is illustrated and discussed further in Appendix~\ref{s:et_and_vf}.

\subsection{Response to very slow loading (quasi-static solution)}\label{analytical_qs}

In the full linear-poroelastic solution above, the porosity and displacement fields (equations~\ref{us-lpe} and \ref{porosity-lpe}) converge to the quasi-static limits ${\phi_{f,\mathrm{lpe}}}(x,t) \to \phi_{f,0}-(1-\phi_{f,0})a(t)$ and $u_{s,\mathrm{lpe}}\to{}a(t)(1-x)$ as $T\to\infty$. This limit is a uniform state of strain in which poroelastic transients are fast relative to the loading period, and are therefore negligible. The fully nonlinear problem can be solved analytically in the same limit by taking $\partial{\phi_f}/\partial{t}\to{}0$. The resulting quasi-static (``$\mathrm{qs}$'') solution is
\begin{equation}
    u_{s,\mathrm{qs}}(x,t)= \frac{a(t)}{1-a(t)} (1-x),
\end{equation}
\begin{equation}\label{phi_qs}
    {\phi_{f,\mathrm{qs}}}(t)= \frac{\phi_{f,0}-a(t)}{1-a(t)},
\end{equation}
and
\begin{equation}\label{vf-qs}
    v_{f,\mathrm{qs}}(x,t)= -\left[\frac{1-\phi_{f,0}}{\phi_{f,0}-a(t)}\right]\left\{\frac{1-x}{[1-a(t)]^2}\right\}\dot{a}(t).
\end{equation}
This solution is kinematically exact for arbitrarily large values of $A$, but is only valid for $T\gg{}1$. Note that this expression for $\Delta{\phi_{f,\mathrm{qs}}}$ is the same as the one in equation~\eqref{phi_avg} for $\langle{\phi_f}\rangle(t)$ because the quasi-static porosity is spatially uniform and must therefore be equal to the average porosity.

\subsection{Scaling quantities}

In the absence of a net flow, fluid motion is directly related to changes in porosity. Hence, we present our solutions and results below in terms of the \textit{change} in porosity with respect to the overall average porosity,
\begin{equation}
    \Delta{\phi_f} \equiv  \phi_f - \langle\overline{{\phi_f}}\rangle.
\end{equation}
This mean change in porosity accounts for the non-zero mean compression.

The various results above suggest simple scaling relationships for the magnitudes of $\Delta{\phi_f}$ and $v_f$ in terms of the dimensionless control parameters $A$ and $T$. In particular, the magnitude of $\Delta{\phi_f}$ is captured by the spatially averaged change in porosity at mid-cycle,
\begin{equation}\label{phi_ast}
    \Delta\phi_f^M= \big|\langle{\Delta \phi_f}\rangle(T/2)\big| =\frac{\phi_{f,0}-A}{1-A}-\langle\overline{\phi_f}\rangle.
\end{equation}
The quasi-static solution suggests that appropriate scales for the magnitude of the solid and fluid velocity are
\begin{equation}\label{vs_star}
    v_s^{\ast}=\frac{2A}{T},
\end{equation} 
and
\begin{equation}\label{vf_star}
    v_f^{\ast}= \bigg( \frac{1-\phi_{f,0}}{\phi_{f,0}} \bigg) \frac{2A}{T}.
\end{equation}
An appropriate scale for the fluid flux is therefore
\begin{equation}\label{qf_star}
    q_f^{\ast}= \phi_{f,0} v_f^{\ast}.
\end{equation}

\section{Numerical solution}\label{numerical-vs-analytical}

We solve the full nonlinear problem (\S\ref{scaling}) numerically in \texttt{MATLAB} using a Chebyshev spectral method in space and an implicit Runge-Kutta method in time, as described in more detail in Appendix~\ref{numerical-method}. In Figure~\ref{fig:example}, we illustrate the basic phenomenology of the response of a high-porosity material ($\phi_{f,0}=0.75$) to periodic loading at large amplitude ($A=0.2$) and moderate period ($T=0.3\pi$) for one cycle in the periodic regime.
\begin{figure}[tp]
    \centering
    \includegraphics[width=0.93\textwidth]{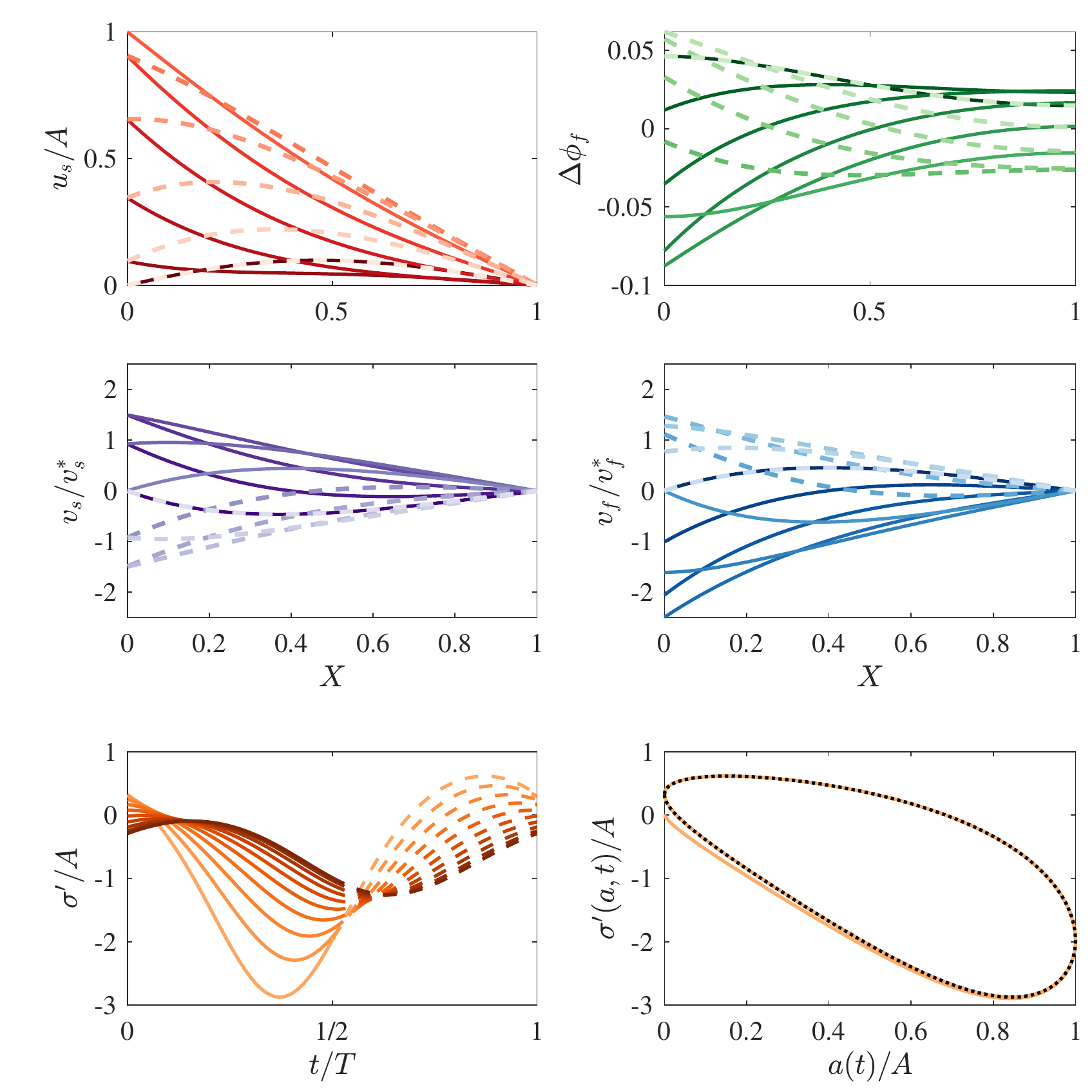}
    \caption{Response of a high-porosity material ($\phi_{f,0}=0.75$) to periodic loading at high amplitude ($A=0.2$) and moderate period ($T=0.3\pi$). In the first two rows, we show the evolution of $u_s$, $\Delta\phi_f$, $v_s$, and $v_f$ (all normalised) at times $t=nT$ to $(n+1)T$ in increments of $0.1T$ (dark to light) during one cycle in the periodic regime, where $n$ is an integer. In the first two rows, we plot all fields against the Lagrangian spatial coordinate $X=x-u_s(x,t)$ for visual clarity. In the left column of the third row, we plot $\sigma^{\prime}$ against $t$ at fixed values of $X$ from $0$ to $1$ (dark to light). In the right column of the third row, we plot a phase portrait of $\sigma^{\prime}(a,t)$ against $a(t)$ for $t=0$ to $t=20T$, emphasising the last cycle (dotted black). In all plots but the lower right panel, we distinguish between the loading half of the cycle ($\dot{a}(t)>0$; solid curves) and the unloading half of the cycle ($\dot{a}(t)<0$; dashed curves). \label{fig:example} }  
\end{figure}

All quantities considered --- displacement, change in porosity, solid velocity, fluid velocity, and effective stress --- are largest in magnitude at the left boundary, where the material is forced, and smallest in magnitude at the right boundary, where the material is stationary, with a magnitude envelope that decreases monotonically from left to right. The displacement and both velocities vanish at the right boundary, as required. Note that these features depend greatly on the boundary conditions for both the solid and the fluid. The flow and deformation will focus toward boundaries where inflow and outflow are permitted, which here is the left side. Reversing the permeability of the two boundaries (\textit{i.e.}, an impermeable moving boundary and a permeable fixed boundary) would instead focus the flow and deformation toward the right side, roughly reversing the spatial profile of $\Delta{\phi_f}$ and producing a much more uniform profile of $v_f$. However, the latter scenario is identical to the present one when viewed from a moving frame that follows the left boundary, $x^\prime=1+a(t)-x$. 

The third row of figure~\ref{fig:example} shows the normalised effective stress $\sigma^{\prime}/A$. The left column shows that the response is out of phase with the loading, with a phase shift that varies with $X$. For example, the stress at the left boundary leads the motion of the left boundary by about $0.15T$ (\textit{i.e.}, the moment of maximum $|\sigma^\prime(a,t)|$ occurs about $0.15T$ before the moment of maximum $a(t)$), whereas the stress at the right boundary lags the motion of the left boundary by a similar amount. In addition, the material near the left boundary experiences strong compression during most of the loading phase ($\sigma^{\prime}<0$ for $\dot{a}>0$) and mild tension during much of unloading ($\sigma^{\prime}>0$ for $\dot{a}<0$). This hysteresis is highlighted by the large area enclosed by the loop in the phase portrait (right column), and it originates in the strong role of viscous dissipation during moderate to fast loading.

Most of the features illustrated in figure~\ref{fig:example} are qualitatively consistent with linear poroelasticity, although the quantitative accuracy of linear poroelasticity depends on the deformation parameters as discussed in the next section. A key qualitative feature introduced by nonlinearity is that the response during loading is not necessarily symmetric with the response during unloading. For example, the minimum values of $\Delta{\phi_f}(a,t)$ and $v_f(a,t)$ are much larger in magnitude than their maximum values and the stress loop is not symmetric about any axis. We explore this asymmetry in more detail in \S\ref{parameter-study}.

\subsection{Comparison with analytical solutions}

\begin{figure}[tp]
    \centering
    \includegraphics[width=0.95\textwidth]{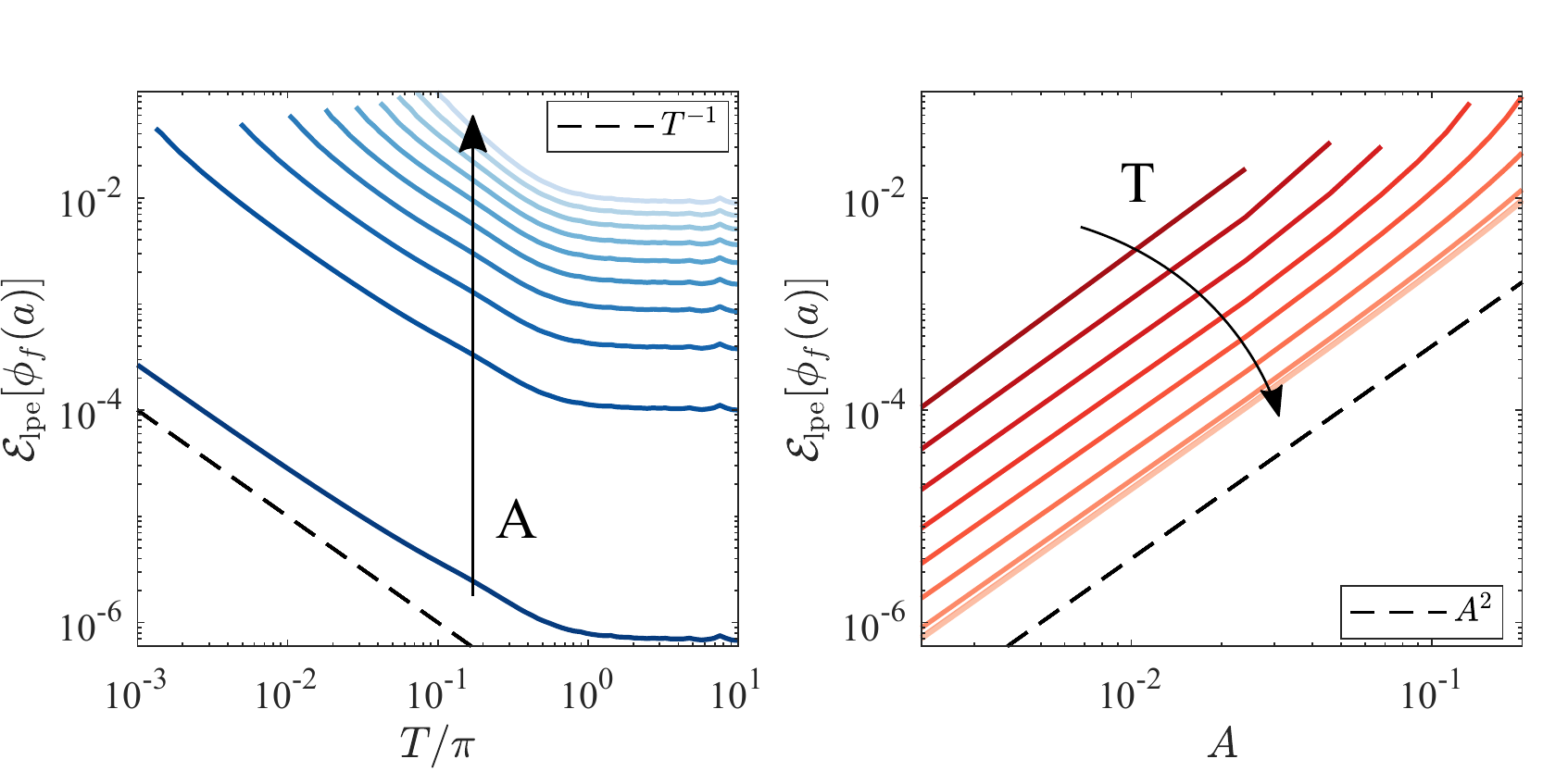}
    \caption{Root-mean-square relative difference between the full numerical solution and the linear-poroelastic analytical solution (from equation \ref{porosity-lpe}) based on $\phi_f(a,t)$ during one cycle in the periodic regime. On the left, we plot the difference against $T$ for fixed values of $A$ ranging from $0.02$ to $0.2$ (dark to light). On the right, we plot the same difference against $A$ for fixed values of $T$ ranging from $0.001\pi$ to $10\pi$ (dark to light). \label{fig:logar_lpe} }
\end{figure}

\begin{figure}[tp]
    \centering
    \includegraphics[width=0.95\textwidth]{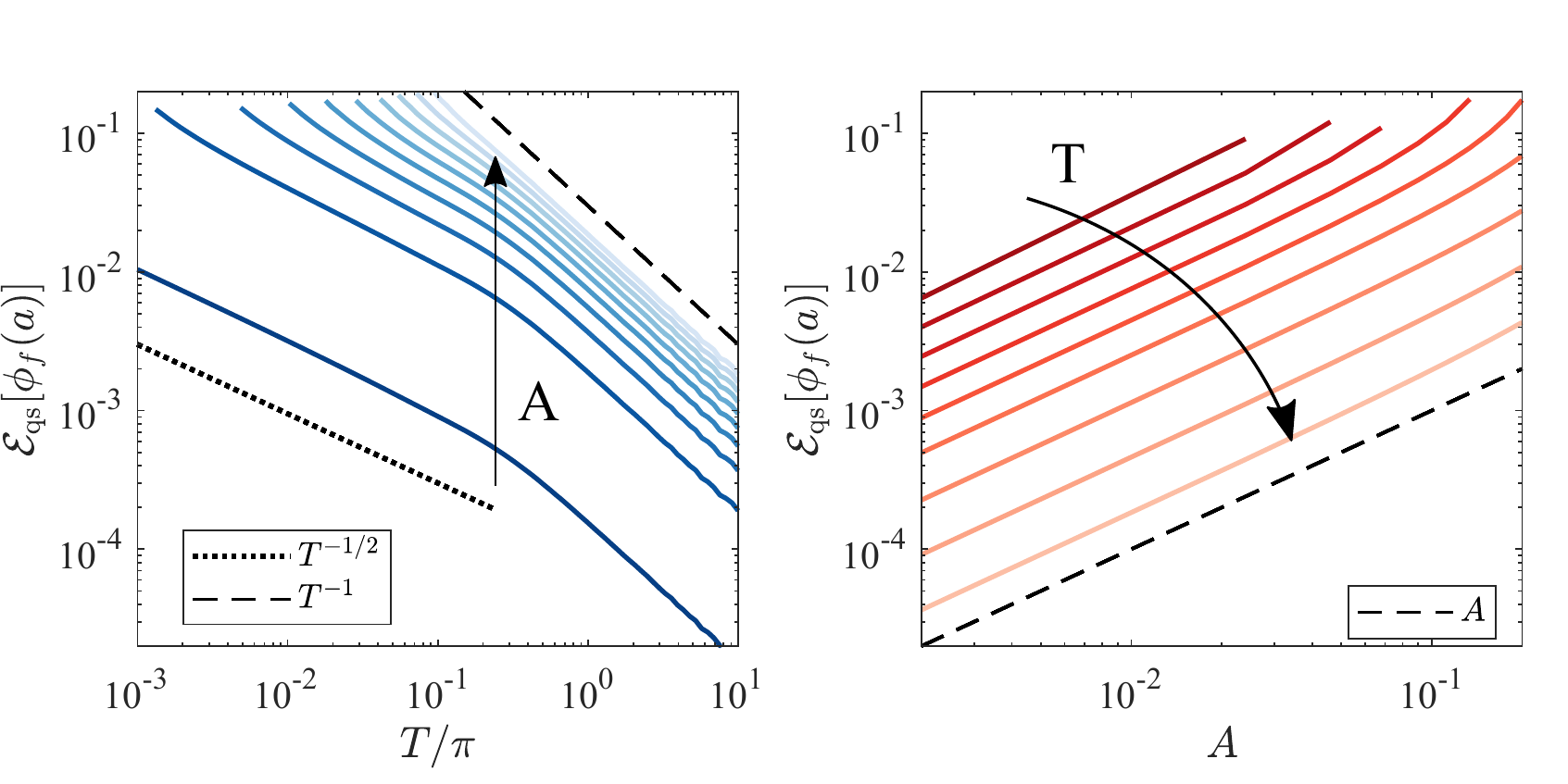}
    \caption{Root-mean-square relative difference between the full numerical solution and the nonlinear quasi-static solution (from equation \ref{phi_qs}) based $\phi_f(a,t)$ during one cycle in the periodic regime. On the left, we plot the difference against $T$ for fixed values of $A$ ranging from $0.02$ to $0.2$ (dark to light). On the right, we plot the same difference against $A$ for fixed values of $T$ ranging from $0.001\pi$ to $10\pi$ (dark to light). \label{fig:logar_qs} }
\end{figure}

We next compare the numerical solution to the linear-poroelastic and nonlinear quasi-static analytical solutions described in \S\ref{analytical-slns}, each of which is appropriate for a specific range of $A$ and $T$. The aim of this comparison is to quantify these ranges of validity and to examine the convergence of the numerical results to each of these special cases. To do so, we calculate all three solutions over a wide range of $A$ and $T$ and then calculate the root-mean-square (RMS) relative difference between the numerical and linear-poroelastic solutions (figure~\ref{fig:logar_lpe}), and between the numerical and nonlinear quasi-static solutions (figure \ref{fig:logar_qs}). We calculate these differences using $\phi_f(a,t)$ during one cycle in the periodic regime. In both figures, we plot these differences against $T$ for several values of $A$ (left panels) and then against $A$ for several values of $T$ (right panels).

As expected, figure~\ref{fig:logar_lpe} shows good agreement between the numerical and linear-poroelastic solutions for small deformations, worsening as $A$ increases. The difference scales with $A^2$ for a given value of $T$ (right panel), as expected from linear poroelasticity, which is first-order in strain. The difference is insensitive to $T$ for $T\gtrsim{}1$, but scales as $T^{-1}$ for faster periods. Decreasing $T$ leads to increasingly strong localisation at the left boundary, and hence increasingly large deformations, even for small $A$, as expected from the early-time and very-fast analyses above (\S\ref{analytical_early}--\ref{analytical_stokes}). We explore this localisation in more detail in \S\ref{parameter-study}.

As expected, figure~\eqref{fig:logar_qs} shows good agreement between the numerical and nonlinear quasi-static solutions for large periods, worsening as $T$ decreases. The difference scales with $T^{-1}$ for $T\gtrsim{}1$ and with $T^{-1/2}$ for shorter periods (left panel); The difference also scales with $A$ (right panel), consistent with the scaling of the non-quasi-static terms in equation~\eqref{porosity-lpe}. Based on these results, we distinguish between ``slow loading'' (SL; $T\lesssim{}0.1\pi$), where spatial variations in porosity are relatively small, and ``fast loading'' (FL; $T\gtrsim{}\pi$), where spatial variations in porosity are relatively large. For very slow loading ($T\gg{}1$), the porosity is uniform and the response is quasi-static (see \S\ref{analytical_qs}). For very fast loading ($T\ll{}1$), the oscillations are localised near the left boundary and the right portion of the material is in a state of static compression (see \S\ref{analytical_stokes}).

\section{Parameter study}\label{parameter-study}

We next examine and compare the poroelastic response for SL and FL as a function of $T$, $A$, and $\phi_{f,0}$. We focus on the evolution of $u_s$, $\phi_f$ and $q_f=\phi_fv_f$ in space and in time, and on the phase behavior of $\sigma^\prime(a,t)$. We conclude by considering the parameter ranges that would be relevant to various biological examples.

\subsection{Impact of loading period}\label{mechanical-fields}

\begin{figure}[tp]
    \centering
    \includegraphics[width=0.9\textwidth]{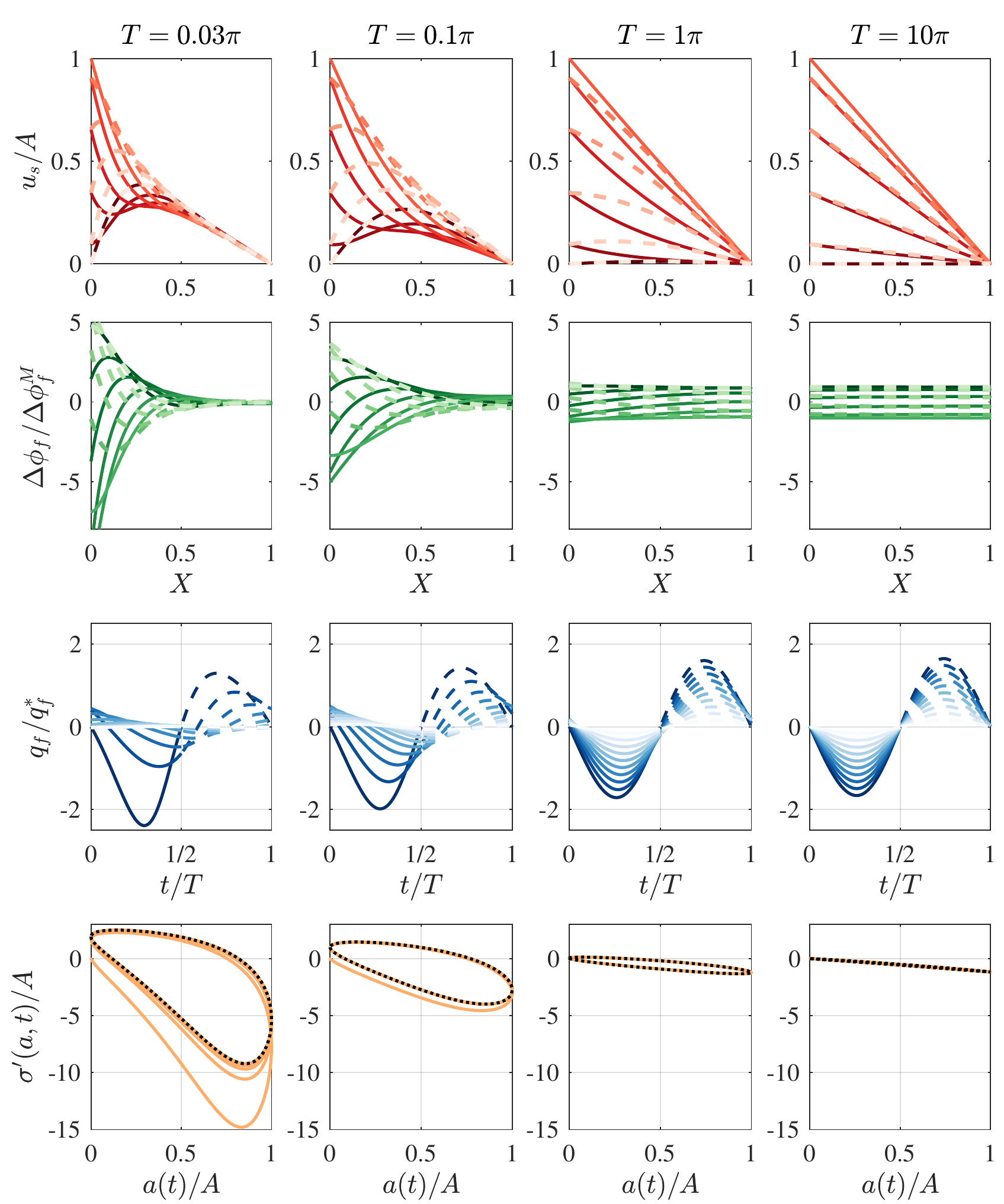}
    \caption{Evolution of normalised $u_s$ (first row) and $\Delta{\phi_f}$ (second row) at times $t=nT$ to $(n+1)T$ in increments of $0.1T$ (dark to light), where $n$ is an integer, during one cycle in the periodic regime for $A=0.1$, $\phi_{f,0}=0.75$, and $T=0.03\pi$ (first column), $0.1\pi$ (second column), $\pi$ (third column), and $10\pi$ (fourth column). As in Figure~\ref{fig:example}, we plot all fields in the first two rows against the Lagrangian spatial coordinate $X=x-u_s(x,t)$ for clarity. In the third row, we plot normalised $q_f$ against $t$ for ten different values of $X$ from $0$ to $1$ (dark to light). We distinguish between the loading half of the cycle ($\dot{a}>0$; solid curves) and the unloading half of the cycle ($\dot{a}<0$; dashed curves). 
    In the last row, plot phase portraits of $\sigma^\prime(a,t)$ against $a(t)$ from $t=0$ to $t=20T$, emphasising the last cycle (dotted black).
    \label{fig:displ-vel-different-T} }
\end{figure}

To visualise the distinct poromechanical responses for SL and FL, and the transition between them, we plot $u_s$, $\Delta{\phi_f}$, and $q_f$ (all normalised) over one cycle in the periodic regime for $\phi_{f,0}=0.75$, $A=0.1$, and four different values of $T$ --- two for SL (left two columns) and two for FL (right two columns) (figure \ref{fig:displ-vel-different-T}).

The first row of figure~\ref{fig:displ-vel-different-T} shows that, for FL (\textit{i.e.}, $T=0.03\pi$ and $0.1\pi$), the displacement is highly nonlinear in $X$ (and in $x$), with substantial differences between the loading and unloading phases. As $T$ increases, transitioning into SL (\textit{i.e.}, $T=1\pi$ and $10\pi$), the displacement is increasingly linear in $X$ and converges toward the quasi-static solution, which is fully determined by the instantaneous value of $a$ and is thus symmetric between loading and unloading. For all values of $T$, the displacement is of characteristic size $A$ at the left and vanishes at the right.

For SL, $\Delta{\phi_f}$ is uniform in space and varies only in time, per the quasi-static solution. The material is in a uniform state of compression, fully determined by the instantaneous value of $a$ and thus symmetric between loading and unloading and fully relaxed at the beginning/end of each cycle. As $T$ decreases, transitioning into FL, $\Delta{\phi_f}$ becomes increasingly localised near the left boundary and also increasingly asymmetric between loading and unloading (figure~\ref{fig:displ-vel-different-T}, second row, left column). For FL, the material experiences a substantial amount of tension in the left portion of the domain during unloading, despite the overall mean compression. Tension emerges for FL because this regime is, by definition, one where the rate of loading is much faster than the poroelastic relaxation time, so the left boundary must \textit{pull} the material to the left during unloading. The right portion of the domain experiences an overall more limited range of porosities and remains compressed throughout the cycle, never reaching a state of tension or even full unloading. The latter feature is also visible in the corresponding displacement fields.

The fluid flux $q_f=\phi_fv_f$ is particularly relevant to the transport of solutes since it drives advection. Fluid leaves the domain during the loading phase of the cycle ($q_f(x=a,t)=q_f(X=0,t)<0$ when $\dot{a}>0$) and enters the domain during unloading. For SL, $q_f$ is entirely in-phase with the loading, but opposite in sign. For FL, the peak value of $q_f$ in the interior exhibits a lag relative to the peak value of $\dot{a}$, and this lag increases with $x$. Note that the fluid flux is orders of magnitude larger for FL than for SL because the rate of loading is orders of magnitude faster, but this variation is largely scaled out by normalisation with $q_f^\ast(\phi_{f,0},A,T)\propto{}T^{-1}$, per equation~\eqref{qf_star} (figure~\ref{fig:displ-vel-different-T}, third row).

The extreme values of $q_f$ at the left boundary are larger in loading than in unloading, but progressively more symmetric as $T$ increases. This asymmetry is a result of the kinematic and constitutive nonlinearity of large deformations. We show in Appendix~\ref{appendix-mechanics-claws} that this asymmetry originates in the nonlinear kinematics of large deformations, meaning that it emerges from the nonlinear model during large deformations even with linear elasticity and constant permeability. This asymmetry is then strongly amplified by deformation-dependent permeability (relative to constant permeability) and slightly suppressed by Hencky elasticity (relative to linear elasticity). The latter occurs because Hencky elasticity stiffens in compression, resulting in a larger poroelastic diffusivity and therefore weaker localisation during loading (see also Appendix~\ref{appendix-constitutive}).

The asymmetry between loading and unloading is also highlighted by the evolution of $\sigma^{\prime}(a,t)$ (figure \ref{fig:displ-vel-different-T}, fourth row). For FL, $\sigma^\prime(a,t)$ exhibits hysteresis: for the same value of $a(t)$, the value of $\sigma^\prime(a,t)$ is considerably higher in magnitude during loading than during unloading (and tensile during much of the unloading phase). This hysteresis decreases as $T$ increases, such that, for SL, $\sigma^\prime(a,t)$ is modest in magnitude, fully compressive, and symmetric between loading and unloading (\textit{i.e.}, non-hysteretic); as a result of the latter, the phase portrait for SL is a single curve (rather than a loop). These phase portraits also illustrate the convergence to the periodic regime: in all cases, the overall response grows less extreme as the transient component decays exponentially over the first $\sim{}T^{-1}$ cycles.

\begin{figure}[tp]
    \includegraphics[width=1\textwidth]{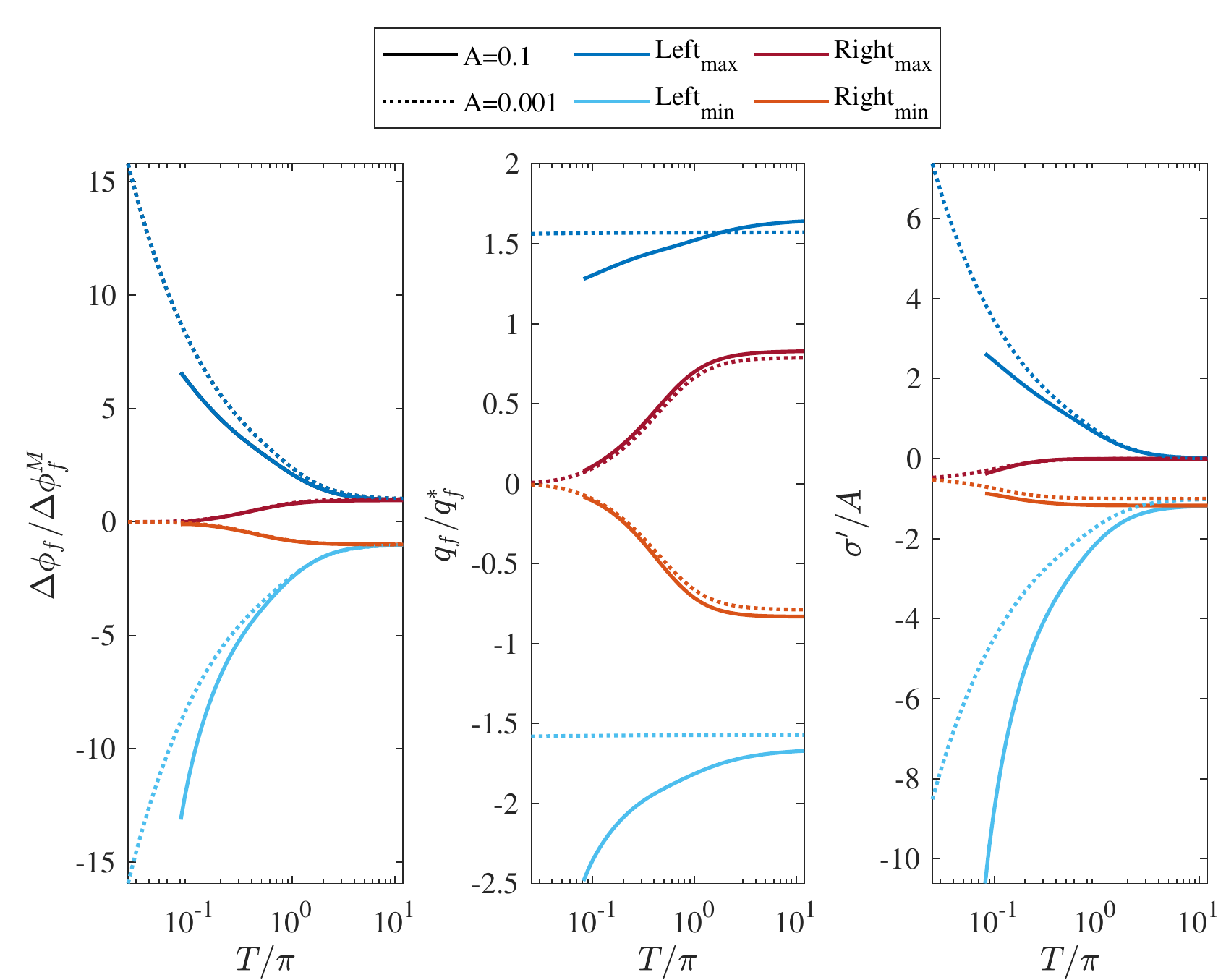}
    \caption{We illustrate the transition from FL to SL as $T$ increases by plotting the normalised maximum and minimum values $\Delta{\phi_f}$ (first column), $q_f$ (second column), and $\sigma^\prime$ (third column) against $T$ for $A=0.1$ (solid lines) and $A=0.001$ (dotted lines) during the periodic regime. We show the maximum values (dark colours) and minimum values (light colours) of all three quantities at the left boundary ($X=0$, blue curves), of $\Delta{\phi_f}$ and $\sigma^\prime$ at the right boundary ($X=1$, red curves), and of $q_f$ at the material midpoint ($X=1/2$, red curves).
    \label{fig:localisation-t} }
\end{figure}

Since a key difference between SL and FL is that the deformation is uniformly distributed in the former and localised toward the left in the latter, we study the emergence of this localisation to further quantify the transition from SL to FL. In figure~\ref{fig:localisation-t}, we plot the normalised maximum and minimum values of $\Delta{\phi_f}$, $q_f$, and $\sigma^\prime$ at the left boundary ($X=0$) and then either at the right boundary ($X=1$) for $\Delta{\phi_f}$ and $\sigma^\prime$ or at the material mid-point ($X=1/2$) for $q_f$. The latter is necessary because $q_f$ vanishes at $X=1$. We consider a range of periods $T$ at two fixed values of $A$ for $\phi_{f,0}=0.75$.

Figure~\ref{fig:localisation-t} shows that, for SL, the deformation is uniform, varying only in time. The maxima and minima of both $\Delta{\phi_f}$ and $\sigma^{\prime}$ remain separate, but their left and right values converge. The flux $q_f$ remains linear in space, even for SL (see \S\ref{analytical_qs}). As $T$ decreases, the deformation is progressively localised at the left, such that the right is increasingly static. At the right, the maximum and minimum values of $\Delta{\phi_f}$ and $q_f$ converge to zero while the maximum and minimum values of $\sigma^\prime$ converge to weak compression, as also illustrated in figure~\ref{fig:displ-vel-different-T}.

We find that a small-amplitude deformation ($A=0.001$)  and a large-amplitude deformation ($A=0.1$) exhibit qualitatively similar features. However, large deformations lead to an increasingly strong asymmetry between the maxima and minima of all quantities at the left boundary as $T$ decreases. In particular, the curves are biased downward, such with higher magnitudes reached in loading (minima) than in unloading (maxima). This asymmetry is does not occur for small deformations, for which the maxima and minimia of all quantities are symmetric in magnitude. Note also that the maxima of $\Delta{\phi_f}$ and $\sigma^{\prime}$ increase as $T$ decreases for both values of $A$, whereas the maximum of $q_f$ is independent of $T$ for small deformations but decreases with $T$ for large deformations, consistent with figure~\ref{fig:displ-vel-different-T}.

\subsection{Impact of loading amplitude}\label{impact_A}

\begin{figure}[tp]
    \centering
    \includegraphics[width=0.95\textwidth]{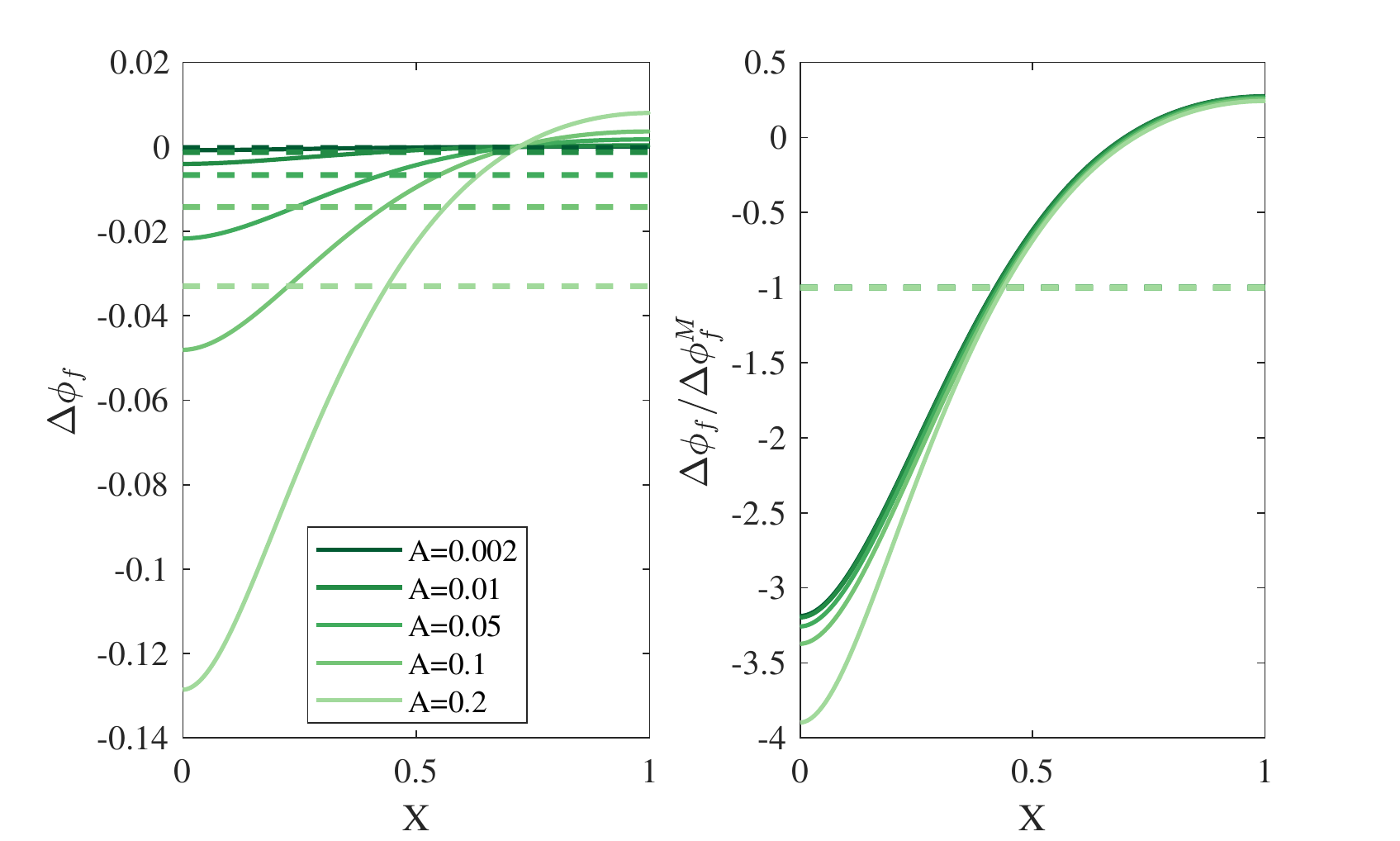}
    \caption{Profiles of $\Delta{\phi_f}$ at mid-cycle during the periodic regime for $\phi_{f,0}=0.75$ and for five values of $A$ ranging from small to large deformations (dark to light), each for two different periods: $T=0.1\pi$ (FL; solid) and $T=10\pi$ (SL; dashed). The left and right panels are non-normalised and normalised, respectively, showing that this normalisation captures the leading-order impact of $A$ on $\Delta{\phi_f}$. \label{fig:A_impact} }
\end{figure}

We next examine the impact of loading amplitude $A$ on the poromechanical response. We first assess the interaction between amplitude and period by plotting $\Delta{\phi_f}$ against $X$ at mid-cycle for two different values of $T$ (one for SL and one for FL) and for five different values of $A$, ranging from small to large deformations ($A=0.002$ to $0.2$; fig.~\ref{fig:A_impact}).
The left panel in figure~\ref{fig:A_impact} shows that, for SL, the non-normalised value of $\Delta{\phi_f}$ is uniform in $X$ and increases with $A$, as expected. For FL, the non-normalised value of $\Delta{\phi_f}$ is increasingly non-uniform with $X$ as $A$ increases. Relative to the SL case, the deformation is increasingly amplified at the left boundary and suppressed at the right boundary (recall that the mean value of $\Delta{\phi_f}$ is independent of $T$). The right panel shows the same results, but now normalised by $\Delta \phi_f^M$ (eq.~\ref{phi_ast}). By definition, all of the SL curves collapse onto $\Delta{\phi_f}/\Delta \phi_f^M=-1$ ($\Delta \phi_f^M$ is the magnitude of the quasi-static change in porosity at mid-cycle). The FL curves also nearly collapse onto a master curve, suggesting that this normalisation captures the leading-order impact of $A$, even for large deformations for FL. The largest deviations from this collapse are near the left boundary, where nonlinearities are particularly pronounced.

\begin{figure}[tp]
    \centering
    \includegraphics[width=0.95\textwidth]{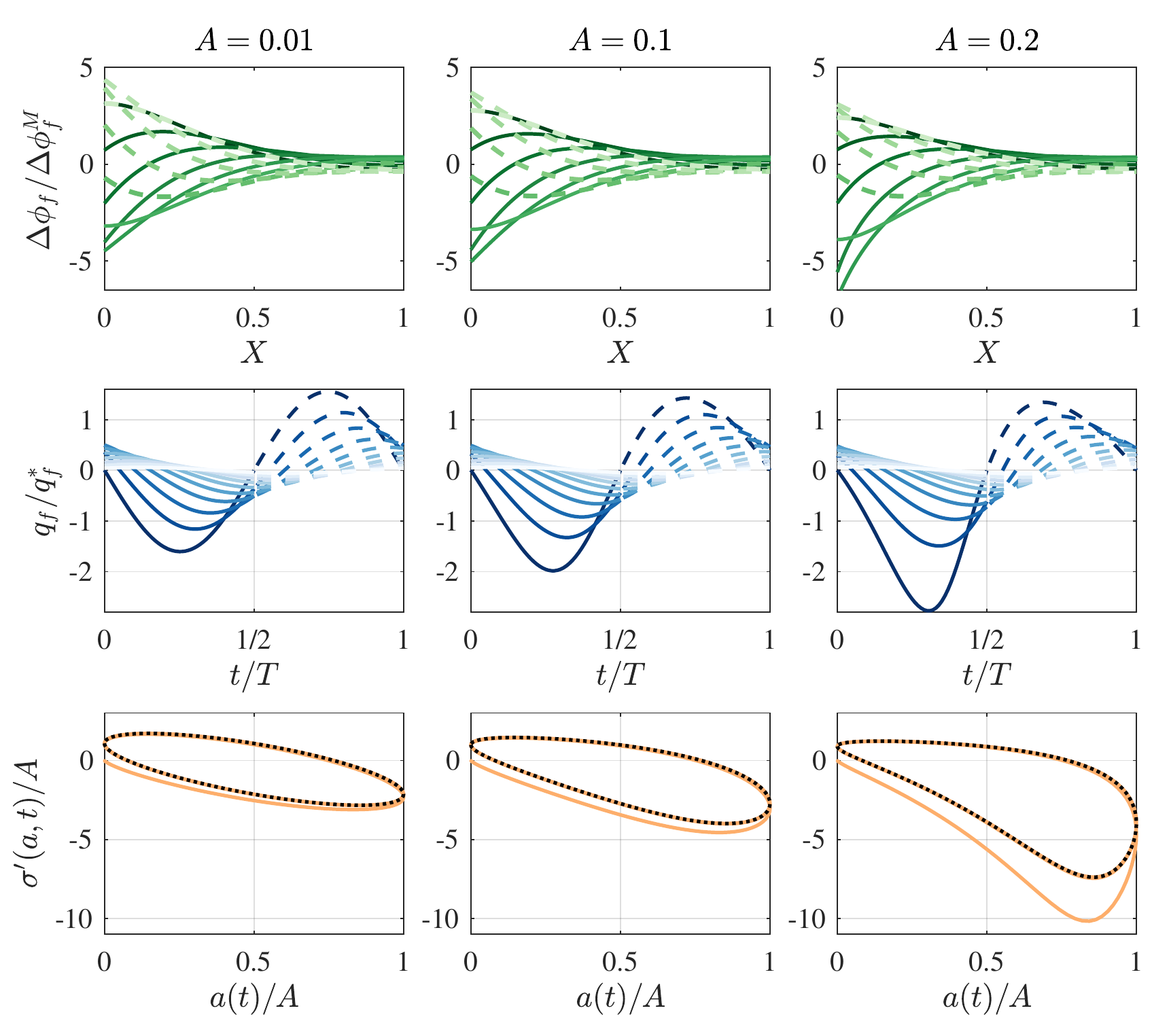}
    \caption{Evolution of normalised $\Delta{\phi_f}$, $q_f$, and $\sigma^\prime(a,t)$ for $\phi_{f,0}=0.75$, $T=0.1\pi$, and $A=0.01$ (first column), $0.1$ (second column), and $0.2$ (third column). In the first row, we plot $\Delta{\phi_f}$ against the Lagrangian spatial coordinate $X=x-u_s$ at times $t=nT$ to $(n+1)T$ in increments of $0.1T$ (dark to light) during one cycle in the periodic regime. In the second row, we plot $q_f$ against $t$ for ten different values of $X$ from $0$ to $1$ (dark to light); we distinguish between loading ($\dot{a}>0$; solid curves) and unloading ($\dot{a}<0$; dashed curves). In the last row, we plot phase portraits of $\sigma^\prime(a,t)$ against $a(t)$ from $t=0$ to $t=20T$, emphasising the last cycle (dotted black lines). \label{fig:por-vel-differentA} }
\end{figure}

In figure~\ref{fig:por-vel-differentA}, we focus on FL by fixing $T=0.1\pi$ and plotting the evolution of $\Delta{\phi_f}$, $q_f$, and $\sigma^\prime(a,t)$ for three different values of $A$, ranging from small to large deformations.
We find that increasing $A$ amplifies the asymmetry in the extreme values of $\Delta{\phi_f}$ and $q_f$ at the left boundary in loading and unloading, as noted in figure~\ref{fig:localisation-t}. The normalisation captures the leading-order impact of $A$ on all of these quantities, as noted in figure~\ref{fig:A_impact}; the non-normalised values of all three quantities would vary by two orders of magnitude across this range of $A$. As $A$ increases, $\sigma^\prime(a,t)$ exhibits more hysteresis and larger normalised magnitudes, in accordance with the amplified asymmetry.

\begin{figure}[tp]
    \centering
    \includegraphics[width=0.95\textwidth]{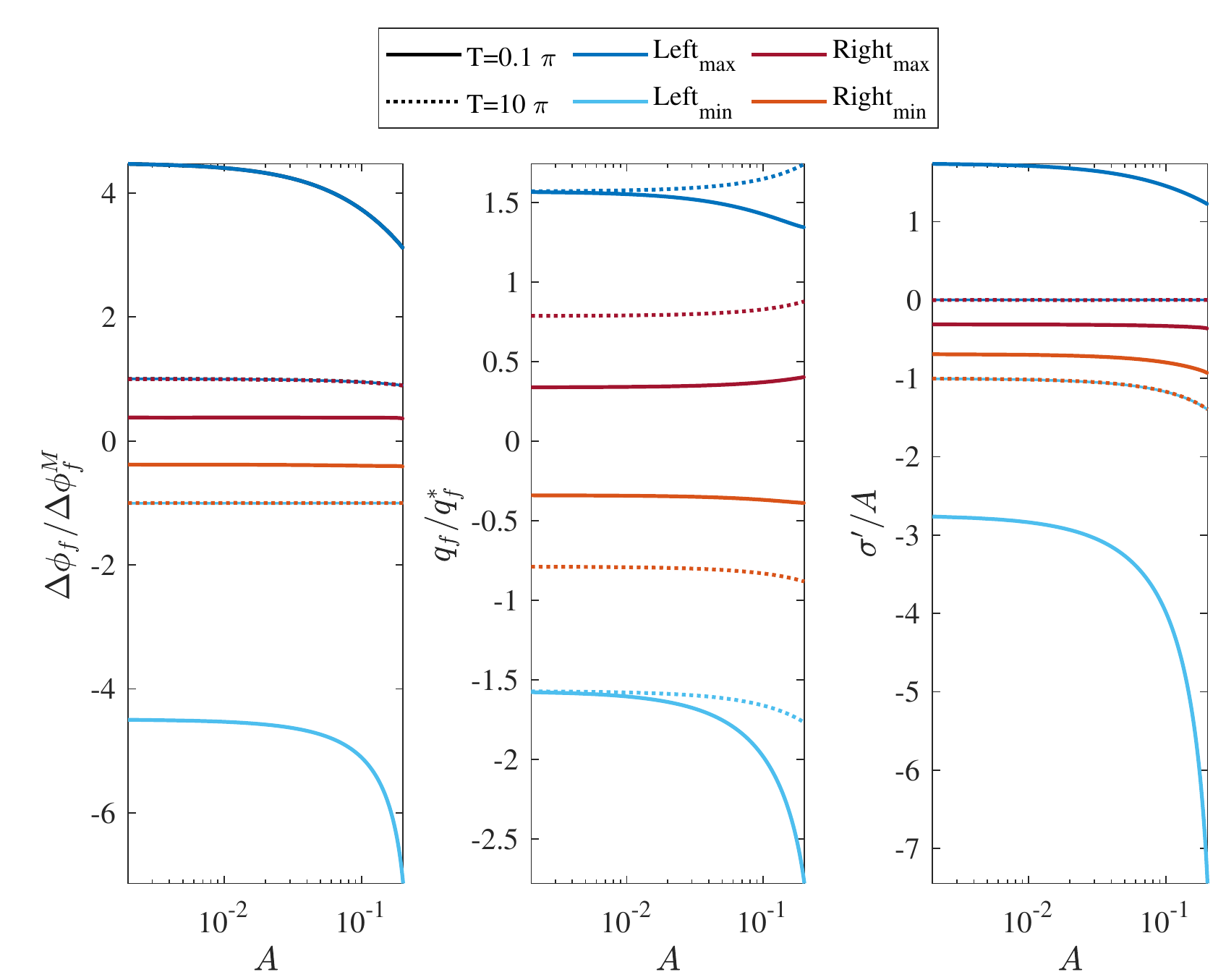}
    \caption{We illustrate the transition from small to large deformations as $A$ increases by plotting the normalised maximum and minimum values $\Delta{\phi_f}$ (first column), $q_f$ (second column), and $\sigma^\prime$ (third column) against $A$ for $T=0.1\pi$ (solid lines) and $T=10\pi$ (dotted lines) during the periodic regime. We show the maximum values (dark colours) and minimum values (light colours) of all three quantities at the left boundary ($X=0$, blue curves), of $\Delta{\phi_f}$ and $\sigma^\prime$ at the right boundary ($X=1$, red curves), and of $q_f$ at the material midpoint ($X=1/2$, red curves). \label{fig:localisationA} }
\end{figure}

We further explore the impact of $A$ in figure~\ref{fig:localisationA} by plotting the normalised maxima and minima of $\Delta{\phi_f}$, $q_f$, and $\sigma^\prime$ at the left and at the right, as in figure~\ref{fig:localisation-t}, but now against $A$ for two different values of $T$, showing the smooth transition from small deformations ($A \lesssim 0.01$)  to large deformations ($A \gtrsim 0.01$).
For small deformations, the normalised maxima and minima of all quantities at the left and at the right become independent of $A$, and the extreme values of $\Delta{\phi_f}$ and $q_f$ become symmetric. For SL, $\Delta{\phi_f}$ and $\sigma^\prime$ become uniform in space, such that their normalised left and right values are equal and depend only weakly on $A$ for the largest deformations shown here (\textit{e.g.}, $A\gtrsim{}0.1$). For FL, the values of all three quantities at the left become increasingly asymmetric and biased downward as $A$ increases.

\subsection{Impact of initial porosity}\label{impact-phi}

\begin{figure}[tp]
    \centering
    \includegraphics[width=0.75\textwidth]{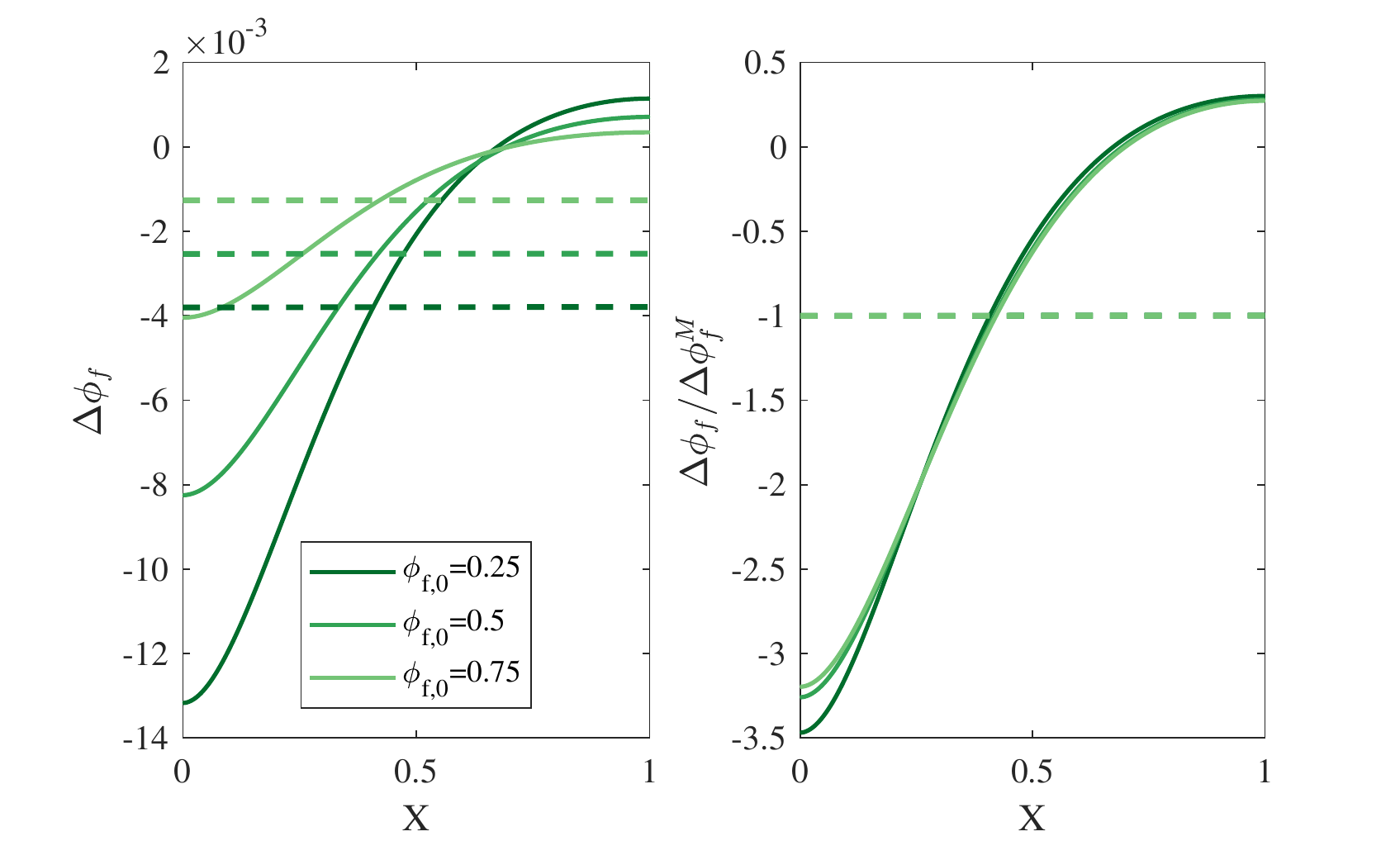}
    \caption{Impact of $\phi_{f,0}$ on $\Delta{\phi_f}$ at mid-cycle in the periodic regime for $A=0.1$ and for two different periods: $T=0.1\pi$ (FL; solid) and $10\pi$ (SL; dashed). We show results for three values of $\phi_{f,0}$ (increasing from dark to light). We plot the non-normalised value of $\Delta{\phi_f}$ on the left and the normalised value on the right. \label{fig:phi0_impact} }
\end{figure}

Finally, we consider the initial porosity $\phi_{f,0}$. In figure~\ref{fig:phi0_impact}, we plot the distribution of $\Delta{\phi_f}$ at mid-cycle for three different values of $\phi_{f,0}$ for fixed $A=0.01$ and for two different values of $T$, one for FL ($T=0.1\pi$) and SL ($T=10\pi$).
For both SL and FL, the non-normalised values of $\Delta{\phi_f}$ (left panel) increase in magnitude as $\phi_{f,0}$ decreases because the same change in total volume is a larger portion of the initial fluid volume. The right panel shows that normalisation by $\Delta{\phi_f^M}$ against captures the leading-order impact of varying $\phi_{f,0}$ on $\Delta{\phi_f}$, again with small deviations at the left boundary for FL, when nonlinearities are most pronounced.

\begin{figure}[tp]
    \centering
    \includegraphics[width=0.95\textwidth]{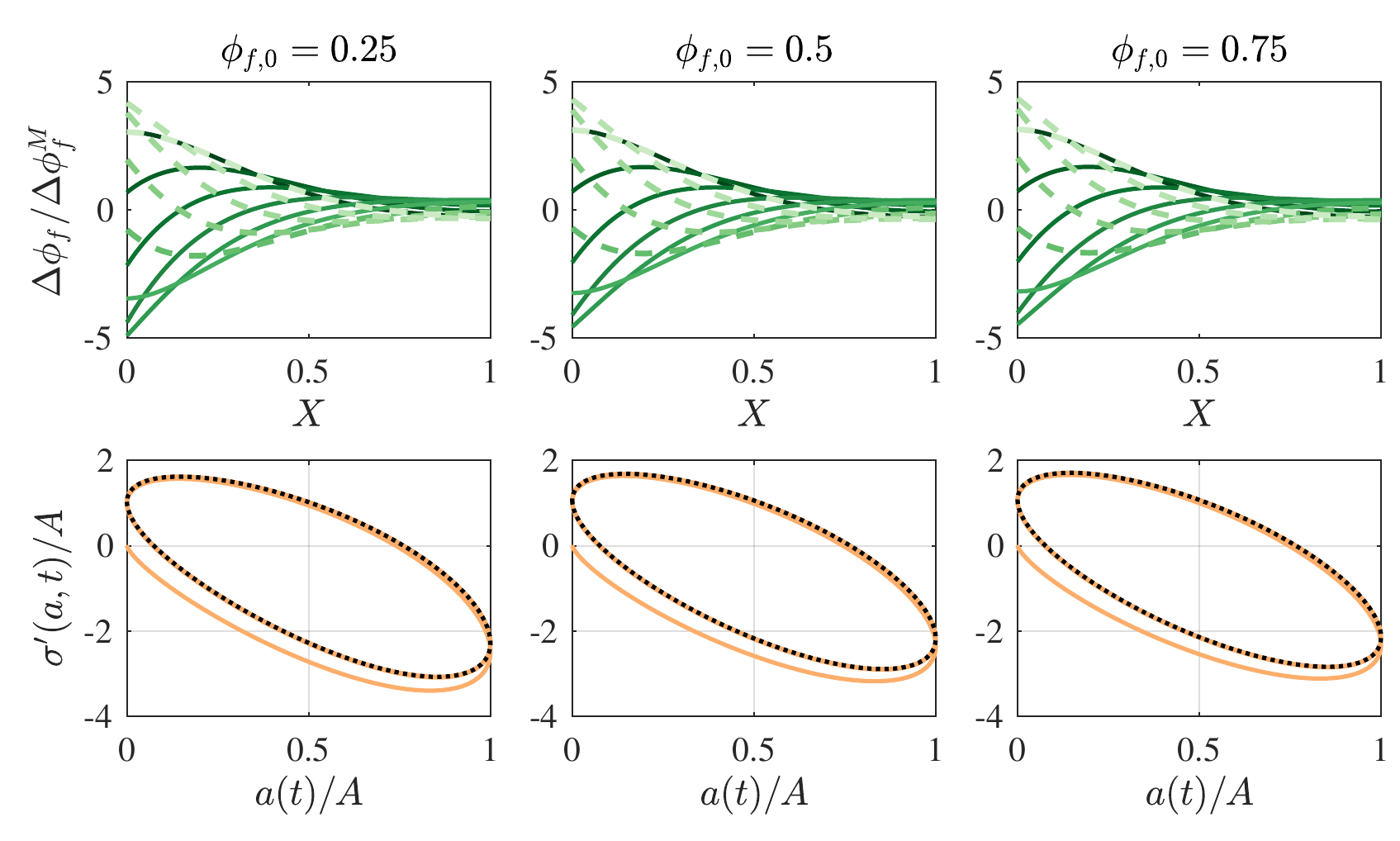}
    \caption{Impact of $\phi_{f,0}$ on the evolution of $\Delta{\phi_f}$ and $\sigma^\prime(a,t)$ for $A=0.01$, $T=0.1\pi$, and $\phi_{f,0}=0.25$ (first column), $0.5$ (second column), and $0.75$ (third column). In the first row, we plot $\Delta{\phi_f}$ against the Lagrangian spatial coordinate $X=x-u_s$ at times $t=nT$ to $(n+1)T$ in increments of $0.1T$ (dark to light) during one cycle in the periodic regime, and we distinguish between loading ($\dot{a}>0$; solid curves) and unloading ($\dot{a}<0$; dashed curves). In the second row, we plot phase portraits of $\sigma^\prime(a,t)$ against $a(t)$ from $t=0$ to $t=20T$, emphasising the last cycle (dotted black lines). \label{fig:por-vel-differentPHi0} }
\end{figure}

We next plot the evolution of $\Delta{\phi_f}$ and $\sigma^\prime$ in time for $A=0.01$ and $T=0.1\pi$ for the same three values of $\phi_{f,0}$ (fig.~\ref{fig:por-vel-differentPHi0}), confirming that normalisation captures the primary impact of $\phi_{f,0}$ on $\Delta{\phi_f}$ and on $\sigma^\prime(a,t)$.

\subsection{FL in biological examples}

We conclude by considering appropriate parameter ranges for several examples of periodic loading in soft biological tissues (Table~\ref{t_pe-values}). Based on these values, we calculate the poroelastic timescale $T_{\mathrm{pe}}$ and the relative dimensionless loading period $\tilde{T}=T/T_{\mathrm{pe}}$ for each example to understand the ranges of relevance for $\tilde A$, $\tilde T$ and $\phi_{f,0}$.
\begin{sidewaystable}
\small{
\begin{tabular}{lcccccccc}
\multicolumn{1}{l|}{Tissue} & \multicolumn{1}{c|}{$L$ [m]} & \multicolumn{1}{c|}{$\tilde{A}=A/L$} & \multicolumn{1}{c|}{$\phi_{f,0}$} & \multicolumn{1}{c|}{$k_0$ [m$^2$]} & \multicolumn{1}{c|}{$\mathcal{M}$ [Pa]} & \multicolumn{1}{c|}{$T_\mathrm{pe}$ [s] } & \multicolumn{1}{c|}{\begin{tabular}[c]{@{}c@{}}Typical loading \\ frequency [Hz]\end{tabular}} & $\tilde{T}=T/T_\mathrm{pe}$ \\ \hline
\multicolumn{1}{l|}{} & \multicolumn{1}{l|}{} & \multicolumn{1}{l|}{} & \multicolumn{1}{l|}{} & \multicolumn{1}{l|}{} & \multicolumn{1}{l|}{} & \multicolumn{1}{l|}{} & \multicolumn{1}{l|}{}& \\
\multicolumn{1}{l|}{Brain ECM \cite{kedarasetti-fbcns-2022}} & \multicolumn{1}{c|}{$2 \times 10^{-4}$} & \multicolumn{1}{c|}{0.1 -- 0.2} &\multicolumn{1}{c|}{$0.2$} &\multicolumn{1}{c|}{$2 \times 10^{-15}$} & \multicolumn{1}{c|}{$2 \times 10^3$} & \multicolumn{1}{c|}{$\approx 10$} & \multicolumn{1}{c|}{0.3 -- 10} & $0.003\pi$ -- $0.1\pi$ \\
\multicolumn{1}{l|}{} & \multicolumn{1}{c|}{} & \multicolumn{1}{c|}{} & \multicolumn{1}{c|}{} & \multicolumn{1}{c|}{} & \multicolumn{1}{c|}{}  & \multicolumn{1}{c|}{} & \multicolumn{1}{c|}{} & \\
\multicolumn{1}{l|}{Cartilage \cite{Ferguson2004}} & \multicolumn{1}{c|}{$2 \times 10^{-3}$}& \multicolumn{1}{c|}{$\approx 0.15 $}& \multicolumn{1}{c|}{$0.8$} & \multicolumn{1}{c|}{$7.5 \times 10^{-18}$} & \multicolumn{1}{c|}{$5 \times 10^6$} & \multicolumn{1}{c|}{$\approx 10^2$} & \multicolumn{1}{c|}{\begin{tabular}[c]{@{}c@{}}0.001 -- 0.1  (sitting)\\ 0.1 -- 1 (running)\end{tabular}} & {\begin{tabular}[c]{@{}c@{}}$0.03\pi$ -- $3\pi$\\ $0.003\pi$ -- $0.03 \pi$\end{tabular}} \\
\multicolumn{1}{l|}{} & \multicolumn{1}{c|}{} & \multicolumn{1}{c|}{} & \multicolumn{1}{c|}{} & \multicolumn{1}{c|}{} & \multicolumn{1}{c|}{}  & \multicolumn{1}{c|}{} & \multicolumn{1}{c|}{} & \\
\multicolumn{1}{l|}{\begin{tabular}[c]{@{}l@{}}Intervertebral \\ Disk\\ (Anulus F.)  \cite{Ferguson2004}\end{tabular}} & \multicolumn{1}{c|}{$ 10^{-2}$}  & \multicolumn{1}{c|}{$\approx 0.15$} & \multicolumn{1}{c|}{$0.7$} & \multicolumn{1}{c|}{$7.5 \times 10^{-19}$} & \multicolumn{1}{c|}{$2.5 \times 10^6$} & \multicolumn{1}{c|}{$  \approx 5 \times 10^4$} & \multicolumn{1}{c|}{\begin{tabular}[c]{@{}c@{}}$2 \times 10^{-5}$ (wake cycle)\\ 0.001 -- 0.1 (sitting)\\ 0.1 -- 1  (running)\end{tabular}} & {\begin{tabular}[c]{@{}c@{}}$0.3\pi$\\ $6\times 10^{-5}\pi$ -- $6\times 10^{-3}\pi$\\ $6\times 10^{-6}\pi$ -- $6\times 10^{-5}\pi$ \end{tabular}} \\
\multicolumn{1}{l|}{} & \multicolumn{1}{l|}{} & \multicolumn{1}{l|}{} & \multicolumn{1}{l|}{} & \multicolumn{1}{l|}{} & \multicolumn{1}{l|}{} & \multicolumn{1}{l|}{}  & \multicolumn{1}{c|}{} &\\
\multicolumn{1}{l|}{\begin{tabular}[c]{@{}l@{}}Cartilage \\ Scaffold\\ (bioreactor) \cite{Sengers2004}\end{tabular}} & \multicolumn{1}{c|}{$ 2 \times 10^{-3} $}  & \multicolumn{1}{c|}{$0-0.15$} & \multicolumn{1}{c|}{$ 0.9$} & \multicolumn{1}{c|}{$10^{-17}$} & \multicolumn{1}{c|}{$ 10^{5} $} & \multicolumn{1}{c|}{$ \approx4 \times 10^{3} $} & \multicolumn{1}{c|}{0.001 -- 1 } & $0.0001\pi$ -- $0.1\pi$ \\
 & \multicolumn{1}{l}{} & \multicolumn{1}{l}{} & \multicolumn{1}{l}{} & \multicolumn{1}{l}{} & \multicolumn{1}{l}{} & \multicolumn{1}{l}{}
\end{tabular}}
\caption{Material and loading pararameters for some examples of biological materials.} \label{t_pe-values}
\end{sidewaystable}

Table~\ref{t_pe-values} shows that, for a variety of soft tissues, deformations are in the range of the dimensionless amplitudes $\tilde{A}$ considered here, with nearly all being near the upper end. The range of $\phi_{f,0}$ is slightly wider than the range considered here, but we showed in \S\ref{impact-phi} that this value has little impact on the normalised mechanical response of the material. The range of poroelastic timescales $T_{\mathrm{pe}}$ and respective loading frequencies suggest that dimensionless loading periods $\tilde{T}$ span a wide range, with many corresponding to FL. These values justify our analysis and underscore the importance of characterising the nonlinearity of large poromechanical deformations during FL in particular.

\section{Conclusions}

We have provided an analysis of the poromechanical coupling between large deformations and fluid flow in a periodically loaded soft porous material. To do so, we used a kinematically rigorous 1D continuum model with Hencky elasticity and a Kozeny-Carman-like permeability law. In particular, we examined the roles of the three dimensionless control parameters: the initial porosity $\phi_{f,0}$, the loading amplitude $A$, and the loading period $T$.

We began by deriving several analytical solutions from linear poroelasticity --- an early-time solution, a full solution, and a solution for the response to very fast loading (Stokes's second problem) --- as well as a quasi-static solution to the fully nonlinear problem. The former are valid for small deformations, which corresponds to $A\ll{}1$ and also, less obviously, to $A\ll{}\sqrt{T}$. The quasi-static solution is valid for very slow loading ($T\gg{}1$) but arbitrarily large amplitudes. We then compared these solutions with our numerical results, highlighting the existence of two mechanical regimes: slow loading (SL), where the loading is much slower than the poroelastic relaxation time $T_\mathrm{pe}$, and fast loading (FL), where the loading is much faster than $T_\mathrm{pe}$.

We then showed that the material response to SL ($T\gtrsim\pi$) is an increasingly uniform deformation throughout the domain, approaching the quasi-static solution for very slow loading. For FL ($T\lesssim0.1\pi$), the deformation is nonuniform and increasingly localised near the left boundary. In the limit of very fast loading, this localisation is such that the left portion of the material oscillates while the right portion is in a state of static compression. We showed that FL is also characterised by asymmetry between loading and unloading, with a larger change in porosity and higher fluid flux magnitudes during loading (when fluid is squeezed out) than during unloading (when fluid is sucked back in). This asymmetry originates in the kinematic nonlinearity of large deformations and is amplified by the localisation of the deformation near the left boundary as $T$ decreases (itself a linear effect) and by the nonlinearity of deformation-dependent permeability as $A$ increases. Thus, this asymmetry emerges from the fast and large deformation of an elastic and initially homogeneous material, and is therefore purely poromechanical; it does not occur during slow loading. Asymmetry between quasi-static loading and unloading can instead result from wall friction in confined geometries and/or constitutive hysteresis due to microstructural changes, as has been observed experimentally for some soft porous materials~\citep[\textit{e.g.},][]{sobac-mecind-2011, hewitt2016, lutz-rsi-2021}.

Through the analysis on the fluid flow at different fixed material points in the domain, we also showed that faster loading leads to an increasing delay in the interior of the domain relative to the motion of the left boundary. Although the motion of the fluid itself is reversible, these features are likely to have an important impact on irreversible phenomena. For example, these results have interesting implications for the deformation-driven transport and mixing of solutes, which we consider in detail in a companion study.

The evolution of the stress at the left boundary as a function of the piston position revealed that, for faster loading and larger deformations, the force required to compress the material is much larger than the force required to pull it back to the same position. This is due to the interaction of the viscous flow through the solid porous skeleton, which instead can be instantaneously squeezed out or re-imbibed in the domain for very slow loading, with no hysteresis.

Finally, we showed that a larger initial porosity $\phi_{f,0}$ leads to much lower fluid fluxes that can be accounted for by normalisation.

Our results elucidate the local and global poromechanical behaviour of soft porous media during periodic loading over a wide range of $\phi_{f,0}$, $A$, and $T$. Having used relatively generic constitutive models, we expect our qualitative insights to be robust across a wide range of materials; however, it is straightforward to adapt our approach to other constitutive models (see Appendix~\ref{appendix-constitutive}).

Declaration of interests: The authors report no conflicts of interest.

\begin{acknowledgments}
This work was supported by the European Research Council (ERC) under the European Union's Horizon 2020 Programme [Grant No. 805469]. S.P. was supported by Start-Up Research Grant (SRG/2021/001269) by the Science and Engineering Research Board, Department of Science and Technology, Government of India. For the purpose of Open Access, the authors have applied a CC BY public copyright licence to any Author Accepted Manuscript (AAM) version arising from this submission.
\end{acknowledgments}

\appendix 

\section{Constitutive laws}\label{appendix-constitutive}

In our model, we consider Hencky elasticity as the constitutive law for the elastic solid. Hencky elasticity is a hyperelastic model commonly used for soft rubbers and polyurethane foams~\cite{Hencky33, Anand1979, xiao2002}, and sometimes for soft biological tissues~\citep[\textit{e.g.},][]{MARCHESSEAU2010185, Fraldi2018}. In this appendix, we compare Hencky elasticity with two other hyperelastic models --- Neo-Hookean and logarithmic Neo-Hookean --- that are more commonly employed for soft tissues~\citep[\textit{e.g.},][]{Ehlers2009, Sengers2004}. This analysis suggests that our qualitative results also apply to these other elasticity laws.

The expressions $\sigma^\prime(\phi_f)$ for Hencky elasticity and for linear elasticity are given in equations~\eqref{sigma-hencky} and \eqref{sigma-linear}, respectively. The appropriate expressions for the Neo-Hookean and logarithmic Neo-Hookean constitutive laws are:
\begin{equation}
     \mathbf{\sigma}^\prime = \mathcal{G} \left(\frac{J^2-1}{J}\right) + {\Lambda} (J-1), 
\end{equation}
and
\begin{equation}
     \mathbf{\sigma}^\prime = \mathcal{G} \left(\frac{J^2-1}{J}\right) + {\Lambda} \frac{\ln(J)}{J}, 
\end{equation}
respectively, where $J=(1-\phi_{f,0})/(1-\phi_f)$ and $\Lambda$ and $\mathcal{G}$ are the Lam\'{e} constants, such that $\mathcal{M}=\Lambda+2\mathcal{G}$.
\begin{figure}[tp]
    \centering
    \includegraphics[width=0.92\textwidth]{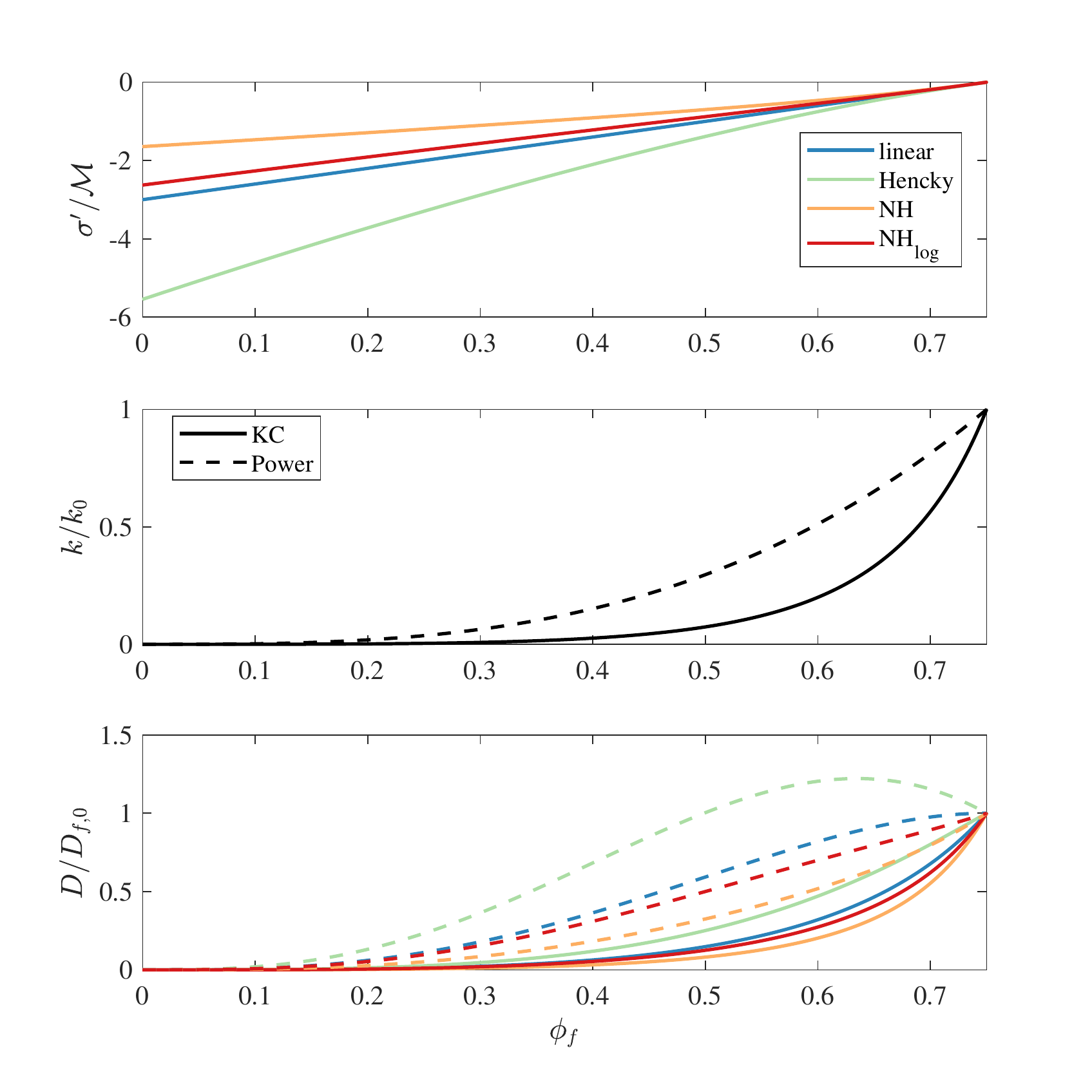}
    \caption{Comparison of different constitutive properties: normalised effective stress $\sigma^\prime(\phi_f)$ (first row), permeability $k(\phi_f)$ (second row), and poroelastic diffusivity $D_f(\phi_f)=(1-\phi_f)(k/\mu)\mathrm{d}\sigma^\prime/\mathrm{d}\phi_f$ (third row) against $\phi_f$ for four different elasticity laws (linear, Hencky, Neo-Hookean, and logatithmic Neo-Hookean) and two different permeability laws (Kozeny-Carman and power-law). The relaxed porosity is $\phi_{f,0}=0.75$ in all cases and we take $\Lambda/\mathcal{G}=0.515$ in the two Neo-Hookean laws based on values reported in ref.~\citep{Ferguson2004}.
    \label{fig:constitutive} }
\end{figure}

In the first row of figure~\ref{fig:constitutive}, we plot $\sigma^\prime/\mathcal{M}$ against $\phi_f$ in compression for all four of these elasticity laws (linear, Hencky, Neo-Hookean and logarithmic Neo-Hookean). Note that the two Neo-Hookean models are characterised by an additional dimensionless parameter in confined compression, the ratio $\Lambda/\mathcal{G}$. All four curves have the same qualitative shape, although Hencky elasticity is noticeably stiffer than the other models during strong compression. However, these quantitative differences are relatively unimportant because the permeability law forces the poroelastic diffusivity smoothly toward zero during strong compression in all cases (see below).

In the second row of figure~\ref{fig:constitutive}, we provide a similar comparison for two different permeability laws: Kozeny-Carman (eq.~\ref{kc-permeability}) and a simpler power-law model given by
\begin{equation}
    k(\phi_f)= k_0\left(\frac{\phi_f}{\phi_{f,0}}\right)^3.
\end{equation}
Both models are commonly used for soft tissues and gels (\textit{e.g.}, Kozeny-Carman in refs.~\cite{Sacco2011, MALANDRINO2014, Rahbari2016, Gao2022} and power-law in refs.~\cite{HOLMES1990, Ehlers2009, Sengers2004}) and the two curves have qualitatively similar shapes. Note that constant permeability $k(\phi_f)= k_0$ is rarely used for soft porous media and is considered non-physical, particularly for moderate to large deformations.

In the third row of figure~\ref{fig:constitutive}, we compare the resulting poroelastic diffusivity $D_f(\phi_f)$ (eq.~\ref{eq:Df}) for all eight combinations of these four elasticity laws with these two permeability laws. Most of these curves have the same qualitative shape, most importantly vanishing smoothly as $\phi_f\to0$. The combination of Hencky elasticity with Kozeny-Carman permeability (solid green line) is roughly in the middle of this family of curves, suggesting that this combination is representative of the poromechanical behavior of many different soft porous materials, thus supporting the qualitative generality of our results.

\section{Early time evolution, penetration length, and periodic response to very fast loading}\label{s:et_and_vf}

\begin{figure}[tp]
    \centering
         \includegraphics[width=\textwidth]{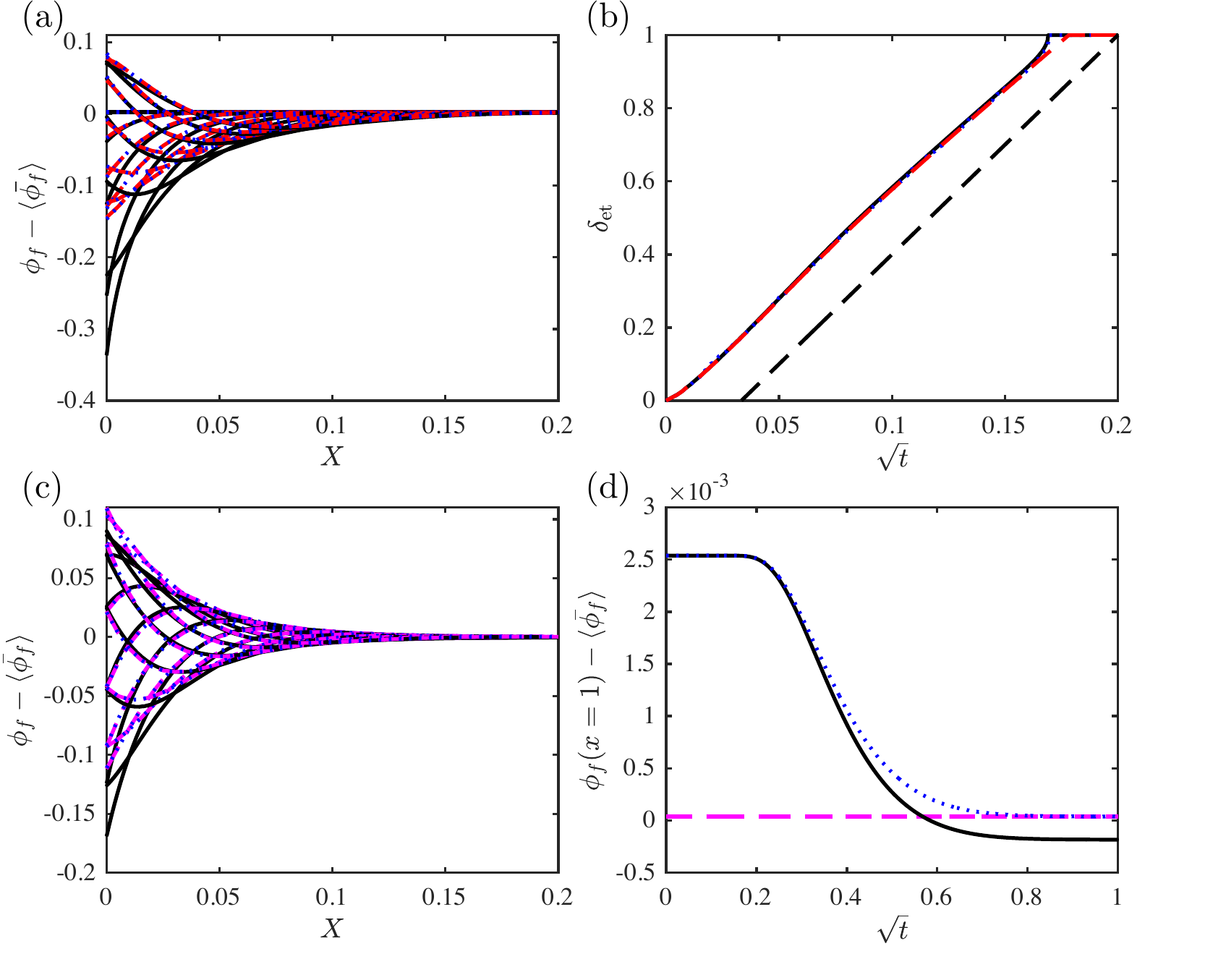}
        \caption{Early time response for $\phi_{f,0}=0.75$ at a small amplitude $A=0.02$ and a very fast period $T=0.001\pi$. We show $\Delta{\phi_f}$ as a function of the Lagrangian spatial coordinate $X$ for several times $t$ during (a)~the first loading cycle and (c)~one cycle in the periodic regime. We also show the evolution of (b)~the penetration distance $\delta_\mathrm{et}$ and (d)~the change in porosity at the right boundary, $\Delta{\phi_f}(1,t)$, both against $\sqrt{t}$. All plots show the numerical solution (solid black) and the full linear-poroelastic solution (dotted blue). The top row includes the early-time linear-poroelastic solution (dashed red). The bottom row includes the very fast linear-poroelastic solution (dashed magenta). The latter is a constant in panel~(d) the bottom left because the very fast solution neglects the initial transient. Panel~(b) also shows a linear trend in $\sqrt{t}$ for reference (dashed black). \label{fig:et_and_vf} }
\end{figure}

The macroscopic strain imposed on the material is always of size $A$. For fast loading, this strain localises toward the left boundary, leading to local strains of size $\sim{}A/\sqrt{T}$ over a region of size $\sqrt{T}$ in the periodic state. Thus, the local strains can become large even when $A$ is small, leading to large deviations from linear poroelasticity. These deviations are particularly large at early times, during which the deformations are localised to an even narrower region of size $\sqrt{t}$~(figure~\ref{fig:et_and_vf}a), but they persist in the periodic state (figure~\ref{fig:et_and_vf}c).

When the loading begins, the deformation propagates into the material diffusively with penetration distance $\delta_\mathrm{et}\sim{}\sqrt{t}$ until the deformation spans the domain~(figure~\ref{fig:et_and_vf}b). In the periodic state, the oscillations will be confined to a region of size $\sqrt{T}$. For very fast loading, $\sqrt{T}\ll1$, the right portion of the material will evolve toward a state of static compression~(figure~\ref{fig:et_and_vf}d).

\section{Transition to the periodic regime}\label{quasi-steady}
 
The linear-poroelastic solution in \S\ref{linear-poro} includes a transient component that decays exponentially and a periodic component with period $T$. In this appendix, we illustrate and quantify the decay of the transient component and hence the convergence to the periodic regime for a large-deformation scenario. To do so, we solve the problem numerically for $\phi_{f,0}=0.75$, $A=0.2$, and for $T=4\pi$, $\pi$, $0.2\pi$, $0.12\pi$, and $0.1\pi$. This amplitude is the largest considered in this study, thus providing an upper bound for difference magnitudes. In each case, we calculate the root-mean-square (RMS) relative difference in $\phi_f(X,t)$ between the solution at time $t$ and at time $t+T$.
\begin{figure}[tp]
    \centering
    \includegraphics[width=0.7\textwidth]{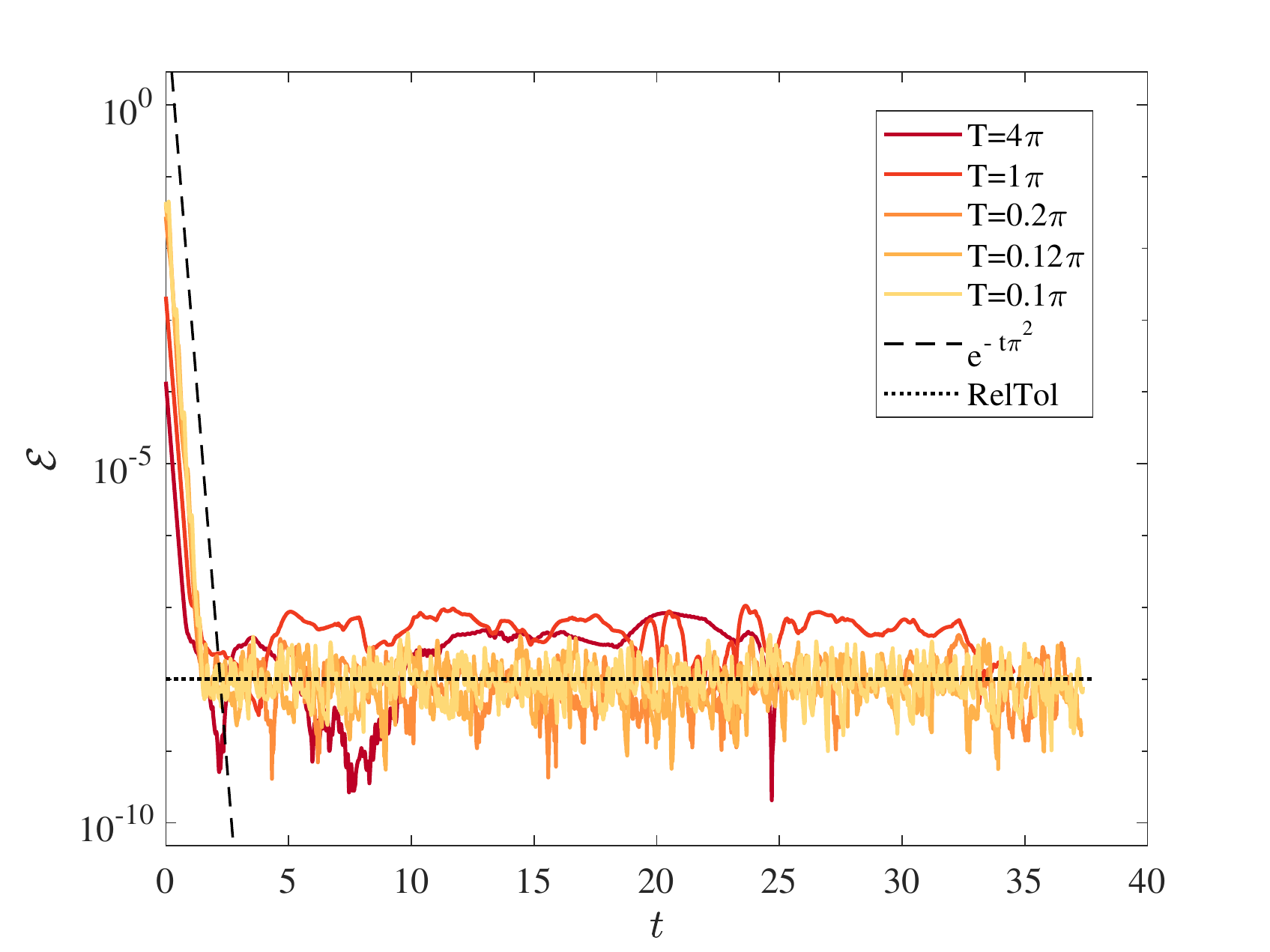}
    \caption{The RMS relative difference between the $\phi_f(X,t)$ at time $t$ and at time $t+T$ for $A=0.2$, $\phi_{f,0}=0.75$,
    and for five values of $T$ (see legend). The dashed black line indicates the exponential decay $e^{-t\pi^2}$ and the dotted black line indicates the relative tolerance selected for time integration, here set to $10^{-8}$ (see figure~\ref{fig:tolerances}). \label{fig:us-1to10} }
\end{figure}

Figure~\ref{fig:us-1to10} confirms the exponential decay of the transient for all five values of $T$ (dashed black dashed). After this initial transition, the RMS relative difference for all curves oscillates around a mean value that is controlled by the relative error tolerance associated with time integration (see~figure \ref{fig:tolerances}).

\section{Numerical method}\label{numerical-method}

We solve our model numerically using Chebyshev spectral differences in space~\cite{weideman-acmtoms-2000} and implicit Runge-Kutta integration in time. We achieve the latter using \texttt{MATLAB}'s built-in solver \texttt{ODE15s}~\cite{shampine-siamjscicomput-1997}. To handle the moving boundary, we rescale the spatial coordinate as
\begin{equation}
    \xi=\frac{x-a}{1-a},
\end{equation}
thus mapping a general conservation law of the form
\begin{equation}\label{pde-xt}
    \frac{\partial{\Phi}}{\partial{t}}+ \frac{\partial}{\partial{x}}[F(\Phi)]=0
\end{equation}
on the domain $a(t)\leq{}x\leq{}1$ to
\begin{equation}
    \frac{\partial{\Phi}}{\partial{t}}-
\left(\frac{1-\xi}{1-a}\right) \dot{a}\frac{\partial{\Phi}}{\partial{\xi}} +\left(\frac{1}{1-a}\right)\frac{\partial}{\partial{\xi}}[F(\Phi)]=0
\end{equation}
on the domain $0\leq{}\xi\leq{}1$. When solving equation~\eqref{conservation-q-scaled}, we then take $\Phi=\phi_f$ and
\begin{equation}
    {F(\phi_f)}=-\tilde{D}_f(\phi_f)\frac{\partial{\phi_f}}{\partial{\tilde{x}}}.
\end{equation}
For our spatial discretisation, we perform a convergence analysis in the number of grid points $N_x$ by calculating the RMS relative difference in $\phi_f(a,t)$ for each solution with respect the solution for $N_x=1000$.
\begin{figure}[tp]
    \centering
    \includegraphics[width=0.99\textwidth]{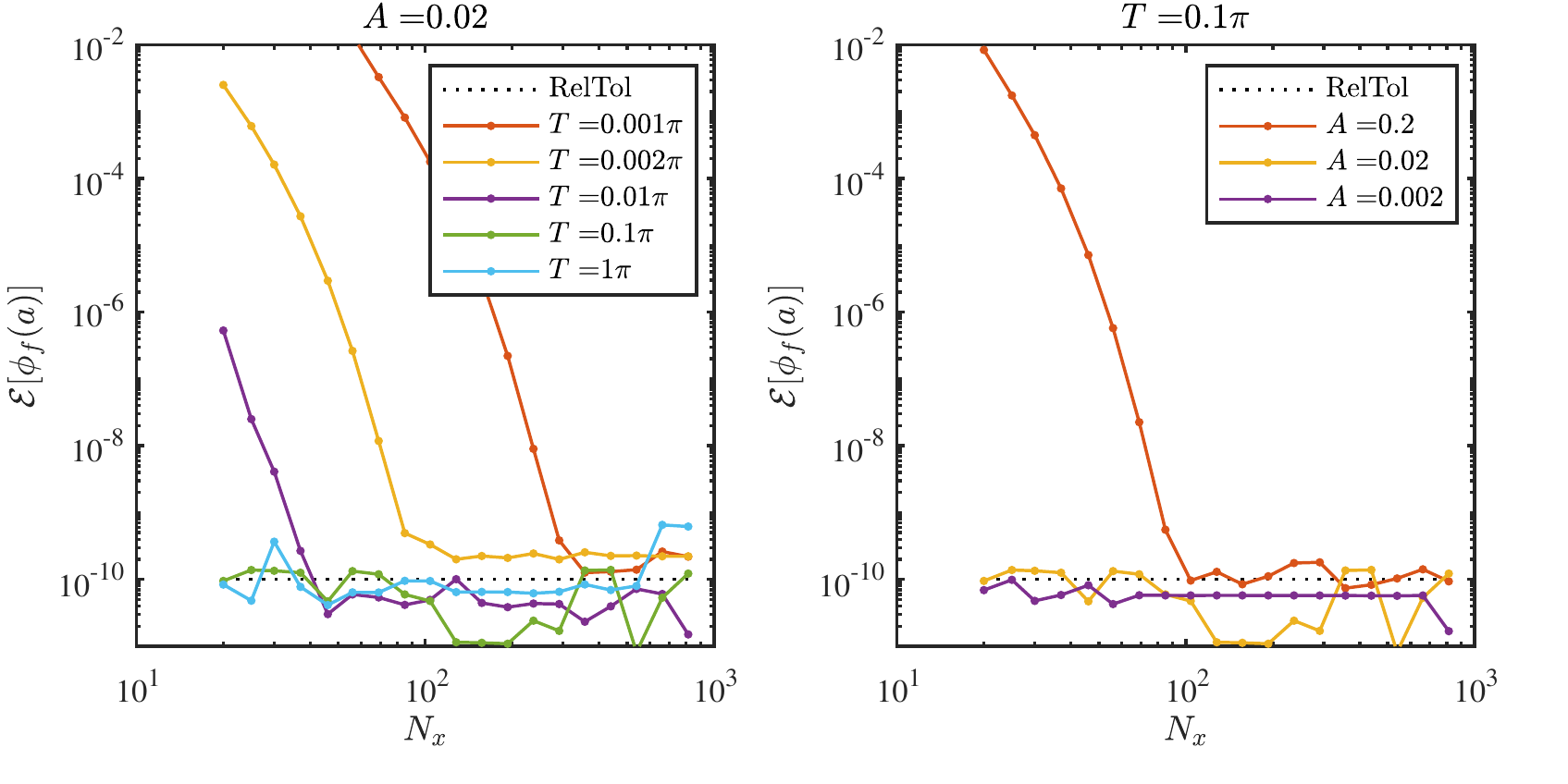}
    \caption{Convergence analysis: RMS relative difference in $\phi_f(a,t)$ relative to the solution for $N_x=1000$. On the left, we fix $A=0.02$ and consider different values of $T$, from very fast to slow. On the right, we fix $T=0.1\pi$ and consider different values of $A$, from small to large. \label{fig:nx} }
\end{figure}

\begin{figure}[tp]
    \centering
    \includegraphics[width=0.7\textwidth]{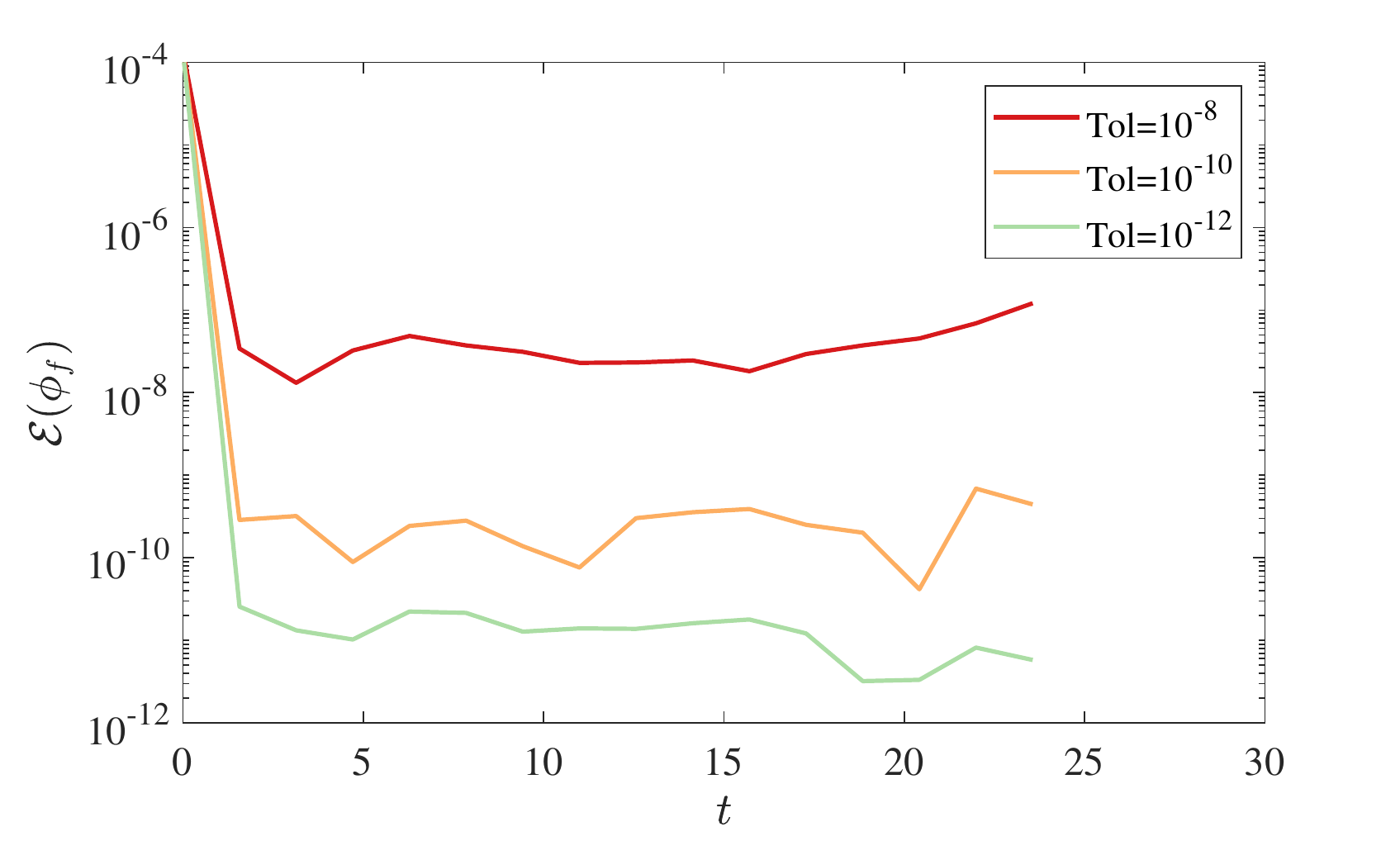}
    \caption{RMS relative difference in $\phi_f(X,t)$ between two consecutive cycles in the periodic regime for $A=0.2$ and $T=4\pi$ and for three values of relative error tolerance for time integration.}
    \label{fig:tolerances}
\end{figure}

Figure~\ref{fig:nx} illustrates the impact of $A$ and $T$ on the spatial accuracy of the numerical solution. Very small amplitudes and very slow periods are characterised by low differences that are on the order of the tolerance chosen for time integration (see figure~\ref{fig:tolerances}). We choose $N_x=300$ for all simulations, associated with a maximum relative difference comparable to that of the relative error tolerance for time integration.

In figure~\ref{fig:tolerances}, we consider the RMS relative difference in $\phi_f(X,t)$ between two consecutive cycles in the periodic regime for $A=0.2$ and $T=4\pi$, and for three values of the relative error tolerance for time integration.
The results confirm that the RMS relative difference is limited by the relative error tolerance of the ODE solver, as expected. Throughout our analysis, we use a relative tolerance of $10^{-8}$.

\section{Impact of elasticity and permeability laws at large amplitudes} \label{appendix-mechanics-claws}

\begin{figure}[tp]
    \centering
    \includegraphics[width=0.85\textwidth]{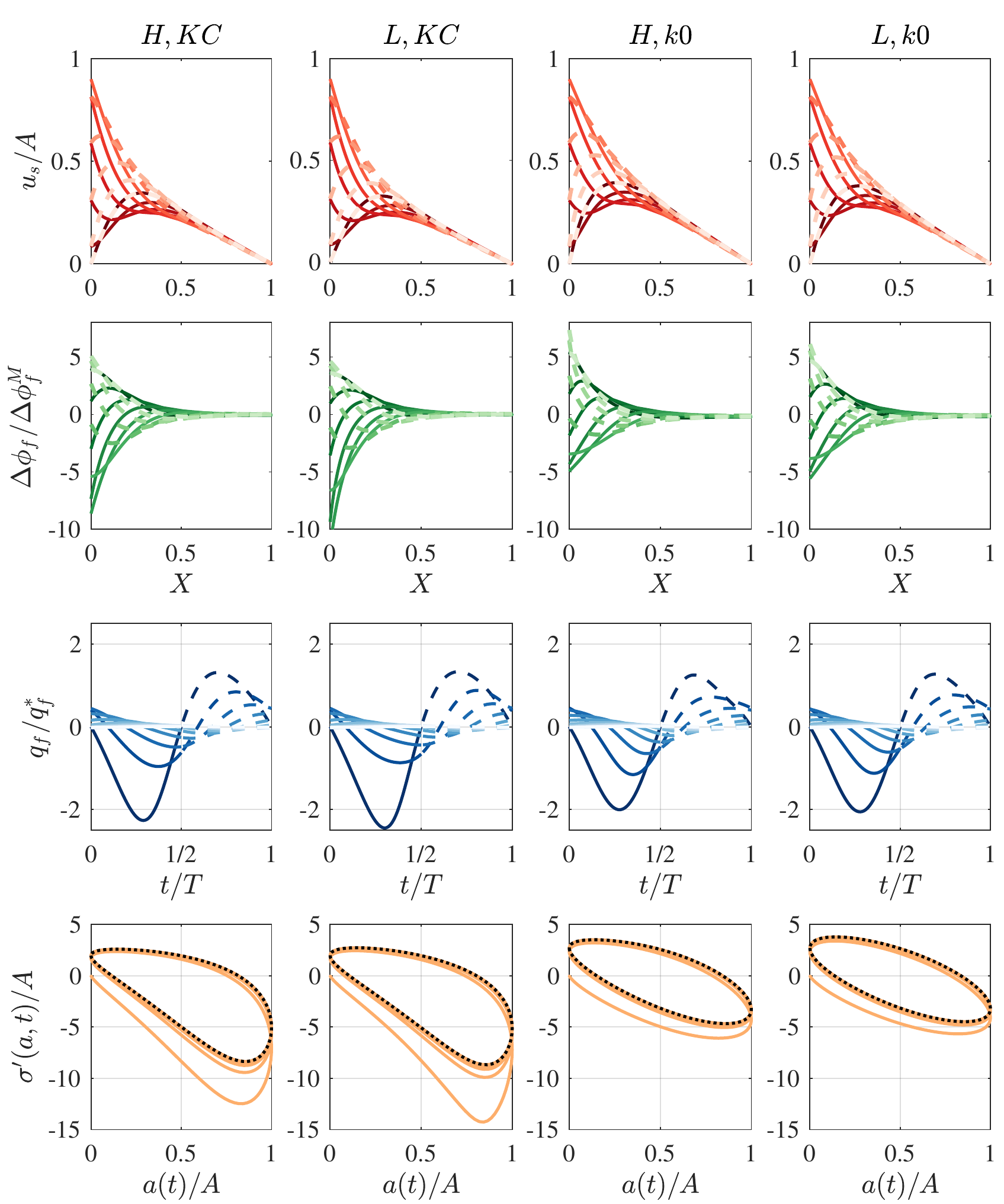}
    \caption{As in figure~\ref{fig:displ-vel-different-T}, but for a slightly lower amplitude ($A=0.09$) and showing four different combinations of constitutive behavior: Hencky elasticity with Kozeny-Carman permeability (first column), linear elasticity with Kozeny-Carman permeability (second column), Hencky elasticity with constant permeability (thrid column), and linear elasticity with constant permeability (fourth column). \label{fig:mechanics-constitutive} }
\end{figure}

In figure~\ref{fig:mechanics-constitutive}, we compare different combinations of elasticity and permeability laws for a scenario involving large deformations and fast loading ($A=0.09$, $T=0.03\pi$). Specifically, we compare four cases: Hencky elasticity with Kozeny-Carman permeability (first column; same as the first column in figure~\ref{fig:displ-vel-different-T}, but for a slightly lower amplitude), linear elasticity with Kozeny-Carman permeability (second column), Hencky elasticity with constant permeability (third column), and linear elasticity with constant permeability (fourth column).
Note that the last column is still kinematically nonlinear because $D_f=1-\phi_f$ (see eq.~\ref{conservation-q-scaled}) and the problem remains a moving-boundary problem. Even for constant permeability and linear elasticity, the fluid flux has a strong asymmetry between loading and unloading. This asymmetry is strongly enhanced by Kozeny-Carman permeability and very gently moderated by Hencky elasticity. The latter occurs because Hencky elasticity is stiffer than linear elasticity in compression.

\clearpage


\begin{thebibliography}{64}%
    \makeatletter
    \providecommand \@ifxundefined [1]{%
     \@ifx{#1\undefined}
    }%
    \providecommand \@ifnum [1]{%
     \ifnum #1\expandafter \@firstoftwo
     \else \expandafter \@secondoftwo
     \fi
    }%
    \providecommand \@ifx [1]{%
     \ifx #1\expandafter \@firstoftwo
     \else \expandafter \@secondoftwo
     \fi
    }%
    \providecommand \natexlab [1]{#1}%
    \providecommand \enquote  [1]{``#1''}%
    \providecommand \bibnamefont  [1]{#1}%
    \providecommand \bibfnamefont [1]{#1}%
    \providecommand \citenamefont [1]{#1}%
    \providecommand \href@noop [0]{\@secondoftwo}%
    \providecommand \href [0]{\begingroup \@sanitize@url \@href}%
    \providecommand \@href[1]{\@@startlink{#1}\@@href}%
    \providecommand \@@href[1]{\endgroup#1\@@endlink}%
    \providecommand \@sanitize@url [0]{\catcode `\\12\catcode `\$12\catcode `\&12\catcode `\#12\catcode `\^12\catcode `\_12\catcode `\%12\relax}%
    \providecommand \@@startlink[1]{}%
    \providecommand \@@endlink[0]{}%
    \providecommand \url  [0]{\begingroup\@sanitize@url \@url }%
    \providecommand \@url [1]{\endgroup\@href {#1}{\urlprefix }}%
    \providecommand \urlprefix  [0]{URL }%
    \providecommand \Eprint [0]{\href }%
    \providecommand \doibase [0]{http://dx.doi.org/}%
    \providecommand \selectlanguage [0]{\@gobble}%
    \providecommand \bibinfo  [0]{\@secondoftwo}%
    \providecommand \bibfield  [0]{\@secondoftwo}%
    \providecommand \translation [1]{[#1]}%
    \providecommand \BibitemOpen [0]{}%
    \providecommand \bibitemStop [0]{}%
    \providecommand \bibitemNoStop [0]{.\EOS\space}%
    \providecommand \EOS [0]{\spacefactor3000\relax}%
    \providecommand \BibitemShut  [1]{\csname bibitem#1\endcsname}%
    \let\auto@bib@innerbib\@empty
    \bibitem [{\citenamefont {Franceschini}\ \emph {et~al.}(2006)\citenamefont {Franceschini}, \citenamefont {Bigoni}, \citenamefont {Regitnig},\ and\ \citenamefont {Holzapfel}}]{Franceschini2006}%
      \BibitemOpen
      \bibfield  {author} {\bibinfo {author} {\bibfnamefont {G.}~\bibnamefont {Franceschini}}, \bibinfo {author} {\bibfnamefont {D.}~\bibnamefont {Bigoni}}, \bibinfo {author} {\bibfnamefont {P.}~\bibnamefont {Regitnig}}, \ and\ \bibinfo {author} {\bibfnamefont {G.A.}\ \bibnamefont {Holzapfel}},\ }\bibfield  {title} {\enquote {\bibinfo {title} {{Brain tissue deforms similarly to filled elastomers and follows consolidation theory}},}\ }\href {\doibase 10.1016/J.JMPS.2006.05.004} {\bibfield  {journal} {\bibinfo  {journal} {Journal of the Mechanics and Physics of Solids}\ }\textbf {\bibinfo {volume} {54}},\ \bibinfo {pages} {2592--2620} (\bibinfo {year} {2006})}\BibitemShut {NoStop}%
    \bibitem [{\citenamefont {Kedarasetti}\ \emph {et~al.}(2020)\citenamefont {Kedarasetti}, \citenamefont {Drew},\ and\ \citenamefont {Costanzo}}]{kedarasetti-fbcns-2022}%
      \BibitemOpen
      \bibfield  {author} {\bibinfo {author} {\bibfnamefont {R.~T.}\ \bibnamefont {Kedarasetti}}, \bibinfo {author} {\bibfnamefont {P.~J.}\ \bibnamefont {Drew}}, \ and\ \bibinfo {author} {\bibfnamefont {F.}~\bibnamefont {Costanzo}},\ }\bibfield  {title} {\enquote {\bibinfo {title} {Arterial vasodilation drives convective fluid flow in the brain: a poroelastic model},}\ }\href@noop {} {\bibfield  {journal} {\bibinfo  {journal} {Fluids and Barriers of the {CNS}}\ }\textbf {\bibinfo {volume} {19}},\ \bibinfo {pages} {34} (\bibinfo {year} {2020})}\BibitemShut {NoStop}%
    \bibitem [{\citenamefont {Bojarskaite}\ \emph {et~al.}(2023)\citenamefont {Bojarskaite}, \citenamefont {Bj{\o}rnstad}, \citenamefont {Vallet}, \citenamefont {{Gullestad Binder}}, \citenamefont {Cunen}, \citenamefont {Heuser}, \citenamefont {Kuchta}, \citenamefont {Mardal},\ and\ \citenamefont {Enger}}]{bojarskaite-natcomms-2023}%
      \BibitemOpen
      \bibfield  {author} {\bibinfo {author} {\bibfnamefont {L.}~\bibnamefont {Bojarskaite}}, \bibinfo {author} {\bibfnamefont {D.~M.}\ \bibnamefont {Bj{\o}rnstad}}, \bibinfo {author} {\bibfnamefont {A.}~\bibnamefont {Vallet}}, \bibinfo {author} {\bibfnamefont {K.~M.}\ \bibnamefont {{Gullestad Binder}}}, \bibinfo {author} {\bibfnamefont {C.}~\bibnamefont {Cunen}}, \bibinfo {author} {\bibfnamefont {K.}~\bibnamefont {Heuser}}, \bibinfo {author} {\bibfnamefont {M.}~\bibnamefont {Kuchta}}, \bibinfo {author} {\bibfnamefont {K.-A.}\ \bibnamefont {Mardal}}, \ and\ \bibinfo {author} {\bibfnamefont {R.}~\bibnamefont {Enger}},\ }\bibfield  {title} {\enquote {\bibinfo {title} {Sleep cycle-dependent vascular dynamics enhance perivascular cerebrospinal fluid flow and solute transport},}\ }\href@noop {} {\bibfield  {journal} {\bibinfo  {journal} {Nature Communications}\ }\textbf {\bibinfo {volume} {14}},\ \bibinfo {pages} {953} (\bibinfo {year} {2023})}\BibitemShut {NoStop}%
    \bibitem [{\citenamefont {Zhang}(2011)}]{ZhangLiHai}%
      \BibitemOpen
      \bibfield  {author} {\bibinfo {author} {\bibfnamefont {L.}~\bibnamefont {Zhang}},\ }\bibfield  {title} {\enquote {\bibinfo {title} {Solute transport in cyclic deformed heterogeneous articular cartilage},}\ }\href@noop {} {\bibfield  {journal} {\bibinfo  {journal} {International Journal of Applied Mechanics}\ }\textbf {\bibinfo {volume} {03}},\ \bibinfo {pages} {507--524} (\bibinfo {year} {2011})}\BibitemShut {NoStop}%
    \bibitem [{\citenamefont {Riches}\ \emph {et~al.}(2002)\citenamefont {Riches}, \citenamefont {Dhillon}, \citenamefont {Lotz}, \citenamefont {Woods},\ and\ \citenamefont {McNally}}]{riches2002}%
      \BibitemOpen
      \bibfield  {author} {\bibinfo {author} {\bibfnamefont {P.~E.}\ \bibnamefont {Riches}}, \bibinfo {author} {\bibfnamefont {N.}~\bibnamefont {Dhillon}}, \bibinfo {author} {\bibfnamefont {J.}~\bibnamefont {Lotz}}, \bibinfo {author} {\bibfnamefont {A.~W.}\ \bibnamefont {Woods}}, \ and\ \bibinfo {author} {\bibfnamefont {D.~S.}\ \bibnamefont {McNally}},\ }\bibfield  {title} {\enquote {\bibinfo {title} {The internal mechanics of the intervertebral disc under cyclic loading},}\ }\href {\doibase https://doi.org/10.1016/S0021-9290(02)00070-2} {\bibfield  {journal} {\bibinfo  {journal} {Journal of Biomechanics}\ }\textbf {\bibinfo {volume} {35}},\ \bibinfo {pages} {1263--1271} (\bibinfo {year} {2002})}\BibitemShut {NoStop}%
    \bibitem [{\citenamefont {Mauck}\ \emph {et~al.}(2003)\citenamefont {Mauck}, \citenamefont {Hung},\ and\ \citenamefont {Ateshian}}]{Mauck2003}%
      \BibitemOpen
      \bibfield  {author} {\bibinfo {author} {\bibfnamefont {R.~L.}\ \bibnamefont {Mauck}}, \bibinfo {author} {\bibfnamefont {C.~T.}\ \bibnamefont {Hung}}, \ and\ \bibinfo {author} {\bibfnamefont {G.~A.}\ \bibnamefont {Ateshian}},\ }\bibfield  {title} {\enquote {\bibinfo {title} {{Modeling of Neutral Solute Transport in a Dynamically Loaded Porous Permeable Gel: Implications for Articular Cartilage Biosynthesis and Tissue Engineering }},}\ }\href@noop {} {\bibfield  {journal} {\bibinfo  {journal} {Journal of Biomechanical Engineering}\ }\textbf {\bibinfo {volume} {125}},\ \bibinfo {pages} {602--614} (\bibinfo {year} {2003})}\BibitemShut {NoStop}%
    \bibitem [{\citenamefont {Sengers}\ \emph {et~al.}(2004)\citenamefont {Sengers}, \citenamefont {Oomens},\ and\ \citenamefont {Baaijens}}]{Sengers2004}%
      \BibitemOpen
      \bibfield  {author} {\bibinfo {author} {\bibfnamefont {B.~G.}\ \bibnamefont {Sengers}}, \bibinfo {author} {\bibfnamefont {C.~W.~J.}\ \bibnamefont {Oomens}}, \ and\ \bibinfo {author} {\bibfnamefont {F.~P.~T.}\ \bibnamefont {Baaijens}},\ }\bibfield  {title} {\enquote {\bibinfo {title} {{An Integrated Finite-Element Approach to Mechanics, Transport and Biosynthesis in Tissue Engineering }},}\ }\href@noop {} {\bibfield  {journal} {\bibinfo  {journal} {Journal of Biomechanical Engineering}\ }\textbf {\bibinfo {volume} {126}},\ \bibinfo {pages} {82--91} (\bibinfo {year} {2004})}\BibitemShut {NoStop}%
    \bibitem [{\citenamefont {Ferguson}\ \emph {et~al.}(2004)\citenamefont {Ferguson}, \citenamefont {Ito},\ and\ \citenamefont {Pyrak-Nolte}}]{Ferguson2004}%
      \BibitemOpen
      \bibfield  {author} {\bibinfo {author} {\bibfnamefont {S.~J.}\ \bibnamefont {Ferguson}}, \bibinfo {author} {\bibfnamefont {K.}~\bibnamefont {Ito}}, \ and\ \bibinfo {author} {\bibfnamefont {L.~J.}\ \bibnamefont {Pyrak-Nolte}},\ }\bibfield  {title} {\enquote {\bibinfo {title} {{Fluid flow and convective transport of solutes within the intervertebral disc}},}\ }\href {\doibase 10.1016/S0021-9290(03)00250-1} {\bibfield  {journal} {\bibinfo  {journal} {Journal of Biomechanics}\ }\textbf {\bibinfo {volume} {37}},\ \bibinfo {pages} {213--221} (\bibinfo {year} {2004})}\BibitemShut {NoStop}%
    \bibitem [{\citenamefont {Schmidt}\ \emph {et~al.}(2010)\citenamefont {Schmidt}, \citenamefont {Shirazi-Adl}, \citenamefont {Galbusera},\ and\ \citenamefont {Wilke}}]{Schmidt2010}%
      \BibitemOpen
      \bibfield  {author} {\bibinfo {author} {\bibfnamefont {H.}~\bibnamefont {Schmidt}}, \bibinfo {author} {\bibfnamefont {A.}~\bibnamefont {Shirazi-Adl}}, \bibinfo {author} {\bibfnamefont {F.}~\bibnamefont {Galbusera}}, \ and\ \bibinfo {author} {\bibfnamefont {H.-J.}\ \bibnamefont {Wilke}},\ }\bibfield  {title} {\enquote {\bibinfo {title} {Response analysis of the lumbar spine during regular daily activities—a finite element analysis},}\ }\href {\doibase https://doi.org/10.1016/j.jbiomech.2010.03.035} {\bibfield  {journal} {\bibinfo  {journal} {Journal of Biomechanics}\ }\textbf {\bibinfo {volume} {43}},\ \bibinfo {pages} {1849 -- 1856} (\bibinfo {year} {2010})}\BibitemShut {NoStop}%
    \bibitem [{\citenamefont {{Di Domenico}}\ \emph {et~al.}(2017)\citenamefont {{Di Domenico}}, \citenamefont {Wang},\ and\ \citenamefont {Bonassar}}]{DiDomenico2017}%
      \BibitemOpen
      \bibfield  {author} {\bibinfo {author} {\bibfnamefont {C.~D.}\ \bibnamefont {{Di Domenico}}}, \bibinfo {author} {\bibfnamefont {Z.~X.}\ \bibnamefont {Wang}}, \ and\ \bibinfo {author} {\bibfnamefont {L.~J.}\ \bibnamefont {Bonassar}},\ }\bibfield  {title} {\enquote {\bibinfo {title} {{Cyclic mechanical loading enhances transport of antibodies into articular cartilage}},}\ }\href {\doibase 10.1115/1.4035265} {\bibfield  {journal} {\bibinfo  {journal} {Journal of Biomechanical Engineering}\ }\textbf {\bibinfo {volume} {139}},\ \bibinfo {pages} {1--7} (\bibinfo {year} {2017})}\BibitemShut {NoStop}%
    \bibitem [{\citenamefont {Cacheux}\ \emph {et~al.}(2022)\citenamefont {Cacheux}, \citenamefont {Ordonez-Miranda}, \citenamefont {Bancaud}, \citenamefont {Jalabert}, \citenamefont {Nomura},\ and\ \citenamefont {Matsunaga}}]{cacheux-arxiv-2022}%
      \BibitemOpen
      \bibfield  {author} {\bibinfo {author} {\bibfnamefont {J.}~\bibnamefont {Cacheux}}, \bibinfo {author} {\bibfnamefont {J.}~\bibnamefont {Ordonez-Miranda}}, \bibinfo {author} {\bibfnamefont {A.}~\bibnamefont {Bancaud}}, \bibinfo {author} {\bibfnamefont {L.}~\bibnamefont {Jalabert}}, \bibinfo {author} {\bibfnamefont {M.}~\bibnamefont {Nomura}}, \ and\ \bibinfo {author} {\bibfnamefont {Y.~T.}\ \bibnamefont {Matsunaga}},\ }\href@noop {} {\enquote {\bibinfo {title} {Asymmetry of tensile vs. compressive elasticity and permeability contributes to the regulation of exchanges in collagen gels},}\ } (\bibinfo {year} {2022}),\ \bibinfo {note} {available at {https://arxiv.org/abs/2212.00915}}\BibitemShut {NoStop}%
    \bibitem [{\citenamefont {Piekarski}\ and\ \citenamefont {Munro}(1977)}]{piekarski-nature-1977}%
      \BibitemOpen
      \bibfield  {author} {\bibinfo {author} {\bibfnamefont {K.}~\bibnamefont {Piekarski}}\ and\ \bibinfo {author} {\bibfnamefont {M.}~\bibnamefont {Munro}},\ }\bibfield  {title} {\enquote {\bibinfo {title} {Transport mechanism operating between blood supply and osteocytes in long bones},}\ }\href@noop {} {\bibfield  {journal} {\bibinfo  {journal} {Nature}\ }\textbf {\bibinfo {volume} {269}},\ \bibinfo {pages} {80--82} (\bibinfo {year} {1977})}\BibitemShut {NoStop}%
    \bibitem [{\citenamefont {Zhang}\ and\ \citenamefont {Cowin}(1994)}]{zhang-jmps-1994}%
      \BibitemOpen
      \bibfield  {author} {\bibinfo {author} {\bibfnamefont {D.}~\bibnamefont {Zhang}}\ and\ \bibinfo {author} {\bibfnamefont {S.~C.}\ \bibnamefont {Cowin}},\ }\bibfield  {title} {\enquote {\bibinfo {title} {Oscillatory bending of a poroelastic beam},}\ }\href@noop {} {\bibfield  {journal} {\bibinfo  {journal} {Journal of the Mechanics and Physics of Solids}\ }\textbf {\bibinfo {volume} {42}},\ \bibinfo {pages} {1575--1599} (\bibinfo {year} {1994})}\BibitemShut {NoStop}%
    \bibitem [{\citenamefont {Manfredini}\ \emph {et~al.}(1999)\citenamefont {Manfredini}, \citenamefont {Cocchetti}, \citenamefont {Maier}, \citenamefont {Redaelli},\ and\ \citenamefont {Montevecchi}}]{MANFREDINI1999}%
      \BibitemOpen
      \bibfield  {author} {\bibinfo {author} {\bibfnamefont {P.}~\bibnamefont {Manfredini}}, \bibinfo {author} {\bibfnamefont {G.}~\bibnamefont {Cocchetti}}, \bibinfo {author} {\bibfnamefont {G.}~\bibnamefont {Maier}}, \bibinfo {author} {\bibfnamefont {A.}~\bibnamefont {Redaelli}}, \ and\ \bibinfo {author} {\bibfnamefont {F.~M.}\ \bibnamefont {Montevecchi}},\ }\bibfield  {title} {\enquote {\bibinfo {title} {Poroelastic finite element analysis of a bone specimen under cyclic loading},}\ }\href {\doibase https://doi.org/10.1016/S0021-9290(98)00162-6} {\bibfield  {journal} {\bibinfo  {journal} {Journal of Biomechanics}\ }\textbf {\bibinfo {volume} {32}},\ \bibinfo {pages} {135--144} (\bibinfo {year} {1999})}\BibitemShut {NoStop}%
    \bibitem [{\citenamefont {Nguyen}\ \emph {et~al.}(2010)\citenamefont {Nguyen}, \citenamefont {Lemaire},\ and\ \citenamefont {Naili}}]{NGUYEN2010}%
      \BibitemOpen
      \bibfield  {author} {\bibinfo {author} {\bibfnamefont {V.-H.}\ \bibnamefont {Nguyen}}, \bibinfo {author} {\bibfnamefont {T.}~\bibnamefont {Lemaire}}, \ and\ \bibinfo {author} {\bibfnamefont {S.}~\bibnamefont {Naili}},\ }\bibfield  {title} {\enquote {\bibinfo {title} {Poroelastic behaviour of cortical bone under harmonic axial loading: A finite element study at the osteonal scale},}\ }\href {\doibase https://doi.org/10.1016/j.medengphy.2010.02.001} {\bibfield  {journal} {\bibinfo  {journal} {Medical Engineering \& Physics}\ }\textbf {\bibinfo {volume} {32}},\ \bibinfo {pages} {384--390} (\bibinfo {year} {2010})}\BibitemShut {NoStop}%
    \bibitem [{\citenamefont {Witt}\ \emph {et~al.}(2014)\citenamefont {Witt}, \citenamefont {Duda}, \citenamefont {Bergmann},\ and\ \citenamefont {Petersen}}]{Witt2014}%
      \BibitemOpen
      \bibfield  {author} {\bibinfo {author} {\bibfnamefont {F.}~\bibnamefont {Witt}}, \bibinfo {author} {\bibfnamefont {G.~N.}\ \bibnamefont {Duda}}, \bibinfo {author} {\bibfnamefont {C.}~\bibnamefont {Bergmann}}, \ and\ \bibinfo {author} {\bibfnamefont {A.}~\bibnamefont {Petersen}},\ }\bibfield  {title} {\enquote {\bibinfo {title} {{Cyclic mechanical loading enables solute transport and oxygen supply in bone healing: An in vitro investigation}},}\ }\href {\doibase 10.1089/ten.tea.2012.0678} {\bibfield  {journal} {\bibinfo  {journal} {Tissue Engineering - Part A}\ }\textbf {\bibinfo {volume} {20}},\ \bibinfo {pages} {486--493} (\bibinfo {year} {2014})}\BibitemShut {NoStop}%
    \bibitem [{\citenamefont {Mauck}\ \emph {et~al.}(2000)\citenamefont {Mauck}, \citenamefont {Soltz}, \citenamefont {Wang}, \citenamefont {Wong}, \citenamefont {Chao}, \citenamefont {Valhmu}, \citenamefont {Hung},\ and\ \citenamefont {Ateshian}}]{Mauck2000}%
      \BibitemOpen
      \bibfield  {author} {\bibinfo {author} {\bibfnamefont {R.~L.}\ \bibnamefont {Mauck}}, \bibinfo {author} {\bibfnamefont {M.~A.}\ \bibnamefont {Soltz}}, \bibinfo {author} {\bibfnamefont {C.~C.~B.}\ \bibnamefont {Wang}}, \bibinfo {author} {\bibfnamefont {D.~D.}\ \bibnamefont {Wong}}, \bibinfo {author} {\bibfnamefont {P.-H.~G.}\ \bibnamefont {Chao}}, \bibinfo {author} {\bibfnamefont {W.~B.}\ \bibnamefont {Valhmu}}, \bibinfo {author} {\bibfnamefont {C.~T.}\ \bibnamefont {Hung}}, \ and\ \bibinfo {author} {\bibfnamefont {G.~A.}\ \bibnamefont {Ateshian}},\ }\bibfield  {title} {\enquote {\bibinfo {title} {{Functional Tissue Engineering of Articular Cartilage Through Dynamic Loading of Chondrocyte-Seeded Agarose Gels }},}\ }\href@noop {} {\bibfield  {journal} {\bibinfo  {journal} {Journal of Biomechanical Engineering}\ }\textbf {\bibinfo {volume} {122}},\ \bibinfo {pages} {252--260} (\bibinfo {year} {2000})}\BibitemShut {NoStop}%
    \bibitem [{\citenamefont {Haj}\ \emph {et~al.}(2009)\citenamefont {Haj}, \citenamefont {Hampson},\ and\ \citenamefont {Gogniat}}]{Haj2009}%
      \BibitemOpen
      \bibfield  {author} {\bibinfo {author} {\bibfnamefont {A.~J.~El}\ \bibnamefont {Haj}}, \bibinfo {author} {\bibfnamefont {K.}~\bibnamefont {Hampson}}, \ and\ \bibinfo {author} {\bibfnamefont {G.}~\bibnamefont {Gogniat}},\ }\enquote {\bibinfo {title} {Bioreactors for connective tissue engineering: Design and monitoring innovations},}\ in\ \href@noop {} {\emph {\bibinfo {booktitle} {Bioreactor Systems for Tissue Engineering}}},\ \bibinfo {editor} {edited by\ \bibinfo {editor} {\bibfnamefont {C.}~\bibnamefont {Kasper}}, \bibinfo {editor} {\bibfnamefont {M.}~\bibnamefont {van Griensven}}, \ and\ \bibinfo {editor} {\bibfnamefont {R.}~\bibnamefont {P{\"o}rtner}}}\ (\bibinfo  {publisher} {Springer Berlin Heidelberg},\ \bibinfo {year} {2009})\ pp.\ \bibinfo {pages} {81--93}\BibitemShut {NoStop}%
    \bibitem [{\citenamefont {Grenier}\ \emph {et~al.}(2005)\citenamefont {Grenier}, \citenamefont {R\'{e}my-Zolghadri}, \citenamefont {Larouche}, \citenamefont {Gauvin}, \citenamefont {Baker}, \citenamefont {Bergeron}, \citenamefont {Dupuis}, \citenamefont {Langelier}, \citenamefont {Rancourt}, \citenamefont {Auger},\ and\ \citenamefont {Germain}}]{grenier2005}%
      \BibitemOpen
      \bibfield  {author} {\bibinfo {author} {\bibfnamefont {G.}~\bibnamefont {Grenier}}, \bibinfo {author} {\bibfnamefont {M.}~\bibnamefont {R\'{e}my-Zolghadri}}, \bibinfo {author} {\bibfnamefont {D.}~\bibnamefont {Larouche}}, \bibinfo {author} {\bibfnamefont {R.}~\bibnamefont {Gauvin}}, \bibinfo {author} {\bibfnamefont {K.}~\bibnamefont {Baker}}, \bibinfo {author} {\bibfnamefont {F.}~\bibnamefont {Bergeron}}, \bibinfo {author} {\bibfnamefont {D.}~\bibnamefont {Dupuis}}, \bibinfo {author} {\bibfnamefont {E.}~\bibnamefont {Langelier}}, \bibinfo {author} {\bibfnamefont {D.}~\bibnamefont {Rancourt}}, \bibinfo {author} {\bibfnamefont {F.~A.}\ \bibnamefont {Auger}}, \ and\ \bibinfo {author} {\bibfnamefont {L.}~\bibnamefont {Germain}},\ }\bibfield  {title} {\enquote {\bibinfo {title} {Tissue reorganization in response to mechanical load increases functionality},}\ }\href {\doibase 10.1089/ten.2005.11.90} {\bibfield  {journal} {\bibinfo  {journal} {Tissue Engineering}\ }\textbf {\bibinfo {volume} {11}},\ \bibinfo {pages} {90--100} (\bibinfo {year} {2005})}\BibitemShut {NoStop}%
    \bibitem [{\citenamefont {Butler}\ \emph {et~al.}(2000)\citenamefont {Butler}, \citenamefont {Goldstein},\ and\ \citenamefont {Guilak}}]{Butler2000}%
      \BibitemOpen
      \bibfield  {author} {\bibinfo {author} {\bibfnamefont {D.~L.}\ \bibnamefont {Butler}}, \bibinfo {author} {\bibfnamefont {S.~A.}\ \bibnamefont {Goldstein}}, \ and\ \bibinfo {author} {\bibfnamefont {F.}~\bibnamefont {Guilak}},\ }\bibfield  {title} {\enquote {\bibinfo {title} {{Functional Tissue Engineering: The Role of Biomechanics }},}\ }\href {\doibase 10.1115/1.1318906} {\bibfield  {journal} {\bibinfo  {journal} {Journal of Biomechanical Engineering}\ }\textbf {\bibinfo {volume} {122}},\ \bibinfo {pages} {570--575} (\bibinfo {year} {2000})}\BibitemShut {NoStop}%
    \bibitem [{\citenamefont {Gauvin}\ \emph {et~al.}(2011)\citenamefont {Gauvin}, \citenamefont {Parenteau-Bareil}, \citenamefont {Larouche}, \citenamefont {Marcoux}, \citenamefont {Bisson}, \citenamefont {Bonnet}, \citenamefont {Auger}, \citenamefont {Bolduc},\ and\ \citenamefont {Germain}}]{GAUVIN2011}%
      \BibitemOpen
      \bibfield  {author} {\bibinfo {author} {\bibfnamefont {R.}~\bibnamefont {Gauvin}}, \bibinfo {author} {\bibfnamefont {R.}~\bibnamefont {Parenteau-Bareil}}, \bibinfo {author} {\bibfnamefont {D.}~\bibnamefont {Larouche}}, \bibinfo {author} {\bibfnamefont {H.}~\bibnamefont {Marcoux}}, \bibinfo {author} {\bibfnamefont {F.}~\bibnamefont {Bisson}}, \bibinfo {author} {\bibfnamefont {A.}~\bibnamefont {Bonnet}}, \bibinfo {author} {\bibfnamefont {F.~A.}\ \bibnamefont {Auger}}, \bibinfo {author} {\bibfnamefont {S.}~\bibnamefont {Bolduc}}, \ and\ \bibinfo {author} {\bibfnamefont {L.}~\bibnamefont {Germain}},\ }\bibfield  {title} {\enquote {\bibinfo {title} {Dynamic mechanical stimulations induce anisotropy and improve the tensile properties of engineered tissues produced without exogenous scaffolding},}\ }\href {\doibase https://doi.org/10.1016/j.actbio.2011.05.034} {\bibfield  {journal} {\bibinfo  {journal} {Acta Biomaterialia}\ }\textbf {\bibinfo {volume} {7}},\ \bibinfo {pages} {3294--3301} (\bibinfo {year} {2011})}\BibitemShut {NoStop}%
    \bibitem [{\citenamefont {Peroglio}\ \emph {et~al.}(2018)\citenamefont {Peroglio}, \citenamefont {Gaspar}, \citenamefont {Zeugolis},\ and\ \citenamefont {Alini}}]{Peroglio2018}%
      \BibitemOpen
      \bibfield  {author} {\bibinfo {author} {\bibfnamefont {M.}~\bibnamefont {Peroglio}}, \bibinfo {author} {\bibfnamefont {D.}~\bibnamefont {Gaspar}}, \bibinfo {author} {\bibfnamefont {D.~I.}\ \bibnamefont {Zeugolis}}, \ and\ \bibinfo {author} {\bibfnamefont {M.}~\bibnamefont {Alini}},\ }\bibfield  {title} {\enquote {\bibinfo {title} {Relevance of bioreactors and whole tissue cultures for the translation of new therapies to humans},}\ }\href@noop {} {\bibfield  {journal} {\bibinfo  {journal} {Journal of Orthopaedic Research}\ }\textbf {\bibinfo {volume} {36}},\ \bibinfo {pages} {10--21} (\bibinfo {year} {2018})}\BibitemShut {NoStop}%
    \bibitem [{\citenamefont {Kim}\ \emph {et~al.}(1999)\citenamefont {Kim}, \citenamefont {Nikolovski}, \citenamefont {Bonadio},\ and\ \citenamefont {Mooney}}]{Kim1999}%
      \BibitemOpen
      \bibfield  {author} {\bibinfo {author} {\bibfnamefont {B.‐S.}\ \bibnamefont {Kim}}, \bibinfo {author} {\bibfnamefont {J.}~\bibnamefont {Nikolovski}}, \bibinfo {author} {\bibfnamefont {J.}~\bibnamefont {Bonadio}}, \ and\ \bibinfo {author} {\bibfnamefont {D.~J.}\ \bibnamefont {Mooney}},\ }\bibfield  {title} {\enquote {\bibinfo {title} {Cyclic mechanical strain regulates the development of engineered smooth muscle tissue},}\ }\href@noop {} {\bibfield  {journal} {\bibinfo  {journal} {Nature Biotechnology}\ }\textbf {\bibinfo {volume} {17}},\ \bibinfo {pages} {979--983} (\bibinfo {year} {1999})}\BibitemShut {NoStop}%
    \bibitem [{\citenamefont {Amrollahi}\ and\ \citenamefont {Tayebi}(2015)}]{amrollahi2016}%
      \BibitemOpen
      \bibfield  {author} {\bibinfo {author} {\bibfnamefont {P.}~\bibnamefont {Amrollahi}}\ and\ \bibinfo {author} {\bibfnamefont {L.}~\bibnamefont {Tayebi}},\ }\bibfield  {title} {\enquote {\bibinfo {title} {Bioreactors for heart valve tissue engineering: a review},}\ }\href@noop {} {\bibfield  {journal} {\bibinfo  {journal} {Journal of Chemical Technology \& Biotechnology}\ }\textbf {\bibinfo {volume} {91}},\ \bibinfo {pages} {847--856} (\bibinfo {year} {2015})}\BibitemShut {NoStop}%
    \bibitem [{\citenamefont {Genna}\ and\ \citenamefont {Cividini}(1989)}]{GENNA1989}%
      \BibitemOpen
      \bibfield  {author} {\bibinfo {author} {\bibfnamefont {F.}~\bibnamefont {Genna}}\ and\ \bibinfo {author} {\bibfnamefont {A.}~\bibnamefont {Cividini}},\ }\bibfield  {title} {\enquote {\bibinfo {title} {Finite element analysis of fluid phase nonlinearity effects on the undrained dynamic behaviour of nearly saturated porous media},}\ }\href {\doibase https://doi.org/10.1016/S0267-7261(89)80020-X} {\bibfield  {journal} {\bibinfo  {journal} {Soil Dynamics and Earthquake Engineering}\ }\textbf {\bibinfo {volume} {8}},\ \bibinfo {pages} {189--201} (\bibinfo {year} {1989})}\BibitemShut {NoStop}%
    \bibitem [{\citenamefont {Li}\ \emph {et~al.}(2004)\citenamefont {Li}, \citenamefont {Borja},\ and\ \citenamefont {Regueiro}}]{Li2004}%
      \BibitemOpen
      \bibfield  {author} {\bibinfo {author} {\bibfnamefont {C.}~\bibnamefont {Li}}, \bibinfo {author} {\bibfnamefont {R.~I.}\ \bibnamefont {Borja}}, \ and\ \bibinfo {author} {\bibfnamefont {R.~A.}\ \bibnamefont {Regueiro}},\ }\bibfield  {title} {\enquote {\bibinfo {title} {{Dynamics of porous media at finite strain}},}\ }\href {\doibase 10.1016/j.cma.2004.02.014} {\bibfield  {journal} {\bibinfo  {journal} {Computer Methods in Applied Mechanics and Engineering}\ }\textbf {\bibinfo {volume} {193}},\ \bibinfo {pages} {3837--3870} (\bibinfo {year} {2004})}\BibitemShut {NoStop}%
    \bibitem [{\citenamefont {Popescu}\ \emph {et~al.}(2006)\citenamefont {Popescu}, \citenamefont {Prevost}, \citenamefont {Deodatis},\ and\ \citenamefont {Chakrabortty}}]{POPESCU2006648}%
      \BibitemOpen
      \bibfield  {author} {\bibinfo {author} {\bibfnamefont {R.}~\bibnamefont {Popescu}}, \bibinfo {author} {\bibfnamefont {J.~H.}\ \bibnamefont {Prevost}}, \bibinfo {author} {\bibfnamefont {G.}~\bibnamefont {Deodatis}}, \ and\ \bibinfo {author} {\bibfnamefont {P.}~\bibnamefont {Chakrabortty}},\ }\bibfield  {title} {\enquote {\bibinfo {title} {Dynamics of nonlinear porous media with applications to soil liquefaction},}\ }\href {\doibase https://doi.org/10.1016/j.soildyn.2006.01.015} {\bibfield  {journal} {\bibinfo  {journal} {Soil Dynamics and Earthquake Engineering}\ }\textbf {\bibinfo {volume} {26}},\ \bibinfo {pages} {648--665} (\bibinfo {year} {2006})}\BibitemShut {NoStop}%
    \bibitem [{\citenamefont {Bonazzi}\ \emph {et~al.}(2021)\citenamefont {Bonazzi}, \citenamefont {Jha},\ and\ \citenamefont {{de Barros}}}]{Bonazzi2021}%
      \BibitemOpen
      \bibfield  {author} {\bibinfo {author} {\bibfnamefont {A.}~\bibnamefont {Bonazzi}}, \bibinfo {author} {\bibfnamefont {B.}~\bibnamefont {Jha}}, \ and\ \bibinfo {author} {\bibfnamefont {F.~P.~J.}\ \bibnamefont {{de Barros}}},\ }\bibfield  {title} {\enquote {\bibinfo {title} {Transport analysis in deformable porous media through integral transforms},}\ }\href {\doibase https://doi.org/10.1002/nag.3150} {\bibfield  {journal} {\bibinfo  {journal} {International Journal for Numerical and Analytical Methods in Geomechanics}\ }\textbf {\bibinfo {volume} {45}},\ \bibinfo {pages} {307--324} (\bibinfo {year} {2021})}\BibitemShut {NoStop}%
    \bibitem [{\citenamefont {Hu}\ \emph {et~al.}(2011)\citenamefont {Hu}, \citenamefont {Zhu},\ and\ \citenamefont {Cheng}}]{HU2011}%
      \BibitemOpen
      \bibfield  {author} {\bibinfo {author} {\bibfnamefont {Y.-J.}\ \bibnamefont {Hu}}, \bibinfo {author} {\bibfnamefont {Y.-Y.}\ \bibnamefont {Zhu}}, \ and\ \bibinfo {author} {\bibfnamefont {C.-J.}\ \bibnamefont {Cheng}},\ }\bibfield  {title} {\enquote {\bibinfo {title} {Transient dynamic response of fluid-saturated soil under a moving cyclic loading},}\ }\href {\doibase https://doi.org/10.1016/j.soildyn.2010.11.004} {\bibfield  {journal} {\bibinfo  {journal} {Soil Dynamics and Earthquake Engineering}\ }\textbf {\bibinfo {volume} {31}},\ \bibinfo {pages} {491--501} (\bibinfo {year} {2011})}\BibitemShut {NoStop}%
    \bibitem [{\citenamefont {Ni}\ \emph {et~al.}(2015)\citenamefont {Ni}, \citenamefont {Indraratna}, \citenamefont {Geng}, \citenamefont {Carter},\ and\ \citenamefont {Chen}}]{Ni2015}%
      \BibitemOpen
      \bibfield  {author} {\bibinfo {author} {\bibfnamefont {J.}~\bibnamefont {Ni}}, \bibinfo {author} {\bibfnamefont {B.}~\bibnamefont {Indraratna}}, \bibinfo {author} {\bibfnamefont {X.-Y.}\ \bibnamefont {Geng}}, \bibinfo {author} {\bibfnamefont {J.~P.}\ \bibnamefont {Carter}}, \ and\ \bibinfo {author} {\bibfnamefont {Y.-L.}\ \bibnamefont {Chen}},\ }\bibfield  {title} {\enquote {\bibinfo {title} {Model of soft soils under cyclic loading},}\ }\href {\doibase 10.1061/(ASCE)GM.1943-5622.0000411} {\bibfield  {journal} {\bibinfo  {journal} {International Journal of Geomechanics}\ }\textbf {\bibinfo {volume} {15}},\ \bibinfo {pages} {04014067} (\bibinfo {year} {2015})}\BibitemShut {NoStop}%
    \bibitem [{\citenamefont {Ni}\ and\ \citenamefont {Geng}(2022)}]{NI2022}%
      \BibitemOpen
      \bibfield  {author} {\bibinfo {author} {\bibfnamefont {J.}~\bibnamefont {Ni}}\ and\ \bibinfo {author} {\bibfnamefont {X.-Y.}\ \bibnamefont {Geng}},\ }\bibfield  {title} {\enquote {\bibinfo {title} {Radial consolidation of prefabricated vertical drain-reinforced soft clays under cyclic loading},}\ }\href {\doibase https://doi.org/10.1016/j.trgeo.2022.100840} {\bibfield  {journal} {\bibinfo  {journal} {Transportation Geotechnics}\ }\textbf {\bibinfo {volume} {37}},\ \bibinfo {pages} {100840} (\bibinfo {year} {2022})}\BibitemShut {NoStop}%
    \bibitem [{\citenamefont {Yamamoto}\ \emph {et~al.}(1978)\citenamefont {Yamamoto}, \citenamefont {Koning}, \citenamefont {Sellmeijer},\ and\ \citenamefont {{van Hijum}}}]{yamamoto-jfm-1978}%
      \BibitemOpen
      \bibfield  {author} {\bibinfo {author} {\bibfnamefont {T.}~\bibnamefont {Yamamoto}}, \bibinfo {author} {\bibfnamefont {H.~L.}\ \bibnamefont {Koning}}, \bibinfo {author} {\bibfnamefont {H.}~\bibnamefont {Sellmeijer}}, \ and\ \bibinfo {author} {\bibfnamefont {E.}~\bibnamefont {{van Hijum}}},\ }\bibfield  {title} {\enquote {\bibinfo {title} {On the response of a poro-elastic bed to water waves},}\ }\href@noop {} {\bibfield  {journal} {\bibinfo  {journal} {Journal of Fluid Mechanics}\ }\textbf {\bibinfo {volume} {87}},\ \bibinfo {pages} {193--206} (\bibinfo {year} {1978})}\BibitemShut {NoStop}%
    \bibitem [{\citenamefont {Madsen}(1978)}]{madsen-geotechnique-1978}%
      \BibitemOpen
      \bibfield  {author} {\bibinfo {author} {\bibfnamefont {O.~S.}\ \bibnamefont {Madsen}},\ }\bibfield  {title} {\enquote {\bibinfo {title} {Wave-induced pore pressures and effective stresses in a porous bed},}\ }\href@noop {} {\bibfield  {journal} {\bibinfo  {journal} {G{\'{e}}otechnique}\ }\textbf {\bibinfo {volume} {28}},\ \bibinfo {pages} {377--393} (\bibinfo {year} {1978})}\BibitemShut {NoStop}%
    \bibitem [{\citenamefont {Karim}\ \emph {et~al.}(2002)\citenamefont {Karim}, \citenamefont {Nogami},\ and\ \citenamefont {Wang}}]{karim-intjsolidstruct-2002}%
      \BibitemOpen
      \bibfield  {author} {\bibinfo {author} {\bibfnamefont {M.~R.}\ \bibnamefont {Karim}}, \bibinfo {author} {\bibfnamefont {T.}~\bibnamefont {Nogami}}, \ and\ \bibinfo {author} {\bibfnamefont {J.~G.}\ \bibnamefont {Wang}},\ }\bibfield  {title} {\enquote {\bibinfo {title} {Analysis of transient response of saturated porous elastic soil under cyclic loading using element-free {Galerkin} method},}\ }\href@noop {} {\bibfield  {journal} {\bibinfo  {journal} {International Journal of Solids and Structures}\ }\textbf {\bibinfo {volume} {39}},\ \bibinfo {pages} {6011--6033} (\bibinfo {year} {2002})}\BibitemShut {NoStop}%
    \bibitem [{\citenamefont {Cheng}(2016)}]{cheng-book-2016}%
      \BibitemOpen
      \bibfield  {author} {\bibinfo {author} {\bibfnamefont {A.~H.-D.}\ \bibnamefont {Cheng}},\ }\href@noop {} {\emph {\bibinfo {title} {Poroelasticity}}},\ \bibinfo {series} {Theory and Applications of Transport in Porous Media}, Vol.~\bibinfo {volume} {27}\ (\bibinfo  {publisher} {Springer},\ \bibinfo {year} {2016})\BibitemShut {NoStop}%
    \bibitem [{\citenamefont {Trefry}\ \emph {et~al.}(2019)\citenamefont {Trefry}, \citenamefont {Lester}, \citenamefont {Metcalfe},\ and\ \citenamefont {Wu}}]{Trefry2019}%
      \BibitemOpen
      \bibfield  {author} {\bibinfo {author} {\bibfnamefont {M.~G.}\ \bibnamefont {Trefry}}, \bibinfo {author} {\bibfnamefont {D.~R.}\ \bibnamefont {Lester}}, \bibinfo {author} {\bibfnamefont {G.}~\bibnamefont {Metcalfe}}, \ and\ \bibinfo {author} {\bibfnamefont {J.}~\bibnamefont {Wu}},\ }\bibfield  {title} {\enquote {\bibinfo {title} {{Temporal Fluctuations and Poroelasticity Can Generate Chaotic Advection in Natural Groundwater Systems}},}\ }\href {\doibase 10.1029/2018WR023864} {\bibfield  {journal} {\bibinfo  {journal} {Water Resources Research}\ }\textbf {\bibinfo {volume} {55}},\ \bibinfo {pages} {3347--3374} (\bibinfo {year} {2019})}\BibitemShut {NoStop}%
    \bibitem [{\citenamefont {Biot}(1956{\natexlab{a}})}]{BiotA}%
      \BibitemOpen
      \bibfield  {author} {\bibinfo {author} {\bibfnamefont {M.~A.}\ \bibnamefont {Biot}},\ }\bibfield  {title} {\enquote {\bibinfo {title} {Theory of propagation of elastic waves in a fluid‐saturated porous solid. i. low‐frequency range},}\ }\href {\doibase 10.1121/1.1908239} {\bibfield  {journal} {\bibinfo  {journal} {The Journal of the Acoustical Society of America}\ }\textbf {\bibinfo {volume} {28}},\ \bibinfo {pages} {168--178} (\bibinfo {year} {1956}{\natexlab{a}})}\BibitemShut {NoStop}%
    \bibitem [{\citenamefont {Biot}(1956{\natexlab{b}})}]{BiotB}%
      \BibitemOpen
      \bibfield  {author} {\bibinfo {author} {\bibfnamefont {M.~A.}\ \bibnamefont {Biot}},\ }\bibfield  {title} {\enquote {\bibinfo {title} {Theory of propagation of elastic waves in a fluid‐saturated porous solid. ii. higher frequency range},}\ }\href {\doibase 10.1121/1.1908241} {\bibfield  {journal} {\bibinfo  {journal} {The Journal of the Acoustical Society of America}\ }\textbf {\bibinfo {volume} {28}},\ \bibinfo {pages} {179--191} (\bibinfo {year} {1956}{\natexlab{b}})}\BibitemShut {NoStop}%
    \bibitem [{\citenamefont {Gajo}\ and\ \citenamefont {Denzer}(2011)}]{Gajo2012}%
      \BibitemOpen
      \bibfield  {author} {\bibinfo {author} {\bibfnamefont {A.}~\bibnamefont {Gajo}}\ and\ \bibinfo {author} {\bibfnamefont {R.}~\bibnamefont {Denzer}},\ }\bibfield  {title} {\enquote {\bibinfo {title} {Finite element modelling of saturated porous media at finite strains under dynamic conditions with compressible constituents},}\ }\href@noop {} {\bibfield  {journal} {\bibinfo  {journal} {International Journal for Numerical Methods in Engineering}\ }\textbf {\bibinfo {volume} {85}},\ \bibinfo {pages} {1705--1736} (\bibinfo {year} {2011})}\BibitemShut {NoStop}%
    \bibitem [{\citenamefont {Liu}\ \emph {et~al.}(2019)\citenamefont {Liu}, \citenamefont {Li}, \citenamefont {Liu},\ and\ \citenamefont {Han}}]{Liu2019}%
      \BibitemOpen
      \bibfield  {author} {\bibinfo {author} {\bibfnamefont {J.}~\bibnamefont {Liu}}, \bibinfo {author} {\bibfnamefont {X.}~\bibnamefont {Li}}, \bibinfo {author} {\bibfnamefont {J.}~\bibnamefont {Liu}}, \ and\ \bibinfo {author} {\bibfnamefont {B.}~\bibnamefont {Han}},\ }\bibfield  {title} {\enquote {\bibinfo {title} {{Numerical Investigation of Transition Mechanism between the Two Kinds of Compressional Waves in Saturated Geotechnical Media}},}\ }\href@noop {} {\bibfield  {journal} {\bibinfo  {journal} {International Journal of Geomechanics}\ }\textbf {\bibinfo {volume} {19}},\ \bibinfo {pages} {1--9} (\bibinfo {year} {2019})}\BibitemShut {NoStop}%
    \bibitem [{\citenamefont {Kameo}\ \emph {et~al.}(2008)\citenamefont {Kameo}, \citenamefont {Adachi},\ and\ \citenamefont {Hojo}}]{KAMEO2008}%
      \BibitemOpen
      \bibfield  {author} {\bibinfo {author} {\bibfnamefont {Y.}~\bibnamefont {Kameo}}, \bibinfo {author} {\bibfnamefont {T.}~\bibnamefont {Adachi}}, \ and\ \bibinfo {author} {\bibfnamefont {M.}~\bibnamefont {Hojo}},\ }\bibfield  {title} {\enquote {\bibinfo {title} {Transient response of fluid pressure in a poroelastic material under uniaxial cyclic loading},}\ }\href {\doibase https://doi.org/10.1016/j.jmps.2007.11.008} {\bibfield  {journal} {\bibinfo  {journal} {Journal of the Mechanics and Physics of Solids}\ }\textbf {\bibinfo {volume} {56}},\ \bibinfo {pages} {1794--1805} (\bibinfo {year} {2008})}\BibitemShut {NoStop}%
    \bibitem [{\citenamefont {Yaogeng}\ \emph {et~al.}(2018)\citenamefont {Yaogeng}, \citenamefont {Wenshuai}, \citenamefont {Shenghu}, \citenamefont {Xu}, \citenamefont {Qun},\ and\ \citenamefont {Xing}}]{chen2018}%
      \BibitemOpen
      \bibfield  {author} {\bibinfo {author} {\bibfnamefont {C.}~\bibnamefont {Yaogeng}}, \bibinfo {author} {\bibfnamefont {W.}~\bibnamefont {Wenshuai}}, \bibinfo {author} {\bibfnamefont {D.}~\bibnamefont {Shenghu}}, \bibinfo {author} {\bibfnamefont {W.}~\bibnamefont {Xu}}, \bibinfo {author} {\bibfnamefont {C.}~\bibnamefont {Qun}}, \ and\ \bibinfo {author} {\bibfnamefont {L.}~\bibnamefont {Xing}},\ }\bibfield  {title} {\enquote {\bibinfo {title} {A multi-layered poroelastic slab model under cyclic loading for a single osteon},}\ }\href@noop {} {\bibfield  {journal} {\bibinfo  {journal} {BioMedical Engineering OnLine}\ }\textbf {\bibinfo {volume} {17}},\ \bibinfo {pages} {97} (\bibinfo {year} {2018})}\BibitemShut {NoStop}%
    \bibitem [{\citenamefont {Gardiner}\ \emph {et~al.}(2007)\citenamefont {Gardiner}, \citenamefont {Smith}, \citenamefont {Pivonka}, \citenamefont {Grodzinsky}, \citenamefont {Frank},\ and\ \citenamefont {Zhang}}]{gardiner2007}%
      \BibitemOpen
      \bibfield  {author} {\bibinfo {author} {\bibfnamefont {B.}~\bibnamefont {Gardiner}}, \bibinfo {author} {\bibfnamefont {D.}~\bibnamefont {Smith}}, \bibinfo {author} {\bibfnamefont {P.}~\bibnamefont {Pivonka}}, \bibinfo {author} {\bibfnamefont {A.}~\bibnamefont {Grodzinsky}}, \bibinfo {author} {\bibfnamefont {E.}~\bibnamefont {Frank}}, \ and\ \bibinfo {author} {\bibfnamefont {L.}~\bibnamefont {Zhang}},\ }\bibfield  {title} {\enquote {\bibinfo {title} {Solute transport in cartilage undergoing cyclic deformation},}\ }\href@noop {} {\bibfield  {journal} {\bibinfo  {journal} {Computer Methods in Biomechanics and Biomedical Engineering}\ }\textbf {\bibinfo {volume} {10}},\ \bibinfo {pages} {265--278} (\bibinfo {year} {2007})}\BibitemShut {NoStop}%
    \bibitem [{\citenamefont {Urciuolo}\ \emph {et~al.}(2008)\citenamefont {Urciuolo}, \citenamefont {Imparato},\ and\ \citenamefont {Netti}}]{Urciuolo2008}%
      \BibitemOpen
      \bibfield  {author} {\bibinfo {author} {\bibfnamefont {F.}~\bibnamefont {Urciuolo}}, \bibinfo {author} {\bibfnamefont {G.}~\bibnamefont {Imparato}}, \ and\ \bibinfo {author} {\bibfnamefont {P.~A.}\ \bibnamefont {Netti}},\ }\bibfield  {title} {\enquote {\bibinfo {title} {{Effect of Dynamic Loading on Solute Transport in Soft Gels Implication for Drug Delivery}},}\ }\href@noop {} {\bibfield  {journal} {\bibinfo  {journal} {AIChE Journal}\ }\textbf {\bibinfo {volume} {54}},\ \bibinfo {pages} {824--834} (\bibinfo {year} {2008})}\BibitemShut {NoStop}%
    \bibitem [{\citenamefont {Vaughan}\ \emph {et~al.}(2013)\citenamefont {Vaughan}, \citenamefont {Galie}, \citenamefont {Stegemann},\ and\ \citenamefont {Grotberg}}]{Vaughan2013}%
      \BibitemOpen
      \bibfield  {author} {\bibinfo {author} {\bibfnamefont {B.~L.}\ \bibnamefont {Vaughan}}, \bibinfo {author} {\bibfnamefont {P.~A.}\ \bibnamefont {Galie}}, \bibinfo {author} {\bibfnamefont {J.~P.}\ \bibnamefont {Stegemann}}, \ and\ \bibinfo {author} {\bibfnamefont {J.~B.}\ \bibnamefont {Grotberg}},\ }\bibfield  {title} {\enquote {\bibinfo {title} {{A poroelastic model describing nutrient transport and cell stresses within a cyclically strained collagen hydrogel}},}\ }\href {\doibase 10.1016/j.bpj.2013.08.048} {\bibfield  {journal} {\bibinfo  {journal} {Biophysical Journal}\ }\textbf {\bibinfo {volume} {105}},\ \bibinfo {pages} {2188--2198} (\bibinfo {year} {2013})}\BibitemShut {NoStop}%
    \bibitem [{\citenamefont {MacMinn}\ \emph {et~al.}(2016)\citenamefont {MacMinn}, \citenamefont {Dufresne},\ and\ \citenamefont {Wettlaufer}}]{Macminn2016}%
      \BibitemOpen
      \bibfield  {author} {\bibinfo {author} {\bibfnamefont {C.~W.}\ \bibnamefont {MacMinn}}, \bibinfo {author} {\bibfnamefont {E.~R.}\ \bibnamefont {Dufresne}}, \ and\ \bibinfo {author} {\bibfnamefont {J.~S.}\ \bibnamefont {Wettlaufer}},\ }\bibfield  {title} {\enquote {\bibinfo {title} {{Large Deformations of a Soft Porous Material}},}\ }\href@noop {} {\bibfield  {journal} {\bibinfo  {journal} {Physical Review Applied}\ }\textbf {\bibinfo {volume} {5}},\ \bibinfo {pages} {044020} (\bibinfo {year} {2016})}\BibitemShut {NoStop}%
    \bibitem [{\citenamefont {Sacco}\ \emph {et~al.}(2014)\citenamefont {Sacco}, \citenamefont {Causin}, \citenamefont {Zunino},\ and\ \citenamefont {Raimondi}}]{Sacco2011}%
      \BibitemOpen
      \bibfield  {author} {\bibinfo {author} {\bibfnamefont {Riccardo}\ \bibnamefont {Sacco}}, \bibinfo {author} {\bibfnamefont {Paola}\ \bibnamefont {Causin}}, \bibinfo {author} {\bibfnamefont {Paolo}\ \bibnamefont {Zunino}}, \ and\ \bibinfo {author} {\bibfnamefont {Manuela~T.}\ \bibnamefont {Raimondi}},\ }\bibfield  {title} {\enquote {\bibinfo {title} {A multiphysics/multiscale 2d numerical simulation of scaffold-based cartilage regeneration under interstitial perfusion in a bioreactor},}\ }\href@noop {} {\bibfield  {journal} {\bibinfo  {journal} {Biomechanics and Modeling in Mechanobiology}\ }\textbf {\bibinfo {volume} {10}},\ \bibinfo {pages} {577--589} (\bibinfo {year} {2014})}\BibitemShut {NoStop}%
    \bibitem [{\citenamefont {Malandrino}\ \emph {et~al.}(2014)\citenamefont {Malandrino}, \citenamefont {Lacroix}, \citenamefont {Hellmich}, \citenamefont {Ito}, \citenamefont {Ferguson},\ and\ \citenamefont {Noailly}}]{MALANDRINO2014}%
      \BibitemOpen
      \bibfield  {author} {\bibinfo {author} {\bibfnamefont {A.}~\bibnamefont {Malandrino}}, \bibinfo {author} {\bibfnamefont {D.}~\bibnamefont {Lacroix}}, \bibinfo {author} {\bibfnamefont {C.}~\bibnamefont {Hellmich}}, \bibinfo {author} {\bibfnamefont {K.}~\bibnamefont {Ito}}, \bibinfo {author} {\bibfnamefont {S.J.}\ \bibnamefont {Ferguson}}, \ and\ \bibinfo {author} {\bibfnamefont {J.}~\bibnamefont {Noailly}},\ }\bibfield  {title} {\enquote {\bibinfo {title} {The role of endplate poromechanical properties on the nutrient availability in the intervertebral disc},}\ }\href@noop {} {\bibfield  {journal} {\bibinfo  {journal} {Osteoarthritis and Cartilage}\ }\textbf {\bibinfo {volume} {22}},\ \bibinfo {pages} {1053--1060} (\bibinfo {year} {2014})}\BibitemShut {NoStop}%
    \bibitem [{\citenamefont {Rahbari}\ \emph {et~al.}(2017)\citenamefont {Rahbari}, \citenamefont {Montazerian}, \citenamefont {Davoodi},\ and\ \citenamefont {Homayoonfar}}]{Rahbari2016}%
      \BibitemOpen
      \bibfield  {author} {\bibinfo {author} {\bibfnamefont {A.}~\bibnamefont {Rahbari}}, \bibinfo {author} {\bibfnamefont {H.}~\bibnamefont {Montazerian}}, \bibinfo {author} {\bibfnamefont {E.}~\bibnamefont {Davoodi}}, \ and\ \bibinfo {author} {\bibfnamefont {S.}~\bibnamefont {Homayoonfar}},\ }\bibfield  {title} {\enquote {\bibinfo {title} {Predicting permeability of regular tissue engineering scaffolds: scaling analysis of pore architecture, scaffold length, and fluid flow rate effects},}\ }\href@noop {} {\bibfield  {journal} {\bibinfo  {journal} {Computer Methods in Biomechanics and Biomedical Engineering}\ }\textbf {\bibinfo {volume} {20}},\ \bibinfo {pages} {231--241} (\bibinfo {year} {2017})}\BibitemShut {NoStop}%
    \bibitem [{\citenamefont {Gao}\ and\ \citenamefont {Cho}(2022)}]{Gao2022}%
      \BibitemOpen
      \bibfield  {author} {\bibinfo {author} {\bibfnamefont {Yiwei}\ \bibnamefont {Gao}}\ and\ \bibinfo {author} {\bibfnamefont {H.~Jeremy}\ \bibnamefont {Cho}},\ }\bibfield  {title} {\enquote {\bibinfo {title} {Quantifying the trade-off between stiffness and permeability in hydrogels},}\ }\href@noop {} {\bibfield  {journal} {\bibinfo  {journal} {Soft Matter}\ }\textbf {\bibinfo {volume} {18}},\ \bibinfo {pages} {7735--7740} (\bibinfo {year} {2022})}\BibitemShut {NoStop}%
    \bibitem [{\citenamefont {Hencky}(1931)}]{hencky}%
      \BibitemOpen
      \bibfield  {author} {\bibinfo {author} {\bibfnamefont {H.}~\bibnamefont {Hencky}},\ }\bibfield  {title} {\enquote {\bibinfo {title} {The law of elasticity for isotropic and quasi‐isotropic substances by finite deformations},}\ }\href@noop {} {\bibfield  {journal} {\bibinfo  {journal} {Journal of Rheology}\ }\textbf {\bibinfo {volume} {2}},\ \bibinfo {pages} {169--176} (\bibinfo {year} {1931})}\BibitemShut {NoStop}%
    \bibitem [{\citenamefont {Hencky}(1933)}]{Hencky33}%
      \BibitemOpen
      \bibfield  {author} {\bibinfo {author} {\bibfnamefont {H.}~\bibnamefont {Hencky}},\ }\bibfield  {title} {\enquote {\bibinfo {title} {{The Elastic Behavior of Vulcanized Rubber}},}\ }\href@noop {} {\bibfield  {journal} {\bibinfo  {journal} {Rubber Chemistry and Technology}\ }\textbf {\bibinfo {volume} {6}},\ \bibinfo {pages} {217--224} (\bibinfo {year} {1933})}\BibitemShut {NoStop}%
    \bibitem [{\citenamefont {Anand}(1979)}]{Anand1979}%
      \BibitemOpen
      \bibfield  {author} {\bibinfo {author} {\bibfnamefont {L.}~\bibnamefont {Anand}},\ }\bibfield  {title} {\enquote {\bibinfo {title} {{On H. Hencky’s Approximate Strain-Energy Function for Moderate Deformations}},}\ }\href@noop {} {\bibfield  {journal} {\bibinfo  {journal} {Journal of Applied Mechanics}\ }\textbf {\bibinfo {volume} {46}},\ \bibinfo {pages} {78--82} (\bibinfo {year} {1979})}\BibitemShut {NoStop}%
    \bibitem [{\citenamefont {Xiao}\ and\ \citenamefont {Chen}(2002)}]{xiao2002}%
      \BibitemOpen
      \bibfield  {author} {\bibinfo {author} {\bibfnamefont {H.}~\bibnamefont {Xiao}}\ and\ \bibinfo {author} {\bibfnamefont {L.~S.}\ \bibnamefont {Chen}},\ }\bibfield  {title} {\enquote {\bibinfo {title} {Hencky's elasticity model and linear stress-strain relations in isotropic finite hyperelasticity},}\ }\href@noop {} {\bibfield  {journal} {\bibinfo  {journal} {Acta Mechanica}\ }\textbf {\bibinfo {volume} {157}},\ \bibinfo {pages} {51--60} (\bibinfo {year} {2002})}\BibitemShut {NoStop}%
    \bibitem [{\citenamefont {Marchesseau}\ \emph {et~al.}(2010)\citenamefont {Marchesseau}, \citenamefont {Heimann}, \citenamefont {Chatelin}, \citenamefont {Willinger},\ and\ \citenamefont {H.}}]{MARCHESSEAU2010185}%
      \BibitemOpen
      \bibfield  {author} {\bibinfo {author} {\bibfnamefont {S.}~\bibnamefont {Marchesseau}}, \bibinfo {author} {\bibfnamefont {T.}~\bibnamefont {Heimann}}, \bibinfo {author} {\bibfnamefont {S.}~\bibnamefont {Chatelin}}, \bibinfo {author} {\bibfnamefont {R.}~\bibnamefont {Willinger}}, \ and\ \bibinfo {author} {\bibfnamefont {Delingette}\ \bibnamefont {H.}},\ }\bibfield  {title} {\enquote {\bibinfo {title} {Fast porous visco-hyperelastic soft tissue model for surgery simulation: Application to liver surgery},}\ }\href@noop {} {\bibfield  {journal} {\bibinfo  {journal} {Progress in Biophysics and Molecular Biology}\ }\textbf {\bibinfo {volume} {103}},\ \bibinfo {pages} {185--196} (\bibinfo {year} {2010})},\ \bibinfo {note} {special Issue on Biomechanical Modelling of Soft Tissue Motion}\BibitemShut {NoStop}%
    \bibitem [{\citenamefont {Fraldi}\ \emph {et~al.}(2018)\citenamefont {Fraldi}, \citenamefont {Palumbo}, \citenamefont {Carotenuto}, \citenamefont {Cutolo}, \citenamefont {Deseri},\ and\ \citenamefont {Pugno}}]{Fraldi2018}%
      \BibitemOpen
      \bibfield  {author} {\bibinfo {author} {\bibfnamefont {M.}~\bibnamefont {Fraldi}}, \bibinfo {author} {\bibfnamefont {S.}~\bibnamefont {Palumbo}}, \bibinfo {author} {\bibfnamefont {A.}~\bibnamefont {Carotenuto}}, \bibinfo {author} {\bibfnamefont {A.}~\bibnamefont {Cutolo}}, \bibinfo {author} {\bibfnamefont {L.}~\bibnamefont {Deseri}}, \ and\ \bibinfo {author} {\bibfnamefont {N.}~\bibnamefont {Pugno}},\ }\bibfield  {title} {\enquote {\bibinfo {title} {Buckling soft tensegrities: Fickle elasticity and configurational switching in living cells},}\ }\href@noop {} {\bibfield  {journal} {\bibinfo  {journal} {Journal of the Mechanics and Physics of Solids}\ }\textbf {\bibinfo {volume} {124}} (\bibinfo {year} {2018})}\BibitemShut {NoStop}%
    \bibitem [{\citenamefont {Ehlers}\ \emph {et~al.}(2009)\citenamefont {Ehlers}, \citenamefont {Karajan},\ and\ \citenamefont {Markert}}]{Ehlers2009}%
      \BibitemOpen
      \bibfield  {author} {\bibinfo {author} {\bibfnamefont {W.}~\bibnamefont {Ehlers}}, \bibinfo {author} {\bibfnamefont {N.}~\bibnamefont {Karajan}}, \ and\ \bibinfo {author} {\bibfnamefont {B.}~\bibnamefont {Markert}},\ }\bibfield  {title} {\enquote {\bibinfo {title} {{An extended biphasic model for charged hydrated tissues with application to the intervertebral disc}},}\ }\href {\doibase 10.1007/s10237-008-0129-y} {\bibfield  {journal} {\bibinfo  {journal} {Biomechanics and Modeling in Mechanobiology}\ }\textbf {\bibinfo {volume} {8}},\ \bibinfo {pages} {233--251} (\bibinfo {year} {2009})}\BibitemShut {NoStop}%
    \bibitem [{\citenamefont {Auton}\ and\ \citenamefont {MacMinn}(2018)}]{Auton2018}%
      \BibitemOpen
      \bibfield  {author} {\bibinfo {author} {\bibfnamefont {L.~C.}\ \bibnamefont {Auton}}\ and\ \bibinfo {author} {\bibfnamefont {C.~W.}\ \bibnamefont {MacMinn}},\ }\bibfield  {title} {\enquote {\bibinfo {title} {{From arteries to boreholes: Transient response of a poroelastic cylinder to fluid injection}},}\ }\href@noop {} {\bibfield  {journal} {\bibinfo  {journal} {Proceedings of the Royal Society A: Mathematical, Physical and Engineering Sciences}\ }\textbf {\bibinfo {volume} {474}} (\bibinfo {year} {2018})}\BibitemShut {NoStop}%
    \bibitem [{\citenamefont {Hewitt}\ \emph {et~al.}(2016)\citenamefont {Hewitt}, \citenamefont {Paterson}, \citenamefont {Balmforth},\ and\ \citenamefont {Martinez}}]{hewitt2016}%
      \BibitemOpen
      \bibfield  {author} {\bibinfo {author} {\bibfnamefont {D.~R.}\ \bibnamefont {Hewitt}}, \bibinfo {author} {\bibfnamefont {D.~T.}\ \bibnamefont {Paterson}}, \bibinfo {author} {\bibfnamefont {N.~J.}\ \bibnamefont {Balmforth}}, \ and\ \bibinfo {author} {\bibfnamefont {D.~M.}\ \bibnamefont {Martinez}},\ }\bibfield  {title} {\enquote {\bibinfo {title} {Dewatering of fibre suspensions by pressure filtration},}\ }\href {\doibase 10.1063/1.4952582} {\bibfield  {journal} {\bibinfo  {journal} {Physics of Fluids}\ }\textbf {\bibinfo {volume} {28}},\ \bibinfo {pages} {063304} (\bibinfo {year} {2016})}\BibitemShut {NoStop}%
    \bibitem [{\citenamefont {Sobac}\ \emph {et~al.}(2011)\citenamefont {Sobac}, \citenamefont {Colombani},\ and\ \citenamefont {Forterre}}]{sobac-mecind-2011}%
      \BibitemOpen
      \bibfield  {author} {\bibinfo {author} {\bibfnamefont {B.}~\bibnamefont {Sobac}}, \bibinfo {author} {\bibfnamefont {M.}~\bibnamefont {Colombani}}, \ and\ \bibinfo {author} {\bibfnamefont {Y.}~\bibnamefont {Forterre}},\ }\bibfield  {title} {\enquote {\bibinfo {title} {On the dynamics of poroelastic foams (\emph{in French})},}\ }\href@noop {} {\bibfield  {journal} {\bibinfo  {journal} {M\'{e}canique \& Industries}\ }\textbf {\bibinfo {volume} {12}},\ \bibinfo {pages} {231--238} (\bibinfo {year} {2011})}\BibitemShut {NoStop}%
    \bibitem [{\citenamefont {Lutz}\ \emph {et~al.}(2021)\citenamefont {Lutz}, \citenamefont {Wilen},\ and\ \citenamefont {Wettlaufer}}]{lutz-rsi-2021}%
      \BibitemOpen
      \bibfield  {author} {\bibinfo {author} {\bibfnamefont {T.}~\bibnamefont {Lutz}}, \bibinfo {author} {\bibfnamefont {L.}~\bibnamefont {Wilen}}, \ and\ \bibinfo {author} {\bibfnamefont {J.}~\bibnamefont {Wettlaufer}},\ }\bibfield  {title} {\enquote {\bibinfo {title} {A method for measuring fluid pressure and solid deformation profiles in uniaxial porous media flows},}\ }\href@noop {} {\bibfield  {journal} {\bibinfo  {journal} {Review of Scientific Instruments}\ }\textbf {\bibinfo {volume} {92}},\ \bibinfo {pages} {025101} (\bibinfo {year} {2021})}\BibitemShut {NoStop}%
    \bibitem [{\citenamefont {Holmes}\ and\ \citenamefont {Mow}(1990)}]{HOLMES1990}%
      \BibitemOpen
      \bibfield  {author} {\bibinfo {author} {\bibfnamefont {M.H.}\ \bibnamefont {Holmes}}\ and\ \bibinfo {author} {\bibfnamefont {V.C.}\ \bibnamefont {Mow}},\ }\bibfield  {title} {\enquote {\bibinfo {title} {The nonlinear characteristics of soft gels and hydrated connective tissues in ultrafiltration},}\ }\href@noop {} {\bibfield  {journal} {\bibinfo  {journal} {Journal of Biomechanics}\ }\textbf {\bibinfo {volume} {23}},\ \bibinfo {pages} {1145--1156} (\bibinfo {year} {1990})}\BibitemShut {NoStop}%
    \bibitem [{\citenamefont {Weideman}\ and\ \citenamefont {Reddy}(2000)}]{weideman-acmtoms-2000}%
      \BibitemOpen
      \bibfield  {author} {\bibinfo {author} {\bibfnamefont {J.~A.}\ \bibnamefont {Weideman}}\ and\ \bibinfo {author} {\bibfnamefont {S.~C.}\ \bibnamefont {Reddy}},\ }\bibfield  {title} {\enquote {\bibinfo {title} {A {MATLAB} differentiation matrix suite},}\ }\href@noop {} {\bibfield  {journal} {\bibinfo  {journal} {ACM Transactions on Mathematical Software (TOMS)}\ }\textbf {\bibinfo {volume} {26}},\ \bibinfo {pages} {465--519} (\bibinfo {year} {2000})}\BibitemShut {NoStop}%
    \bibitem [{\citenamefont {Shampine}\ and\ \citenamefont {Reichelt}(1997)}]{shampine-siamjscicomput-1997}%
      \BibitemOpen
      \bibfield  {author} {\bibinfo {author} {\bibfnamefont {L.~F.}\ \bibnamefont {Shampine}}\ and\ \bibinfo {author} {\bibfnamefont {M.~W.}\ \bibnamefont {Reichelt}},\ }\bibfield  {title} {\enquote {\bibinfo {title} {The {MATLAB} {ODE} suite},}\ }\href@noop {} {\bibfield  {journal} {\bibinfo  {journal} {SIAM Journal on Scientific Computing}\ }\textbf {\bibinfo {volume} {18}},\ \bibinfo {pages} {1--2} (\bibinfo {year} {1997})}\BibitemShut {NoStop}%
    \end{thebibliography}
 
%
    
\end{document}